\documentclass[%
 reprint,
 superscriptaddress,
 amsmath,amssymb,
 aps,
]{revtex4-2}

\usepackage{amsmath}
\usepackage{amsthm}
\usepackage{quantikz}
\usepackage{graphicx}
\usepackage[svgnames]{xcolor}
\usepackage{dcolumn}
\usepackage{bm}
\usepackage{braket}
\usepackage{chemformula}
\usepackage{afterpage}
\usepackage{makecell}
\usepackage{multirow}
\usepackage{booktabs}
\usepackage{algpseudocode}
\usepackage{algorithm}
\usepackage{siunitx}
\usepackage{enumitem} 
\usepackage{placeins}
\usepackage{longtable}

\let\ce\ch

\usepackage[
     colorlinks        = true,
     unicode           = true,
     pdfstartview      = FitV,
     linktocpage       = true,
     linkcolor         = RoyalBlue,
     citecolor         = RoyalBlue,
     urlcolor          = RoyalBlue,
     bookmarks         = true,
     bookmarksnumbered = true,
     breaklinks=true,
     pdftitle={unitary_weight_concentration},
     pdfauthor={}
]{hyperref}
\newcommand{\fujitsu}{Quantum Laboratory, Fujitsu Research, Fujitsu Limited, 4-1-1 Kamikodanaka, Nakahara, Kawasaki, Kanagawa 211-8588, Japan}
\newcommand{\osaka}{Fujitsu Quantum Computing Joint Research Division, Center for Quantum Information and Quantum Biology, Osaka University, 1-2 Machikaneyama, Toyonaka, Osaka, 565-8531, Japan}

\newcommand{\abs}[1]{\left| #1 \right|}

\begin{document}

\preprint{APS/123-QED}

\title{Enabling Chemically Accurate Quantum Phase Estimation in the Early Fault-Tolerant Regime via Unitary Weight Concentration}

\author{Shota Kanasugi}
 \email{kanasugi.shota@fujitsu.com}
 \affiliation{\fujitsu}
 \affiliation{\osaka}
 
\author{Riki Toshio}%
 \affiliation{\fujitsu}
 \affiliation{\osaka}

\author{Kazunori Maruyama}%
 \affiliation{\fujitsu}
 \affiliation{\osaka}

\author{Hirotaka Oshima}%
 \affiliation{\fujitsu}
 \affiliation{\osaka}

\date{\today}

\begin{abstract}
Quantum simulation of molecular electronic structure is one of the most promising applications of quantum computing. However, achieving chemically accurate predictions for strongly correlated systems requires quantum phase estimation (QPE) on fault-tolerant quantum computing (FTQC) devices. Existing resource estimates for typical FTQC architectures suggest that such calculations demand millions of physical qubits, thereby placing them beyond the reach of near-term devices. 
Here, we investigate the feasibility of performing QPE for chemically relevant molecular systems in an early-FTQC regime, characterized by partial fault tolerance, constrained qubit budgets, and limited circuit depth.  
Our framework is based on single-ancilla, Trotter-based QPE implementations combined with partially randomized time evolution. Within this framework, we develop a novel Hamiltonian optimization strategy, termed unitary weight concentration, that reduces algorithmic cost by reshaping linear-combination-of-unitaries representations.
Applying this framework to active-space models of iron–sulfur clusters, cytochrome P450 active sites, and \ce{CO2}-utilization catalysts, we perform end-to-end resource estimation using the latest version of the space-time efficient analog rotation (STAR) architecture. 
Our results indicate that ground-state energy estimation for active spaces of approximately 20--50 spatial orbitals—well beyond the reach of classical full configuration interaction—is achievable using $\sim 10^5$ physical qubits per processor, with runtimes on the order of days to weeks. 
These findings demonstrate that while full-fledged fault-tolerant quantum computers will be required for even larger molecular simulations, chemically meaningful quantum chemistry problems are already within reach in an experimentally relevant, early-FTQC regime.
\end{abstract}

\maketitle


\section{Introduction \label{sec:intro}}
Quantum simulation of molecular electronic structures represents one of the most compelling applications of quantum computing.
At its core, this promise stems from the fact that strongly correlated molecules often require a multireference description that treats a large set of partially occupied orbitals on an equal footing.
In such regimes, complete active-space approaches must contend with an exponentially growing configuration space to capture near-degeneracy and many-body entanglement effects~\cite{jensen2017introduction}.
This exponential scaling ultimately limits the size of active spaces that can be treated exactly on classical computers, restricting full configuration interaction (full-CI) calculations to roughly 20 electrons in 20 (spatial) orbitals~\cite{Vogiatzis2017-fc,Gao2024-ac}, a scale often insufficient for chemically relevant transition-metal complexes and catalytic intermediates.

To extend applicability beyond this regime, classical heuristic approaches such as coupled-cluster theory~\cite{bartlett2007coupled}, selected configuration interaction methods such as heat-bath configuration interaction (HCI)~\cite{Holmes2016-jd,Sharma2017-sg}, and tensor-network methods like the density matrix renormalization group (DMRG)~\cite{chan2011density,baiardi2020density} have been developed. 
However, their accuracy often relies on heuristic assumptions about wavefunction structure, and the error remaining at finite truncation is generally assessed through convergence studies or extrapolation rather than being rigorously bounded, which limits their reliability at the level of chemical accuracy required for predicting reaction energetics and catalytic processes~\footnote{In this work, we use the term chemical accuracy in the conventional sense of achieving energy errors of $1.6\, \mathrm{mHa} \simeq 1\, \mathrm{kcal/mol}$.}.
These limitations motivate the exploration of quantum-computational approaches that offer a path toward controllable and certifiable accuracy beyond the classical frontier~\cite{cao2019quantum,mcardle2020quantum,bauer2020quantum}.

Recent progress in noisy intermediate-scale quantum (NISQ) hardware has enabled proof-of-principle demonstrations of quantum–classical hybrid algorithms, including the variational quantum eigensolver~\cite{Peruzzo2014-tu} and quantum-selected configuration interaction~\cite{Kanno2023-ek} (also known as sample-based quantum diagonalization~\cite{Robledo-Moreno2025-ve}). These approaches aim to circumvent the exponential scaling of classical exact methods by leveraging quantum state preparation and measurement. However, their achievable accuracy strongly depends on ansatz design, optimization heuristics, and hardware noise. Moreover, their scalability to larger systems remains limited due to issues such as barren plateaus~\cite{Larocca2025-gn}, measurement overhead~\cite{Reinholdt2025-ge}, and increasing circuit complexity~\cite{Hagelueken2026-jn}. Crucially, they generally do not provide certified guarantees of achieving chemical accuracy, making it difficult to establish a definitive quantum advantage for high-precision quantum chemistry. 

In contrast, quantum phase estimation (QPE)~\cite{kitaev1995quantum} provides a systematically controllable framework for ground-state energy estimation. By projecting onto exact eigenstates within a chosen active space, QPE enables energy calculations with explicitly quantifiable error bounds. 
As such, QPE offers a principled route toward chemical accuracy for a chosen Hamiltonian: the target error enters as an algorithmic parameter, whereas in classical heuristic approaches the remaining error is assessed rather than bounded. This distinction is material in regimes where quantitatively reliable energy predictions are required~\cite{Lee2023-rd}.
However, implementing QPE at scale requires fault-tolerant quantum computing (FTQC) that demands millions of physical qubits and substantial space–time overhead for simulating chemically relevant molecular systems, judging from existing resource estimates for typical FTQC architectures.
For instance, prior analyses of large transition-metal active sites, such as the iron-molybdenum co-factor (FeMoco), suggest total resource requirements exceeding $10^6$ physical qubits under standard surface-code assumptions~\cite{Reiher2017-bd,Berry2019-sy, Von_Burg2021-du,Lee2021-tz,Rocca2024-yx,Caesura2025-tc,Low2025-tb}. This places such calculations firmly beyond the reach of near-term quantum hardware.

Recent architectural proposals and experimental advances in quantum error correction suggest the emergence of an intermediate regime of early fault tolerance, bridging NISQ and full-fledged FTQC devices~\cite{Piveteau2021-tc,Suzuki2022-fz,Akahoshi2024-hj,Toshio2025-nn,Akahoshi2025-ra}. In this regime, quantum processors support encoded logical qubits with imperfect error suppression~\cite{Piveteau2021-tc,Suzuki2022-fz}. Rather than relying on large-scale magic-state distillation and deep circuit decompositions, these architectures aim to reduce space–time overhead through tailored state-preparation protocols and hardware-aware implementations of non-Clifford operations~\cite{Akahoshi2024-hj,Toshio2025-nn,Akahoshi2025-ra,Chung2026-uc}. Although such devices do not yet achieve full-fledged fault-tolerant operation, they represent a qualitatively distinct computational regime beyond purely noisy hardware.

A central open question is whether QPE-based electronic-structure calculations for chemically relevant molecules can become practical within this early-FTQC regime, and what resource trade-offs such feasibility would entail. Early-FTQC architectures operate under stringent constraints on logical qubit counts, circuit depth, and tolerable logical-error accumulation. Consequently, feasibility cannot be inferred by simply rescaling resource estimates derived for full-fledged FTQC architectures. Instead, assessing practicality in this regime necessitates an explicit co-design of algorithms, Hamiltonian representations, and architectural constraints imposed by error correction. Determining the parameter regimes in which these elements can be reconciled constitutes the central objective of the present study. 

To this end, we introduce \emph{unitary weight concentration} (UWC), a Hamiltonian-optimization technique, and demonstrate the feasibility of performing ground-state energy estimation for chemically relevant molecular active-space models on early-FTQC devices. 
The key constraint in the architecture considered here is not only the available physical-qubit count but also the maximum spacetime volume of a single coherent execution. Since non-Clifford rotations are error-mitigated rather than implemented using distilled magic states, individual circuits must remain short. Within this regime, our estimates require $\mathcal{O}(10^{5})$ physical qubits per QPU.
The UWC method restructures the linear-combination-of-unitaries (LCU) Hamiltonian representation via \textit{spectrally-invariant} Hamiltonian transformations, concentrating the coefficient distribution while leaving the eigenvalue spectrum in the target symmetry sector unchanged. 
In practice, this is achieved by combining orbital optimization (OO)~\cite{Koridon2021-xm,Ollitrault2024-ko} and block-invariant symmetry shift (BLISS)~\cite{loaiza2013_bliss,Patel2025-np} transformations.
This leads to reductions in both overall gate cost and circuit depth when combined with \emph{partially randomized} product-formula evolution~\cite{Ouyang2020-aw,Hagan2023-qe,Jin2025-ht,Rajput2022-xi,Gunther2025-zk}. 
Across the molecular systems considered, UWC consistently provides an additional order-of-magnitude reduction in total simulation cost beyond partial randomization alone.
In this study, we adopt single-ancilla, Trotter-based QPE implementations~\cite{Dobsicek2007-bf,Wiebe2016-rp,Somma2019-fc,Lin2022-wm,Ding2023-bo,Ni2023-it} compatible with limited qubit budgets, bounded circuit depth, and partial fault tolerance. These implementations therefore depart from the qubitization-based approaches~\cite{Low2019-op,Babbush2018-bs,Berry2019-sy,Von_Burg2021-du,Lee2021-tz,Rocca2024-yx,Caesura2025-tc,Low2025-tb} typically assumed in resource estimation studies for full-fledged FTQC architectures.

We apply this framework to molecular models of practical relevance, including iron-sulfur clusters~\cite{Beinert1997-ok,Lee2023-rd,Ollitrault2024-ko}, cytochrome P450 active-site models~\cite{Goings2022-so}, and ruthenium-based catalytic complexes for \ce{CO2} utilization~\cite{Von_Burg2021-du}. For each system, we construct UWC-optimized Hamiltonian representations and perform end-to-end resource estimation assuming execution on the \emph{space-time efficient analog rotation} (STAR) architecture~\cite{Akahoshi2024-hj,Toshio2025-nn,Akahoshi2025-ra,Chung2026-uc}, incorporating the latest improvements of logical rotation gate operations, termed \emph{STAR-magic mutation} (SMM)~\cite{Toshio2026}. For each molecular instance, we quantify physical qubit requirements, maximum per-shot runtime, total time-to-solution, and achievable parallelism under fixed qubit-budget assumptions. Our results indicate that tractable active spaces range from 20 to 50 orbitals using $\sim10^5$ physical qubits, with total runtimes on the order of days to weeks, while larger systems remain beyond reach in this regime. 
Importantly, this qubit scale appears to fall within the anticipated single-cryostat range discussed in recent superconducting architectural proposals~\cite{Mohseni2024-ji}, potentially avoiding distributed FTQC overheads~\cite{Caleffi2024-kw}.
Although molecular systems in this active-space range remain smaller than those typically envisioned for full-fledged FTQC~\cite{Babbush2025-id,Goings2022-so,Reiher2017-bd,Berry2019-sy,Von_Burg2021-du,Lee2021-tz,Rocca2024-yx,Caesura2025-tc,Low2025-tb}, they already extend into a regime that is not only inaccessible to classical full-CI methods~\cite{Vogiatzis2017-fc,Gao2024-ac} but also encompasses chemically and industrially relevant molecular systems~\cite{Beinert1997-ok,Goings2022-so,Von_Burg2021-du,Bellonzi2024-ku,Zhou2025-tb,Gunther2025-td}. Therefore, they represent a scientifically meaningful target for early fault-tolerant quantum simulation. Together, these results delineate a practically meaningful region between classical exact computations and millions-of-qubit full-fledged FTQC and provide a quantitative feasibility map that clarifies the architectural and algorithmic targets for early-FTQC in quantum chemistry.

The rest of the paper is organized as follows. In Sec.~\ref{sec:single_ancilla_prpe}, we introduce single-ancilla QPE and partially randomized product formulas as background for our work. In Sec.~\ref{sec:hamiltonian}, we describe Hamiltonian representations and our optimization strategy, UWC, for minimizing the partially randomized simulation cost for molecular electronic structure Hamiltonians. In Sec.~\ref{sec:numerical}, we present numerical results demonstrating the effectiveness of UWC in reducing the simulation cost of partially randomized QPE for chemically relevant molecular models. In Sec.~\ref{sec:resource_estimation}, we assess the feasibility of practical chemistry simulation on early-FTQC hardware by performing comprehensive resource estimation for a range of molecular models on an improved variant of the STAR architecture, SMM. Finally, in Sec.~\ref{sec:conclusion}, we conclude our discussion and outline future directions toward realizing quantum chemistry simulation in the era of early fault tolerance.

\section{Single-ancilla quantum phase estimation and partially randomized time evolution\label{sec:single_ancilla_prpe}}
In this section, we briefly review the single-ancilla QPE framework and the partially randomized time-evolution schemes upon which our work is built, largely following the presentation in Ref.~\cite{Gunther2025-zk}.

\subsection{Single-ancilla quantum phase estimation}
\label{subsec:single_ancilla_rpe}

First, we review a single-ancilla QPE framework tailored for early fault-tolerant quantum computers. We consider Hamiltonians of the form
\begin{equation}
  \hat{H} = \sum_{\ell=1}^{L} c_\ell \hat{P}_\ell,
  \label{eq:pauli_ham}
\end{equation}
where $\hat{P}_\ell \in \{I, X, Y, Z\}^{\otimes N_q}$ are $N_q$-qubit Pauli string operators and $c_\ell \in \mathbb{R}$. 
Although our discussion can be generalized to other LCU representations (e.g., the double-factorized Hamiltonian representation~\cite{Motta2021-kh}), we focus on this Pauli LCU representation throughout this work. 
This is because the Trotterized time evolution under the Pauli LCU representation can be implemented using small-angle Pauli rotations that are well suited to the early-FTQC architecture considered in Sec.~\ref{sec:resource_estimation}, whereas alternative LCU forms typically involve larger-angle operations, such as Givens rotations in the double-factorized Trotterization~\cite{Motta2021-kh}.
For electronic structure Hamiltonians expressed in a basis of $N$ spatial orbitals, the number of qubits is $N_q = 2N$ and the number of terms is $L = \mathcal{O}(N^4)$. In the context of Hamiltonians specified by Eq.~\eqref{eq:pauli_ham}, we define the total weight or $\ell_1$-norm as
\begin{equation}
  \lambda \coloneqq \sum_{\ell=1}^{L} |c_\ell|. 
  \label{eq:lambda_def}
\end{equation}
Let $\{E_k, \ket{\psi_k}\}$ denote the eigenpairs of $\hat{H}$, ordered such that $E_0 \le E_1 \le \cdots \le E_{d}$, where $d=2^{N_q}$. 
We assume access to an \emph{initial state} $\ket{\psi}$ with a nonzero overlap with the ground state: $p_0 \coloneqq|\braket{\psi_0 | \psi}|^2 \geq \eta$ for some known lower bound $\eta>0$. 
The preparation of such a suitable initial state is a crucial research area in its own right~\cite{Tubman2018-lg,Lee2023-rd,Fomichev2024-us,Ollitrault2024-ko,Morchen2024-ts,Erakovic2025-cl,Berry2025-dq}. In this work, however, we focus on the implementation cost of single-ancilla QPE itself and do not detail a specific state preparation protocol.

The single-ancilla QPE formulation we employ is based on the Hadamard test~\cite{Dobsicek2007-bf,Wiebe2016-rp,Somma2019-fc,Lin2022-wm,Ding2023-bo,Ni2023-it}. Starting from the state $\ket{\psi}$ on the data register and $\ket{0}$ on an ancilla qubit, the Hadamard test applies a Hadamard gate $\mathrm{H}$ to the ancilla, a controlled time-evolution unitary $e^{-it\hat{H}}$ acting on the data register, and a final rotation and measurement on the ancilla qubit (see Fig.~\ref{fig:measurement_circuit}). The expectation value of the complex-valued outcome of the Hadamard test $\bm{Z}(t)$ is then expressed as
\begin{equation}
  g(t) \coloneqq \mathbb{E}[\bm{Z}(t)] = \bra{\psi} e^{-it\hat{H}} \ket{\psi}
  = \sum_k p_k e^{-it E_k},
  \label{eq:g_t_def}
\end{equation}
where $p_k = |\braket{\psi_k|\psi}|^2$. Thus, $g(t)$ is a complex signal whose frequency components correspond to the eigenenergies $\{ E_k \}$. By sampling $\bm{Z}(t)$ at a set of times $\{t_m\}$ and applying an appropriate signal-processing routine, one can estimate $E_0$ to precision $\epsilon$ with Heisenberg-limited scaling in the total evolution time~\cite{Lin2022-wm,Ding2023-bo,Ni2023-it}.

\begin{figure}[tbp]
    \begin{quantikz}[thin lines]
      \lstick{{$\ket{0}$}} & \gate{\mathrm{H}} & \ctrl{1} & \gate{W} & \gate{\mathrm{H}} & \meter{} \\
      \lstick{{$\ket{\psi}$}} & \qwbundle{N_q} & \gate{e^{-it\hat{H}}} & & & \\
    \end{quantikz}
    \vspace{0.2cm}
    \caption{Circuit for the Hadamard test to measure the expectation value $\bra{\psi}e^{-it\hat{H}}\ket{\psi}$. The real part is obtained by setting $W=I$ (identity) and measuring the ancilla in the $X$ basis. The imaginary part is obtained by setting $W=S^{\dag}$ and measuring in the $Y$ basis.}
    \label{fig:measurement_circuit}
\end{figure}

In this work, we specifically use the \emph{robust phase estimation} (RPE) scheme of Ref.~\cite{Ni2023-it}, as reviewed and analyzed in Ref.~\cite{Gunther2025-zk}. RPE chooses a sequence of evolution times $t_m = 2^m$ for $m=0,1,\dots,M$, with $M = \lceil \log_2(\epsilon^{-1}) \rceil$, and uses the outcomes of the Hadamard test at each time to iteratively refine an estimate of $E_0$. The details of the procedure are summarized in Appendix~\ref{append:rpe}. For a constant ground-state overlap $\eta$ above a known threshold $4-2\sqrt{3}\simeq 0.54$, this scheme achieves Heisenberg-limited scaling in its maximal evolution time $T_{\max} = \max{\{t_m\}}= \mathcal{O}(\epsilon^{-1})$ and total evolution time across all circuit executions, $T_{\mathrm{total}} = \sum_m N_m t_m= \mathcal{O}(\epsilon^{-1})$~\cite{Ni2023-it}. Here, $N_m$ denotes the number of samples for the $m$-th round of the Hadamard test measurement.

Crucially for our purposes, when the initial state has a high fidelity to the ground state, i.e., $\eta = 1 - \omega$ with $\omega\ll 1$, RPE admits a trade-off between maximal and total evolution times parameterized by $\xi = \mathcal{O}(\omega)$ as follows~\cite{Ni2023-it}:
\begin{equation}
  T_{\max} = \mathcal{O}\!\left(\frac{\xi}{\epsilon}\right),
  \quad
  T_{\mathrm{total}} = \mathcal{O}\!\left(\frac{1}{\xi\,\epsilon}\right).
  \label{eq:rpe_trade-off}
\end{equation}
Therefore, decreasing $\xi$ (enabled by a higher-fidelity initial state) reduces the maximum required evolution time at the cost of increasing the total evolution time. In our early-FTQC setting, $T_{\max}$ is directly related to the maximum circuit depth per shot, while $T_{\mathrm{total}}$ controls the total number of gate layers accumulated across all shots. 
This parameter $\xi$ thus acts as a tunable knob that determines the trade-off between per-shot circuit depth and overall sampling overhead. 
The precise theoretical guarantees and algorithmic procedure we use are provided in Appendix~\ref{append:rpe}.

\subsection{Partially randomized time evolution}
\label{subsec:pr_time_evolution}

The RPE framework described above assumes access to the exact time evolution unitary $e^{-it\hat{H}}$.
In practice, we approximate this unitary using product formulas.
Following Ref.~\cite{Gunther2025-zk}, we distinguish three approaches:
(i) deterministic product formulas,
(ii) fully randomized product formulas, and
(iii) \emph{partially randomized} product formulas that interpolate between these extremes.

\subsubsection{Deterministic product formula \label{subsubsec:det}}
Deterministic product formulas, often called Trotterization~\cite{Trotter1959-uc,Suzuki1990-pr}, approximate the time-evolution unitary for a small timestep $\delta$ as
\begin{equation}
  e^{-i\delta\hat{H}} \approx \hat{S}_p(\delta),
\end{equation}
where $\hat{S}_p(\delta)$ is a $p$-th-order Suzuki-Trotter formula constructed from exponentials of the individual terms $\hat{H}_\ell:=c_\ell \hat{P}_\ell$. For instance, the second-order Suzuki-Trotter product formula is 
\begin{align}
    \hat{S}_{2}(\delta) &= \prod_{\ell=1}^{L}e^{-i\frac{\delta}{2}\hat{H}_\ell} \prod_{\ell=L}^{1}e^{-i\frac{\delta}{2} \hat{H}_{\ell}},
    \label{eq:2nd_order_trotter}
\end{align}
where the terms in the first product are ordered from $\ell=1$ to $L$, and in the reverse order for the second product. To simulate the evolution for a total time $t$, this formula is applied $r$ times with a step size of $\delta = t/r$:
\begin{equation}
  e^{-it\hat{H}} = (e^{-i\delta\hat{H}})^r \approx \hat{S}_p(\delta)^r.
  \label{eq:trotter_step}
\end{equation}
The resulting simulation error can be characterized either by its operator norm or, more relevantly for QPE, by the induced bias in the ground-state energy~\cite{Gunther2025-zk,Babbush2015-ua,Reiher2017-bd}:
\begin{align}
    |E_0-E_{\mathrm{eff},0}| \leq C_{\rm gs}\delta^{p}.
    \label{eq:trotter_error_energy}
\end{align}
Here, $E_{\mathrm{eff},0}$ is the ground-state energy of an effective Hamiltonian $\hat{H}_{\rm eff}$ defined by $\hat{S}_p(\delta)=e^{-i\delta\hat{H}_{\rm eff}}$, and $C_{\rm gs}=C_{\rm gs}(p,\{\hat{H}_\ell\})$ is a Trotter constant that depends on the order $p$ and the chosen permutation of the terms $\{\hat{H}_\ell\}$. While rigorous operator-norm bounds~\cite{Childs2021-qk} suggest $C_{\rm gs} = \mathcal{O}(\lambda^{p+1})$, it is well known that such worst-case bounds significantly overestimate the actual Trotter error in many practical scenarios~\cite{Sahinoglu2021-si,Chen2024-fd,Zhao2022-kc}. To avoid such overestimation, our resource estimates in Secs.~\ref{sec:numerical} and~\ref{sec:resource_estimation} utilize empirical values for the Trotter error constant $C_{\rm gs}$ reported in Ref.~\cite{Gunther2025-zk} (see Appendix~\ref{append:prnd} for details).

To estimate the ground-state energy $E_0$ with precision $\epsilon$, the Trotter-induced energy bias in Eq.~\eqref{eq:trotter_error_energy} must be suppressed to $\mathcal{O}(\epsilon)$, which requires a step size $\delta=\mathcal{O}((\epsilon/C_{\rm gs})^{1/p})$. The cost per Trotter step is $\mathcal{O}(L)$ gates for a Hamiltonian with $L$ terms. Combining these factors with the RPE time requirements from Eq.~\eqref{eq:rpe_trade-off}, the maximum gate count per shot (determined by $T_{\max}$) and the total gate count across all shots (determined by $T_{\mathrm{total}}$) can be estimated. Specifically, the maximum gate count per shot is $\mathcal{O}(L \cdot T_{\max}/\delta) = \mathcal{O}(L\xi C_{\rm gs}^{1/p}\epsilon^{-1-1/p})$, and the total gate count is $\mathcal{O}(L \cdot T_{\mathrm{total}}/\delta) = \mathcal{O}(L\xi^{-1} C_{\rm gs}^{1/p}\epsilon^{-1-1/p})$.

\subsubsection{Randomized product formula \label{subsubsec:rand}}
Randomized product formulas~\cite{Campbell2019-of,Chen2021-gs,Wan2022-hz,Kiss2023-fg,Granet2024-gq,Kiumi2025-qm}, such as the quantum stochastic drift (qDRIFT) protocol~\cite{Campbell2019-of}, randomized Taylor expansion (RTE)~\cite{Wan2022-hz}, and time-evolution by probabilistic angle interpolation (TE-PAI)~\cite{Kiumi2025-qm}, simulate Hamiltonian evolution by stochastically sampling Hamiltonian terms $\{\hat{H}_\ell\}$ according to a probability distribution derived from their weights $\{ |c_\ell| \}$. This approach leads to gate counts that scale with the total weight $\lambda$ given in Eq.~\eqref{eq:lambda_def}, rather than the number of terms $L$.

Specifically, the Hamiltonian is rewritten to define a probability distribution $\{b_\ell\}$:
\begin{align}
    \hat{H} = \lambda\sum_{\ell=1}^{L}b_\ell \hat{P}_\ell', \quad b_\ell = \frac{|c_\ell|}{\lambda},
\end{align}
where the sign of each coefficient $c_\ell$ is absorbed into the corresponding Pauli string, $\hat{P_\ell}' \to \mathrm{sgn}(c_\ell)\hat{P_\ell}$. The qDRIFT protocol approximates the time evolution $e^{-it\hat{H}}$ by composing $r$ stochastic steps. For a total time $t$, this corresponds to simulating the rescaled evolution $(e^{-i\tau(\hat{H}/\lambda)})^r$, where $\tau = \lambda t/r$ is an effective timestep.
The key idea is to implement a unitary $\hat{U}_\ell$ in each step, sampled with probability $b_\ell$, such that the expectation value over the sampling correctly reproduces the target short-time evolution. That is, we require
\begin{align}
    \mathbb{E}[\hat{U}_\ell] = \sum_{\ell=1}^{L} b_\ell \hat{U}_\ell \approx e^{-i\tau(\hat{H}/\lambda)}.
\end{align}
A common choice is $\hat{U}_\ell = e^{i\phi\hat{P}_\ell'}$, where $\phi$ is a fixed rotation angle~\cite{Campbell2019-of}. The expectation value then becomes
\begin{align}
    \sum_{\ell=1}^{L} b_\ell e^{i\phi\hat{P}_\ell'} &= \sum_{\ell=1}^{L} b_\ell (\cos\phi \,\hat{I} + i\sin\phi\, \hat{P}_\ell') \nonumber\\
    &= \cos\phi \,\hat{I} + i\sin\phi \,(\hat{H}/\lambda),
\end{align}
where $\hat{I}$ is the identity operator. 
By comparing this with the Taylor expansion $e^{-i\tau(\hat{H}/\lambda)} = I - i\tau(\hat{H}/\lambda) + \mathcal{O}(\tau^2)$, we see that the two expressions become proportional if we set $\phi = -\arctan(\tau)$~\cite{Campbell2019-of,Wan2022-hz,Gunther2025-zk}. The full evolution is then approximated by applying a sequence of $r$ randomly selected unitaries. Specifically, for a sequence of random indices $\vec{\ell} = (\ell_1, \dots, \ell_r)$ sampled with the probability $q_{\vec{\ell}}=b_{\ell_1}\cdots b_{\ell_r}$, the corresponding unitary is $\hat{W}_{\vec{\ell}}=e^{i\phi\hat{P}'_{\ell_1}}\cdots e^{i\phi\hat{P}'_{\ell_r}}$.

When this stochastic evolution is used in the Hadamard test, the expectation value of the measurement outcome $\bar{\bm{Z}}_r(t)$ is given by
\begin{align}
    \bar{g}_r(t) := \mathbb{E}[\bar{\bm{Z}}_r(t)] &= \sum_{\vec{\ell}} q_{\vec{\ell}} \bra{\psi}\hat{W}_{\vec{\ell}} \ket{\psi}. \label{eq:hadamard_randomized}
\end{align}
As shown in Ref.~\cite{Gunther2025-zk}, this expression evaluates to
\begin{align}
    \bar{g}_r(t)
    &= (1+\tau^2)^{-r/2}\bra{\psi}\left(\hat{I} - i\tau(\hat{H}/\lambda) \right)^{r} \ket{\psi} \nonumber\\
    &= (1+\tau^2)^{-r/2}\sum_k p_k\left(1-i\tau( E_k/\lambda) \right)^r \nonumber\\
    &\approx \sum_{k}p_k e^{-\frac{1}{2}r\tau^2(1-(E_k/\lambda)^2)} e^{-it E_k},
    \label{eq:qdrift_signal}
\end{align}
where the approximation holds for small $\tau$ (and thus small $\phi$). The resulting signal $\bar{g}_r(t)$ is similar to the exact signal $g(t)$ in Eq.~\eqref{eq:g_t_def}, but is suppressed by an eigenvalue-dependent damping factor. To prevent the sampling overhead from increasing exponentially with time, this damping must be kept constant, i.e., $\tau^2 r = \mathcal{O}(1)$. This condition implies that the number of required Pauli rotations is $r=\mathcal{O}(\lambda^2t^2)$, with each rotation having a fixed angle of $\phi \approx -\tau = \mathcal{O}((\lambda t)^{-1})$.

Other randomized methods, such as RTE~\cite{Wan2022-hz} and TE-PAI~\cite{Kiumi2025-qm}, also exhibit quadratic scaling in cost with $\lambda t$. Their primary advantage over qDRIFT is that the resulting signal $\bar{g}_r(t)$ is unbiased, and has a damping factor that is independent of the eigenvalues $E_k$, i.e., $\bar{g}_r(t) \approx B(t)^{-1} g(t)$~\cite{Gunther2025-zk,Kiumi2025-qm}. The sampling overhead $B(t)^2$ can be suppressed to $\mathcal{O}(1)$ by choosing $r=\mathcal{O}(\lambda^2 t^2)$, leading to the same overall cost scaling with qDRIFT.

While randomized product formulas remove the explicit dependence on $L=\mathcal{O}(N^4)$ seen in Trotterization, their cost scales quadratically with the simulation time $t$. This leads to an unfavorable scaling for QPE. Specifically, applying the results of Ref.~\cite{Gunther2025-zk}, RPE based on randomized product formulas results in a maximum gate count per shot of $\mathcal{O}(\lambda^2\xi^2\epsilon^{-2})$ and a total gate count of $\mathcal{O}(\lambda^2\epsilon^{-2})$.

\subsubsection{Partially randomized product formula \label{subsubsec:part_rand}}
Partially randomized product formulas~\cite{Gunther2025-zk,Ouyang2020-aw,Hagan2023-qe,Jin2025-ht,Rajput2022-xi} are designed to exploit the empirical structure of electronic structure Hamiltonians. In these Hamiltonians, a relatively small number of terms with large coefficients typically dominate the total weight $\lambda$, while the remaining terms form a long tail of smaller contributions (see Fig.~\ref{fig:coeff_distribution} for an example). The core idea is to treat the dominant, large-weight terms deterministically using Trotterization, while stochastically sampling the numerous small-weight terms. This approach aims to combine the favorable scaling of deterministic methods with respect to evolution time and the insensitivity of randomized methods to the number of terms $L$. The analysis in Ref.~\cite{Gunther2025-zk} shows that, when combined with RPE, this hybrid scheme can yield significantly improved performance compared to both purely deterministic and fully randomized approaches, especially for molecular Hamiltonians with highly inhomogeneous coefficient distributions.

The partially randomized simulation framework begins by partitioning the Hamiltonian~\eqref{eq:pauli_ham} into a deterministic part $\hat{H}_D$ and a randomized part $\hat{H}_R$:
\begin{align}
    \hat{H} = \underbrace{\sum_{\ell=1}^{L_D}\hat{H}_\ell}_{=:\hat{H}_D} + \underbrace{\sum_{\ell=L_D+1}^{L}\hat{H}_\ell}_{=:\hat{H}_R},
    \label{eq:ham_partially_randomized}
\end{align}
where the terms $\hat{H}_\ell (= c_\ell \hat{P}_\ell)$ are assumed to be sorted in descending order of their coefficient magnitudes, $|c_1| \ge |c_2| \ge \cdots \ge |c_L|$. Here, $\hat{H}_D$ comprises the $L_D$ terms with the largest weights, while $\hat{H}_R$ contains the remaining small-weight terms.

The evolution is then approximated using a Trotter formula applied to this partitioned Hamiltonian. For example, using a second-order formula, a target time evolution $e^{-it\hat{H}}$ is approximated as
\begin{align}
    e^{-it\hat{H}} \approx \hat{S}_{2}(\delta)^r = \left(\prod_{\ell=1}^{L_D}e^{-i\frac{\delta}{2}\hat{H}_\ell} e^{-i\delta \hat{H}_R} \prod_{\ell=L_D}^{1}e^{-i\frac{\delta}{2}\hat{H}_{\ell}}\right)^r,
    \label{eq:2nd_order_part_rand}
\end{align}
where $\delta=t/r$. 
In this scheme, the evolution $e^{-i\delta \hat{H}_R}$ corresponding to the randomized part is implemented stochastically, as described in Sec.~\ref{subsubsec:rand}.
This introduces a stochastic element into the simulation. As with the fully randomized method, performing a Hadamard test yields a signal that reproduces the desired expectation value $\bra{\psi} \hat{S}_2(\delta)^{r} \ket{\psi}$, but with a damping factor that introduces additional sampling overhead. 
This overhead can be suppressed to $\mathcal{O}(1)$ by choosing the number of steps to scale as $r = \mathcal{O}(\lambda_R^2 \delta^2)$~\cite{Gunther2025-zk}, where $\lambda_R$ is the total weight of the randomized part:
\begin{align}
    \lambda_R := \sum_{\ell=L_D+1}^{L}|c_\ell|.
    \label{eq:lambda_rnd}
\end{align}
The overall Trotter error from the deterministic part imposes an additional constraint on the step size $\delta$. Combining these considerations, the total cost inherits features from both methods.

The overall simulation costs for partially randomized RPE are a hybrid of the deterministic and randomized costs. 
As derived in Ref.~\cite{Gunther2025-zk}, the maximum number of Pauli rotations per shot and the total number of rotations across all shots are given by
\begin{align*}
    \text{Max per-shot cost:} \quad &\mathcal{O}(L_D\xi C_{\rm gs}^{1/p}\epsilon^{-1-1/p}+\lambda_R^2\xi^2\epsilon^{-2}), \\
    \text{Total cost:} \quad &\mathcal{O}(L_D\xi^{-1} C_{\rm gs}^{1/p}\epsilon^{-1-1/p}+\lambda_R^2\epsilon^{-2}).
\end{align*}
These expressions clearly show that partial randomization is effective when a partition exists such that both $L_D \ll L$ and $\lambda_R \ll \lambda$. For comparison, the simulation costs for the three approaches are summarized in Table~\ref{tab:part_rand}.

\begin{table*}[tbp]
\caption{Comparison of asymptotic simulation costs for the RPE algorithm using different time-evolution methods: deterministic ($p$-th order Suzuki-Trotter), fully randomized, and partially randomized product formulas, according to Ref.~\cite{Gunther2025-zk}. The costs are expressed in terms of Pauli rotations per shot and total Pauli rotations across all shots, indicating their dependencies on Hamiltonian parameters ($L$, $L_D$, $\lambda$, $\lambda_R$, $C_{\rm gs}$), RPE parameters ($\xi$, $\epsilon$), and Trotter order ($p$).
}
\label{tab:part_rand} 
\begin{ruledtabular}
\begin{tabular}{lcc}
    Method & \makecell{Maximum per-shot simulation cost} & \makecell{Total simulation cost} \\
    \midrule
    Deterministic ($p$-th order Suzuki-Trotter) & $\mathcal{O}(L\xi C_{\rm gs}^{1/p}\epsilon^{-1-1/p})$ & $\mathcal{O}(L\xi^{-1} C_{\rm gs}^{1/p}\epsilon^{-1-1/p})$  \\
    Randomized & $\mathcal{O}(\lambda^2\xi^2\epsilon^{-2})$ & $\mathcal{O}(\lambda^2\epsilon^{-2})$  \\
    Partially randomized & $\mathcal{O}(L_D\xi C_{\rm gs}^{1/p}\epsilon^{-1-1/p}+\lambda_R^2\xi^2\epsilon^{-2})$ & $\mathcal{O}(L_D\xi^{-1} C_{\rm gs}^{1/p}\epsilon^{-1-1/p}+\lambda_R^2\epsilon^{-2})$
    \end{tabular}
\end{ruledtabular}
\end{table*}

\subsubsection{Algorithmic cost model for partially randomized robust phase estimation}
\label{subsec:alg_cost_model}

In the remainder of this work, we shift our focus from the asymptotic precision guarantees of RPE to the concrete circuit cost of implementing the time-evolution unitaries that appear in each shot. Following Ref.~\cite{Gunther2025-zk}, we formulate this cost at the level of the logical circuit, in terms of the number of elementary gate applications required by the partially randomized RPE scheme.

The optimal partitioning of the Hamiltonian into deterministic and randomized parts (Eq.~\eqref{eq:ham_partially_randomized}) is found by minimizing the simulation cost of the resulting circuit. For partially randomized RPE, the gate cost for a single RPE iteration at time $t_m=2^m$ is given by
\begin{align}
  G_m
  &= \min_{L_D} \left[ g^{(\rm det)}_m L_D  +  g^{(\rm rand)}_m \lambda_R^2 \right] \nonumber \\
  &= \min_{L_D} \left[ g^{(\rm det)}_m L_D  +  g^{(\rm rand)}_m \left( \sum_{\ell=L_D+1}^{L}|c_\ell| \right)^2 \right] .
  \label{eq:abstract_cost_model}
\end{align}
The factors $g^{(\rm det)}_m$ and $g^{(\rm rand)}_m$ encapsulate algorithm-specific details and are provided in Appendix~\ref{append:prnd}. The gate cost $G_m$ quantifies the number of elementary gates that are considered the primary bottleneck for a given quantum computing architecture. For the early fault-tolerant setting discussed in Sec.~\ref{sec:resource_estimation}, this bottleneck is the implementation of non-Clifford Pauli rotation gates. In this context, $G_m$ represents the number of non-Clifford gates required to implement the evolution for time $t_m$.

Since RPE evaluates the Hadamard test at a sequence of evolution times $\{t_m\}$, the total logical gate cost of the entire procedure is the sum over all iterations:
\begin{equation}
  G_{\mathrm{total}} =
  \sum_{m=0}^{M}e N_{m} G_m ,
  \label{eq:total_cost}
\end{equation}
where $N_{m}$ is the number of shots required for the time step $t_m$, and $e \simeq 2.72$ is a constant overhead factor inherent to the randomized protocol. 
As detailed in Appendix~\ref{append:prnd}, the overall cost $G_{\mathrm{total}}$ is heavily dominated by the contribution from the final iteration ($m=M$) and can be approximated as $G_{\mathrm{total}}\simeq e N_M G_M$.

In Sec.~\ref{sec:hamiltonian}, we will consider the total gate cost $G_{\mathrm{total}}$ as a function of the Hamiltonian representation—specifically, of the Pauli coefficient distribution $\{ c_\ell \}$. The UWC method introduced there is formulated as an optimization problem that seeks to find Hamiltonian representations that minimize this total cost, subject to preserving the energy spectrum of the underlying electronic structure Hamiltonian.

\section{Hamiltonian representation and optimization \label{sec:hamiltonian}}
As described in Sec.~\ref{sec:single_ancilla_prpe}, the partially randomized time evolution leads to an algorithmic simulation cost that depends on the structure of its coefficient distribution $\{c_\ell\}$ and on the number of large-weight terms promoted into the deterministic segment of the time evolution. In this setting, different representations of the same Hamiltonian can lead to distinct values of the total simulation cost, even when their spectra are identical.

Motivated by this observation, we do not restrict ourselves to a single representation of $\hat{H}$, but instead treat the Hamiltonian representation itself as an optimization degree of freedom. In particular, we consider Hamiltonian transformations that preserve the eigenvalue spectrum within a relevant symmetry sector, while allowing the coefficient distribution to be reshaped. We refer to the resulting optimization problem as UWC. This nomenclature reflects the conceptual goal of promoting concentrated Pauli coefficient distributions, which are characterized by a small number of dominant terms that significantly contribute to the total weight.

\subsection{Electronic structure Hamiltonian \label{subsec:molecular_hamil}}
We consider active-space electronic structure Hamiltonians for molecules in the second-quantized form:
\begin{align}
    \hat{H} &= \sum_{pq}^{N}\sum_{\sigma=\uparrow,\downarrow} h_{pq}\hat{a}_{p\sigma}^{\dagger}\hat{a}_{q\sigma}
    + \frac{1}{2}\sum_{pqrs}^{N}\sum_{\sigma\tau} g_{pqrs} \hat{a}_{p\sigma}^{\dagger}\hat{a}_{r\tau}^{\dagger}\hat{a}_{s\tau}\hat{a}_{q\sigma} \nonumber \\
    &\eqqcolon \sum_{pq}^N k_{pq}\hat{F}_{pq} + \frac{1}{2}\sum_{pqrs}^{N} g_{pqrs} \hat{F}_{pq} \hat{F}_{rs}, \label{eq:elec_ham}
\end{align}
where $\hat{a}_{p\sigma}$ and $\hat{a}_{p\sigma}^{\dagger}$ are fermionic annihilation and creation operators for a $p$-th orbital electron with spin $\sigma$, $N$ is the number of spatial orbitals in the active space, and $h_{pq}$ and $g_{pqrs}$ are the corresponding one- and two-electron integrals. For convenience, we have introduced $\hat{F}_{pq}\coloneqq\sum_{\sigma}\hat{a}_{p\sigma}^{\dagger}\hat{a}_{q\sigma}$ and $k_{pq}\coloneqq h_{pq} - \frac{1}{2}\sum_{r}g_{prrq}$ in Eq.~\eqref{eq:elec_ham}.

After a fermion-to-qubit mapping, such as the Jordan-Wigner transformation, $\hat{H}$ takes the Pauli LCU form of Eq.~\eqref{eq:pauli_ham}. For the electronic structure Hamiltonian, this Pauli LCU representation can be written as a sum of products of Majorana operators~\cite{Koridon2021-xm}:
\begin{align}
    \hat{H} &= \frac{i}{2}\sum_{pq,\sigma}\left( k_{pq} + \sum_r g_{pqrr} \right) \hat{\gamma}_{p\sigma}\hat{\bar{\gamma}}_{q\sigma} \nonumber\\
    &+ \frac{1}{4} \sum_{p>r,s>q}\sum_\sigma (g_{pqrs} - g_{psrq})\hat{\gamma}_{p\sigma}\hat{\gamma}_{r\sigma}\hat{\bar{\gamma}}_{q\sigma}\hat{\bar{\gamma}}_{s\sigma} \nonumber\\
    &+ \frac{1}{4}\sum_{pqrs}g_{pqrs}\hat{\gamma}_{p\uparrow}\hat{\gamma}_{r\downarrow}\hat{\bar{\gamma}}_{q\uparrow}\hat{\bar{\gamma}}_{s\downarrow} + \mathrm{const.} ,
    \label{eq:majorana_ham}
\end{align}
where the Majorana operators are defined as
\begin{align}
    \hat{\gamma}_{p\sigma}= \hat{a}_{p\sigma} + \hat{a}_{p\sigma}^{\dagger}, \quad
    \hat{\bar{\gamma}}_{p\sigma}= -i(\hat{a}_{p\sigma} - \hat{a}_{p\sigma}^{\dagger}) .
    \label{eq:majorana_op}
\end{align}
Each product of Majorana operators in Eq.~\eqref{eq:majorana_ham} corresponds to one of the Pauli strings $\{P_\ell \}$ in the Pauli LCU representation~\eqref{eq:pauli_ham}. Although the explicit form of the Pauli string depends on the chosen fermion-to-qubit mapping, the Pauli coefficient distribution $\{c_\ell\}$ is solely determined by the electron integrals $h_{pq}$ and $g_{pqrs}$, as expressed in Eq.~\eqref{eq:majorana_ham}, irrespective of the choice of the fermion-to-qubit mapping.

\subsection{Spectrally-invariant Hamiltonian transformations \label{subsec:hamil_transform}}

Before introducing UWC, we define a structured family of Hamiltonian transformations that will serve as its foundation. We refer to these transformations as \emph{spectrally-invariant} transformations, as they preserve the eigenvalue spectrum within a specified target symmetry sector while modifying the Hamiltonian representation. In this work, this family is characterized by the following two complementary layers.

\subsubsection{Orbital optimization \label{subsubsec:OO}}

The first layer is an OO transformation~\cite{Koridon2021-xm,Ollitrault2024-ko}, which consists of a unitary rotation in the one-particle basis before fermion-to-qubit mapping. This unitary transformation preserves the entire eigenvalue spectrum of $\hat{H}$, while altering the Pauli coefficient distribution $\{ c_\ell \}$. OO is realized through the rotation of molecular orbital bases as
\begin{align}
    \hat{a}'_{i\sigma} = \sum_{p}^N (U_{\mathrm{oo}})_{ip} \hat{a}_{p\sigma}, 
    \label{eq:orbital_rotation}
\end{align}
where $U_{\mathrm{oo}} := e^{-K}$ is an orbital rotation unitary, and $K$ is a skew-symmetric matrix (i.e., $K^{\rm T}=-K$) defined as
\begin{align}
    K := 
    \begin{pmatrix}
        0 & \kappa_{12} & \kappa_{13} & \cdots & \kappa_{1N} \\
        -\kappa_{12} & 0 & \kappa_{23} & \cdots & \kappa_{2N} \\
        -\kappa_{13} & -\kappa_{23} & 0 & \cdots & \kappa_{3N} \\
        \vdots & \vdots & \vdots & \ddots & \vdots \\
        -\kappa_{1N} & -\kappa_{2N} & -\kappa_{3N} & \cdots & 0
    \end{pmatrix}.
    \label{eq:OO_matrix}
\end{align}
For real-valued molecular orbitals, the parameters $\kappa_{ij}$ can be assumed real without loss of generality, leading to $N(N-1)/2$ controllable parameters in total. Denoting these parameters as $\vec{\kappa}=\{ \kappa_{ij} \mid  i<j \}$, the OO-transformed Hamiltonian is written as
\begin{align}
    \hat{H}'(\vec{\kappa}) 
    &\coloneqq  \sum_{pq}^{N} k'_{pq}(\vec{\kappa}) \hat{F}_{pq} 
    + \frac{1}{2}\sum_{pqrs}^{N} g'_{pqrs}(\vec{\kappa}) \hat{F}_{pq} \hat{F}_{rs}, \label{eq:elec_ham_OO}
\end{align}
where 
\begin{align}
    h'_{pq}(\vec{\kappa}) &= \sum_{ij}^{N}[e^{-K}]_{ip}h_{ij}[e^{-K}]_{jq}, \\
    g'_{pqrs}(\vec{\kappa}) &= \sum_{ijkl}^{N}[e^{-K}]_{ip}[e^{-K}]_{jq}g_{ijkl}[e^{-K}]_{kr}[e^{-K}]_{ls}, 
    \label{eq:integrals_OO}
\end{align}
with $k'_{pq}(\vec{\kappa}) \coloneqq h'_{pq}(\vec{\kappa})-\frac{1}{2}\sum_{r}g'_{prrq}(\vec{\kappa})$. In practice, the success of OO is sensitive to the choice of the initial orbitals~\cite{Koridon2021-xm,Ollitrault2024-ko}. In this work, among the initial orbitals (specifically, those provided in the literature and those rotated along Cholesky vectors~\cite{Gunther2025-zk}), the one with the smallest $\ell_1$-norm is selected as the initial orbital for OO (see Appendix~\ref{append:uwc_details} for details).

\subsubsection{Block-invariant symmetry shift \label{subsubsec:BLISS}}
The second layer is a BLISS transformation~\cite{loaiza2013_bliss,Patel2025-np}, which applies block-diagonal shift transformations within fixed symmetry sectors of the Hamiltonian (e.g., particle number, spin, and relevant point-group irreducible representations). BLISS preserves the spectrum within a specific target block, while possibly changing the spectra of other blocks. The definition of the BLISS transformation depends on the target symmetry sector and corresponding symmetry operation.

For instance, when we wish to preserve the eigenvalue spectrum in a fixed particle-number sector with $N_e$ electrons, the corresponding BLISS operator is defined as~\cite{loaiza2013_bliss,Patel2025-np}
\begin{align}
    \hat{T}(\vec{\mu},\vec{\xi}) &:= \mu_1(\hat{N}_e-N_e)+\mu_2(\hat{N}_e^2-N_e^2) \nonumber\\
    &\quad +\sum_{pq}^{N}\xi_{pq}\hat{F}_{pq}(\hat{N}_e-N_e), \label{eq:bliss_operator}
\end{align}
where $\mu_1,\mu_2, \xi_{pq} \in \mathbb{R}$ with $\xi_{pq}=\xi_{qp}$, and $\hat{N}_e \coloneqq \sum_{p,\sigma}\hat{a}_{p\sigma}^{\dagger}\hat{a}_{p\sigma}$ is the particle number operator. The parameters are denoted as $\vec{\mu}:=\{\mu_1,\mu_2\} $ and $\vec{\xi}:=\{ \xi_{pq} \mid p\leq q\}$, leading to $2+N(N+1)/2$ independent parameters for optimization. The Hamiltonian $\hat{H}$ is transformed as
\begin{align}
    \hat{H}'(\vec{\mu},\vec{\xi}) &\coloneqq \hat{H}-\hat{T}(\vec{\mu},\vec{\xi}) \nonumber\\
    &= \sum_{pq}^N k'_{pq}(\vec{\mu},\vec{\xi})\hat{F}_{pq} + \frac{1}{2}\sum_{pqrs}^N g'_{pqrs}(\vec{\mu},\vec{\xi})\hat{F}_{pq}\hat{F}_{rs} \nonumber\\
    &\quad + \mu_1N_e + \mu_2N_e^2, 
    \label{eq:bliss_transformation}
\end{align}
where 
\begin{align}
   h'_{pq}(\vec{\mu},\vec{\xi}) &= h_{pq} - (\mu_1+\mu_2)\delta_{pq} + (N_e-1)\xi_{pq}, \\
    g'_{pqrs}(\vec{\mu},\vec{\xi}) &= g_{pqrs} - 2\mu_2\delta_{pq}\delta_{rs} - (\xi_{pq}\delta_{rs}+\delta_{pq}\xi_{rs}),  
\end{align}
with $k'_{pq}(\vec{\mu},\vec{\xi}) \coloneqq h'_{pq}(\vec{\mu},\vec{\xi})-\frac{1}{2}\sum_{r}g'_{prrq}(\vec{\mu},\vec{\xi})$. Since $\hat{T}(\vec{\mu},\vec{\xi})\ket{\psi} = 0$ for any $N_e$-electron state $\ket{\psi}$ by definition, the BLISS transformation in Eq.~\eqref{eq:bliss_transformation} preserves the eigenvalue spectrum in the target $N_e$-electron particle-number sector. The BLISS operator $\hat{T}(\vec{\mu},\vec{\xi})$ is Hermitian and has the same fermionic polynomial degrees as $\hat{H}$. These properties are desired constraints for BLISS, as they simplify analysis and optimization~\cite{loaiza2013_bliss}. In this work, BLISS operations are also employed for spin symmetry, as detailed in Appendix~\ref{append:BLISS_spin}.

\subsection{Unitary weight concentration \label{subsec:uwc}}
Here, we describe our Hamiltonian optimization strategy based on spectrally invariant transformations to minimize the gate cost of partially randomized QPE. 

Let $\mathcal{F}(\hat{H})$ denote a set of spectrally-invariant transformations applicable to the original Hamiltonian $\hat{H}$, such as OO and BLISS. We then express a possible Hamiltonian representation under spectrally-invariant transformations as
\begin{align}
    \hat{H}'(\vec{\chi}) &= f_{\vec{\chi}}(\hat{H}), \quad f_{\vec{\chi}} \in \mathcal{F}(\hat{H})
    \label{eq:ham_trans_general}
\end{align}
where $\vec{\chi}$ is a set of parameters characterizing the Hamiltonian transformation $f_{\vec{\chi}}$. In this work, the $f_{\vec{\chi}}$ consists of the OO and BLISS transformations. These OO and BLISS transformations alter only the coefficient distribution $\{ c_\ell \}$, without introducing Majorana polynomials not included in the original Hamiltonian~\eqref{eq:majorana_ham}, and preserve the Pauli string set $\{ P_\ell \}$. The Pauli LCU representation for the transformed Hamiltonian $\hat{H}(\vec{\chi})$ is then given by
\begin{align}
    \hat{H}'(\vec{\chi}) &= \sum_\ell c'_\ell(\vec{\chi}) \hat{P}_\ell. 
    \label{eq:ham_trans_pauli}
\end{align}
The transformed coefficients $\{ c'_\ell(\vec{\chi}) \}$ are determined by the transformed electron integrals $h'_{pq}(\vec{\chi})$ and $g'_{pqrs}(\vec{\chi})$, as expressed in Eq.~\eqref{eq:majorana_ham}, and therefore $c'_\ell(\vec{\chi})$ is a functional of these transformed intagrals, i.e., $c'_\ell(\vec{\chi})\coloneqq c'_\ell[\{h'_{pq}(\vec{\chi}),g'_{pqrs}(\vec{\chi})\}]$. 
With this transformed coefficient distribution $\{ c'_\ell(\vec{\chi}) \}$, we evaluate the total algorithmic cost $G_{\mathrm{total}}$ (Eq.~\eqref{eq:total_cost}), which was introduced in Sec.~\ref{subsec:alg_cost_model}. 

In this setup, our goal is to solve the following optimization problem: 
\begin{align}
    \vec{\chi}^*
    &=\arg\min_{\vec{\chi}} G_{\mathrm{total}}(\hat{H}'(\vec{\chi})) ,
    \label{eq:uwc_objective}
\end{align}
where we denote $G_{\mathrm{total}}$ as a function of the Hamiltonian representation $\hat{H}'(\vec{\chi})$. As described in Sec.~\ref{subsec:alg_cost_model}, the total simulation cost $G_{\mathrm{total}}$ is dominated by the contribution from the last round $m=M$. Thus, the optimization problem~\eqref{eq:uwc_objective} can be approximated as
\begin{align}
    \vec{\chi}^*
    \approx \arg\min_{\vec{\chi}} G_{M}(\hat{H}'(\vec{\chi})). 
    \label{eq:uwc_obejective_approx}
\end{align}
We note that, in principle, one could perform a round-dependent optimization and determine a distinct Hamiltonian representation for each $m$. However, this would increase the classical optimization overhead by roughly a factor of $M$. Since the total computational cost of RPE is dominated by the final round ($m=M$), with earlier rounds contributing only marginally, optimizing $G_M$ alone captures the dominant effect on resource reduction.

However, solving this optimization problem is notably challenging due to two primary factors. First, estimating the objective function $G_M$ for a given Hamiltonian representation $\hat{H}'(\vec{\chi})$ is itself an optimization problem with respect to the deterministic/randomized partitioning point $L_D$, as described in Eq.~\eqref{eq:abstract_cost_model}. This bilevel optimization structure significantly increases computational cost, rendering exact solutions for the problem~\eqref{eq:uwc_obejective_approx} computationally intractable in practice. Second, the embedded sorting operation for $\{ c'_\ell(\vec{\chi}) \}$, which is necessary to determine the optimal deterministic/randomized partitioning, introduces severe non-differentiability, non-smoothness, and combinatorial complexity, thus precluding the use of traditional gradient-based optimization methods. 

To circumvent such difficulties, the UWC method instead aims to solve the following softened optimization problem: 
\begin{widetext}
\begin{align}
    & \vec{\chi}^* = \arg\min_{\vec{\chi}} G_{\rm soft}(\hat{H}'(\vec{\chi})) , \nonumber\\
    & \mathrm{for} \quad G_{\rm soft}(\hat{H}'(\vec{\chi})) := g^{(\rm det)}_M \sum_{\ell=1}^{L}\varsigma\left(\frac{|c'_\ell(\vec{\chi})|-w_{\rm soft}}{\epsilon_{\rm soft}}\right) + g^{(\rm rand)}_M \left( \sum_{\ell=1}^{L} |c'_{\ell}(\vec{\chi})|\left[ 1-\varsigma\left(\frac{|c'_\ell(\vec{\chi})|-w_{\rm soft}}{\epsilon_{\rm soft}}\right)\right] \right)^2, 
    \label{eq:uwc_def}
\end{align}
\end{widetext}
where $\varsigma(x):=1/(1+\exp(-x))$ is the sigmoid function and $\epsilon_{\rm soft} >0$ is a constant used for the sigmoid regularization. Here, $w_{\rm soft}$ is defined as the mean coefficient weight at the optimal deterministic/randomized partitioning point for the initial Hamiltonian $\hat{H}$: 
\begin{align}
    w_{\rm soft} &:= \frac{|c_{L_D^\ast}|+|c_{L_D^\ast+1}|}{2}, \label{eq:w_soft} \\
    L_D^\ast &:= \arg\min_{L_D} \left[ g^{(\rm det)}_M L_D  +  g^{(\rm rand)}_M \lambda_R^2 \right]. \label{eq:LD_opt}
\end{align}
In Eq.~\eqref{eq:uwc_def}, the bilevel structure of the original problem~\eqref{eq:uwc_obejective_approx} is simplified by incorporating information about the deterministic/randomized partitioning solely from the initial coefficient distribution $\{ c_\ell \}$ as the hyperparameter $w_{\rm soft}$. Furthermore, discontinuous behaviors arising from the sorting operations are mitigated by introducing a sigmoid regularization with a smoothness hyperparameter $\epsilon_{\rm soft}$: the objective function $G_{\rm soft}$ approaches the original cost function $G_M$ as $\epsilon_{\rm soft} \to 0$. In practice, UWC is performed as an iterative optimization of combined OO and BLISS transformations, adjusting the hyperparameter $w_{\rm soft}$ at each iteration. Iterations are terminated when the relative improvement in the cost function $G_{\rm soft}$ falls below a prescribed threshold. 
Neither hyperparameter is freely tunable. The parameter $w_{\rm soft}$ approximates a threshold determined by the ratio of the cost coefficients in Eq.~\eqref{eq:abstract_cost_model}, while $\epsilon_{\rm soft}$ must be sufficiently small for the softened partition to resolve the underlying exact partition, as shown in Appendix~\ref{append:soft_cost_params}. Moreover, each iteration is accepted only if the exact cost $G_M$ does not increase. Consequently, the returned representation never requires more gates than the input representation, regardless of how accurately $G_{\rm soft}$ approximates $G_M$ or how well the inner minimizations converge (Appendix~\ref{append:uwc_algorithm}). Details of the numerical optimization procedure are provided in Appendix~\ref{append:uwc_details}.

It is worth noting that both OO and BLISS were originally proposed for minimizing the $\ell_{1}$-norm $\lambda$~\cite{Koridon2021-xm, loaiza2013_bliss}, which is the appropriate target for the algorithms they were developed for: the query complexity of qubitization scales as $\mathcal{O}(\lambda)$~\cite{Lee2021-tz,Von_Burg2021-du,Low2025-tb,Low2019-op} and the gate count of fully randomized product formulas as $\mathcal{O}(\lambda^{2})$, so in both cases minimizing $\lambda$ is equivalent to minimizing the cost. For partially randomized simulation, this equivalence fails. The cost $G_{M}$ depends on the entire profile of tail sums $\sum_{\ell>k}|c_{\ell}|$ ($k=0,1,\cdots,L$), whereas $\lambda$ is the value of that profile at the single point $k=0$; as shown in Appendix~\ref{append:uwc_l1_comparison}, the two coincide only when the optimal partition is trivial, $L_{D}^{*}=0$, that is, in the fully randomized regime. Outside it---which includes every instance studied here---the $\ell_{1}$-norm constrains the gate count only through the bound $G_{M}\le g^{(\mathrm{rand})}_{M}\lambda^{2}$, which is loose by a factor growing in proportion to $\lambda$; the representation with the smallest $\ell_{1}$-norm is therefore not, in general, the one with the smallest gate count. Targeting the gate count directly is therefore necessary rather than merely convenient, and doing so yields reductions of up to $60\%$ relative to $\ell_{1}$-norm optimization over the same family of transformations (see Appendix~\ref{append:uwc_l1_comparison}).

\section{Numerical validation and benchmark results of UWC \label{sec:numerical}}
In this section, we present numerical results validating the effectiveness of UWC and quantify its impact on the cost of partially randomized QPE for chemically relevant molecular models. 
First, using a representative molecular instance, we demonstrate how UWC reshapes the coefficient distribution of Pauli-LCU Hamiltonians and alters the balance between deterministic and randomized segments.
Subsequently, we assess the practical consequences of this reshaping by quantifying gate-count reductions across representative molecular active-space models.

\subsection{Representative molecular models \label{subsec:models}}
To assess the practical impact of UWC under early-FTQC constraints, we consider a set of molecular active-space models drawn from three chemically and industrially relevant families: iron-sulfur clusters~\cite{Beinert1997-ok,Lee2023-rd,Ollitrault2024-ko}, cytochrome P450 active-site models~\cite{Goings2022-so}, and ruthenium-based catalysts for transforming \ce{CO2} to high-value chemical methanol~\cite{Von_Burg2021-du}.
All molecular Hamiltonians and active-space definitions used in this work were adopted from established prior studies~\cite{Ollitrault2024-ko,Goings2022-so,Von_Burg2021-du}, and were not constructed specifically for the present analysis. 
Consequently, the benchmark set comprises chemically motivated and well-studied electronic-structure models that have been extensively analyzed in the quantum computing for chemistry literature.

For each molecular family, we select representative active-space models spanning approximately 20 to 40 spatial orbitals. The information for each active-space model is summarized in Table~\ref{tab:mol_info_representative}. 
These active spaces, while possessing Hilbert-space dimensions approaching or exceeding the practical limits of classical full-CI computation (typically $\sim 10^{12}$~\cite{Vogiatzis2017-fc,Gao2024-ac}), are nevertheless expected to be potentially accessible on early-FTQC hardware. 
The chosen models capture chemically relevant valence and near-valence orbitals that govern bonding, spin character, and catalytic reactivity, following active-space constructions proposed in the original references (see Appendix~\ref{append:models_detail} for details).

Across all benchmark systems, the electronic Hamiltonians are expressed in second-quantization form (Eq.~\eqref{eq:elec_ham}), and their corresponding Pauli-LCU representation is represented using the electron integrals $h_{pq}$ and $g_{pqrs}$, as presented in Eq.~\eqref{eq:majorana_ham}. 
This uniform treatment enables a controlled comparison of deterministic, partially randomized, and UWC-optimized evolutions across different molecular families and active-space sizes. 

Detailed information regarding the source references, active-space definitions, and symmetry sectors for each benchmark system is provided in Appendix~\ref{append:models_detail}. In the following subsections, we use these models to examine how UWC reshapes Hamiltonian coefficient distributions and how this reshaping translates into reductions in the cost of partially randomized QPE. 

\begin{table}[tbp]
\caption{Representative molecular models considered in the main text. They present models from chemically and industrially relevant areas, including bioinorganic clusters, enzymatically active sites, and homogeneous catalysts for \ce{CO2} utilization. The active spaces are specified in terms of the number of correlated electrons and orbitals. The ``Molecule ID" identifies the molecule species and active space size, as detailed in Appendix~\ref{append:models_detail}. 
}
\label{tab:mol_info_representative} 
\begin{ruledtabular}
\begin{tabular}{lccc}
    Molecule ID & Orbitals & Electrons & \makecell{Hilbert space\\dimension} \\
    \midrule
    \multicolumn{4}{c}{(a) Iron-sulfur clusters~\cite{Ollitrault2024-ko} }\\
    \midrule
    $[2\ce{Fe}\mathchar`-2\ce{S}]^{-3}$ & 20 & 31 &  $7.51\times 10^{7}$ \\
    $[2\ce{Fe}\mathchar`-2\ce{S}]^{-2}$ & 20 & 30 & $2.40\times10^8$ \\
    $[4\ce{Fe}\mathchar`-4\ce{S}]^{-2}$ & 36 & 54 & $8.86\times10^{15}$ \\
    $[4\ce{Fe}\mathchar`-4\ce{S}]$ & 36 & 52 & $6.46\times 10^{16}$ \\
    \midrule
    \multicolumn{4}{c}{(b) Cytochrome P450~\cite{Goings2022-so} }\\ 
    \midrule
    P450-Cpd I (D) & 23 & 15 & $5.61\times 10^{11}$ \\
    P450-Cpd I (E) & 31 & 33 & $3.74\times 10^{16}$ \\
    P450-Cpd I (F) & 41 & 45 & $2.77\times 10^{22}$ \\
    P450-Cpd I (G) & 43 & 47 & $4.43\times 10^{23}$ \\
    \midrule
    \multicolumn{4}{c}{(c) Ruthenium-based catalysts for \ce{CO2} utilization~\cite{Von_Burg2021-du} }\\
    \midrule
    Ru-XVIII (Md) & 20 &  28 &  $1.50\times 10^{9}$ \\
    Ru-IX (Md) &  26 &  32 &  $2.82\times 10^{13}$ \\
    Ru-II-III (Md) & 29 &  38 &  $4.01\times 10^{14}$ \\
    Ru-VIII (Md) & 29 &  40 & $1.00\times 10^{14}$ \\
    \end{tabular}
\end{ruledtabular}
\end{table}

\subsection{Concentrated coefficient distribution and partially randomized simulation cost \label{subsec:uwc_coefficients}}

We first examine how UWC reshapes the coefficient distribution for the Pauli-LCU representation of electronic structure Hamiltonians. To isolate this effect and provide a clear mechanistic picture, we focus on a single representative molecular active-space model of intermediate size from the benchmark set. Specifically, we choose a model of the compound I (Cpd I) in the P450 catalytic cycle, characterized by an intermediate-sized 31-orbital active space E~\cite{Goings2022-so} (see Appendix~\ref{append:models_detail} for chemical details of this active space). 

\begin{figure}[tbp]
  \centering
  \includegraphics[width=0.48\textwidth]{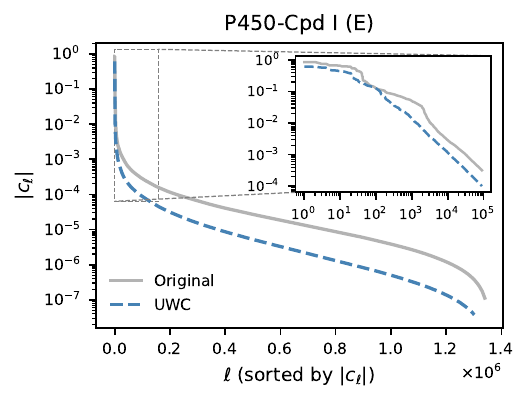}
  \caption{\label{fig:coeff_distribution}
  Demonstration of Pauli coefficient concentration induced by UWC for the P450-Cpd I (E) model. The main panel shows the magnitudes of Pauli-LCU coefficients $\{ |c_\ell |\}$, sorted in descending order, for the original (gray solid line) and UWC-optimized (blue dashed line) Hamiltonian representations. The inset presents a magnified view of the large-weight region. }
\end{figure}
\begin{figure}[tbp]
  \centering
  \includegraphics[width=0.48\textwidth]{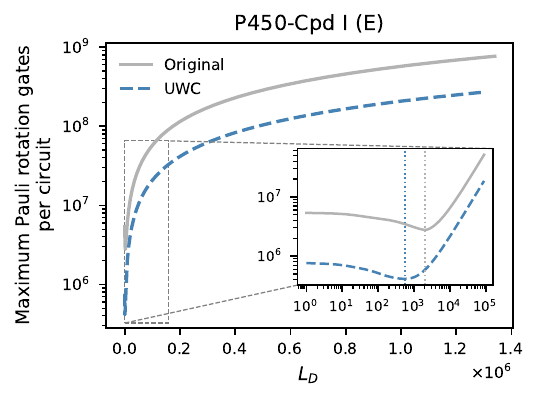}
  \caption{\label{fig:pr_cost} 
  Maximum per-shot Pauli-rotation gate count of partially randomized RPE with $\xi=0.01$ as a function of the number of deterministic terms, $L_D$ (up to a maximum of $L$), for the P450-Cpd I (E) model. Results are shown for the original (gray solid line) and UWC-optimized (blue dashed line) Hamiltonian representations. UWC lowers the overall cost and shifts the optimal deterministic partitioning to a smaller $L_D$, as shown in the inset, reflecting the more concentrated coefficient distribution in Fig.~\ref{fig:coeff_distribution}. }
\end{figure}
Figure~\ref{fig:coeff_distribution} compares the Pauli coefficient distribution before and after applying UWC. 
The computational details for obtaining the UWC-optimized Hamiltonian are summarized in Appendix~\ref{append:uwc_details}. 
Although the two Hamiltonian representations are spectrally equivalent within each symmetry sector, the UWC-optimized representation exhibits a markedly more concentrated coefficient distribution: a larger fraction of the total weight is carried by a smaller number of leading terms, while simultaneously, the weight of the small-weight tail terms is significantly suppressed. Such a concentrated weight profile is particularly relevant for the partially randomized QPE scheme, as explained in Sec.~\ref{subsec:alg_cost_model}.  

\begin{figure*}[tbp]
  \centering
  \includegraphics[width=1.0\textwidth]{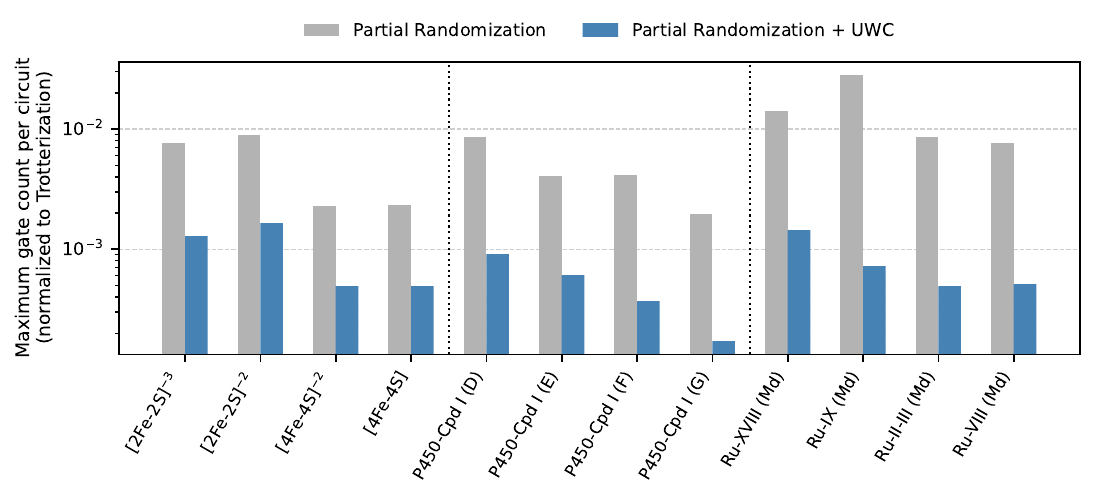}
  \caption{\label{fig:aggregate_gates}
    Normalized maximum per-shot Pauli-rotation gate counts for representative molecular active-space models. Results are shown for partially randomized time evolution (gray bars) and for partially randomized time evolution combined with UWC (blue hatched bars), for the initial-state parameter $\xi=0.01$. The values are normalized to the gate count for deterministic second-order Trotterization. 
}
\end{figure*}

We next examine how this coefficient concentration induced by UWC translates into changes in the cost structure of partially randomized RPE. In the partially randomized framework, the Hamiltonian is partitioned into a deterministic segment, consisting of the $L_D$ largest Pauli terms, and a randomized segment formed from the remaining terms. The total gate cost depends nontrivially on the choice of $L_D$, as described in Sec.~\ref{subsec:alg_cost_model}, reflecting a trade-off between the cost of deterministic evolution and the variance contributed by the randomized tail. In all the following resource estimations for partially randomized RPE, we assume that the deterministic terms are handled by the second-order Suzuki-Trotter product formula, while the randomized part is implemented via the qDRIFT protocol, unless otherwise mentioned. 

Figure~\ref{fig:pr_cost} shows the maximum Pauli rotation gate count per circuit as a function of the number of deterministically treated terms, $L_D$, for the P450-Cpd I (E) model. Here, the initial state parameter $\xi$ for RPE is fixed at $\xi=0.01$, corresponding to $p_0 > 0.99$. 
For the original Hamiltonian representation, the cost curve exhibits a sharp minimum at relatively small $L_D (\ll L)$, indicating that the Pauli-LCU coefficient distribution of this molecular Hamiltonian is concentrated and partial randomization is effective in reducing the product-formula simulation cost~\cite{Gunther2025-zk}.   
When UWC is applied, the cost curve is uniformly lowered and its minimum shifts to smaller values of $L_D$. 
This shift directly reflects the enhanced dominance of the leading coefficients observed in Fig.~\ref{fig:coeff_distribution}: owing to a greater concentration of the total weight in fewer terms, a more efficient partially randomized treatment can be achieved with a smaller deterministic fragment. 

Having established the mechanistic origin of coefficient concentration and its impact on the cost structure of partially randomized time evolution, we now assess the aggregate performance of UWC across the full benchmark set of molecular active-space models. This analysis allows us to quantify how consistently the benefits observed in the representative example extend to chemically diverse systems and a range of active-space sizes.

Figure~\ref{fig:aggregate_gates} summarizes normalized gate counts for the representative molecular models in Table~\ref{tab:mol_info_representative} under partially randomized time evolution and UWC, with all values reported relative to deterministic second-order Trotterization.
Across all models, partial randomization alone reduces the gate count to approximately $10^{-2}$ of the deterministic Trotter baseline, while the additional application of UWC further suppresses the cost to the $10^{-3}$ level.
This additional reduction reflects the systematic enhancement of coefficient concentration induced by UWC, as demonstrated in Figs.~\ref{fig:coeff_distribution} and~\ref{fig:pr_cost}. 
Although the absolute magnitude of the UWC-induced improvement varies across molecular instances, the qualitative trend is robust: UWC consistently provides an additional order-of-magnitude reduction beyond partial randomization alone.

Taken together, these results demonstrate that partially randomized time evolution delivers the dominant improvement over deterministic Trotterization, while UWC acts as a complementary optimization that further refines the Hamiltonian representation and yields substantial additional reductions in gate cost. In the following section, we incorporate these UWC-optimized Hamiltonians into end-to-end resource estimates to assess the feasibility of QPE on early-FTQC hardware.


\subsection{Scaling with system size \label{subsec:scaling}}

The reductions reported above are absolute gate counts for a set of molecular active-space models whose sizes span a limited range and whose electronic structures differ considerably from one another. They therefore do not by themselves reveal whether UWC improves the \emph{scaling} of the cost with system size or only its prefactor. To separate the two, we turn to a systematic series of one-dimensional hydrogen chains H$_{N}$ with $N=10$--$80$ atoms at a fixed interatomic separation of $0.74\,\mathrm{\AA}$ in a STO-6G minimal basis~\cite{Lee2021-tz}, for which the active space grows with $N$ at fixed composition, and the cost can be followed over nearly one decade in system size.
 
\begin{figure*}[t]
  \includegraphics[width=\textwidth]{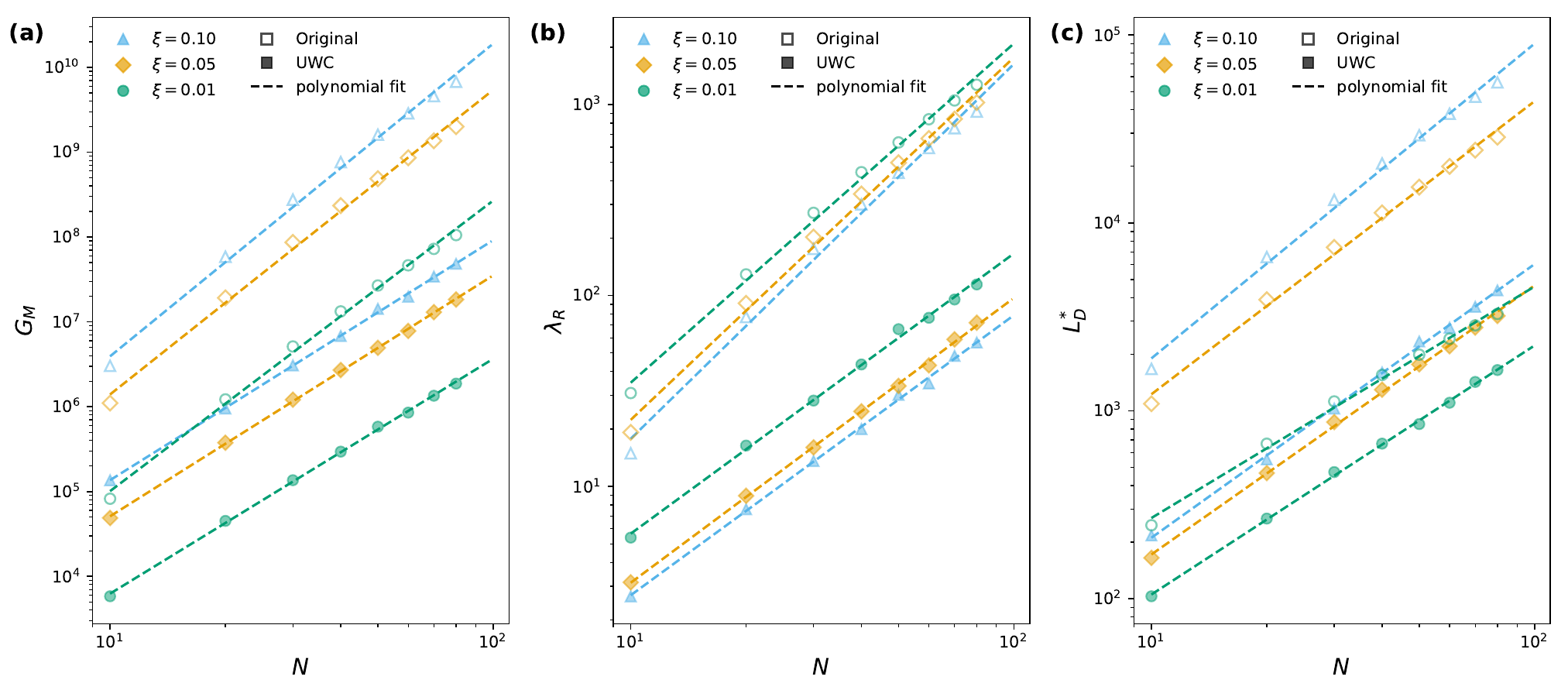}
  \caption{Scaling with system size for hydrogen chains H$_{N}$,
    $N=10$--$80$. (a) Gate count $G_{M}$ of
Eq.~\eqref{eq:abstract_cost_model}; (b) randomized
    $\ell_{1}$-norm $\lambda_{R}$; (c) size of the deterministic segment
    $L_{D}^{*}$. Open symbols denote the original representation and
    filled symbols the UWC-optimized one, for
    $\xi=0.10$ (triangles), $0.05$ (diamonds) and $0.01$ (circles);
    dashed lines are power-law fits. Fitted exponents are listed in
    Table~\ref{tab:exponents}.}
  \label{fig:scaling}
\end{figure*}

Figure~\ref{fig:scaling} shows the resulting gate count $G_{M}$ together with the two quantities that compose it in Eq.~\eqref{eq:abstract_cost_model}, the randomized $\ell_{1}$-norm $\lambda_{R}$ and the size of the deterministic segment $L_{D}^{*}$, for the original and UWC-optimized representations at three values of $\xi$. All six data sets in each panel are well described by power laws in $N$ over the range studied ($R^{2}>0.99$). We write $\beta_{G}$, $\beta_{\lambda_{R}}$ and $\beta_{L_{D}}$ for the exponents obtained by fitting $G_{M}$, $\lambda_{R}$ and $L_{D}^{*}$ to $AN^{\beta}$, and $\Delta\beta$ for the reduction in an exponent achieved by UWC, so that e.g., $\Delta\beta_{G}=\beta_{G}^{\rm orig}-\beta_{G}^{\rm UWC}$. The fitted values are collected in Table~\ref{tab:exponents}.
 
\begin{table*}[t]
  \caption{Power-law fits $Y=A N^{\beta}$ of the gate count $G_{M}$, the
    randomized $\ell_{1}$-norm $\lambda_{R}$ and the size of the
    deterministic segment $L_{D}^{*}$ against the number of active
    orbitals, over the range $N=10$--$80$, for the original and
    UWC-optimized representations at three values of $\xi$.}
  \label{tab:exponents}
  \begin{ruledtabular}
  \begin{tabular}{llccccccccc}
    & & \multicolumn{3}{c}{$G_{M}$}
      & \multicolumn{3}{c}{$\lambda_{R}$}
      & \multicolumn{3}{c}{$L_{D}^{*}$}\\
    \cline{3-5}\cline{6-8}\cline{9-11}
    $\xi$ & representation
      & $\beta$ & $A$ & $R^{2}$
      & $\beta$ & $A$ & $R^{2}$
      & $\beta$ & $A$ & $R^{2}$\\
    \colrule
    0.10 & original & 3.69 & $8.10\times10^{2}$ & 0.9953
                     & 1.96 & 0.193  & 0.9927
                     & 1.68 & 40.2   & 0.9946\\
         & UWC       & 2.83 & $2.02\times10^{2}$ & 0.9994
                     & 1.47 & 0.0917 & 0.9985
                     & 1.46 & 7.38   & 0.9989\\
    \colrule
    0.05 & original & 3.59 & $3.58\times10^{2}$ & 0.9959
                     & 1.90 & 0.282  & 0.9941
                     & 1.56 & 33.8   & 0.9950\\
         & UWC       & 2.84 & $7.39\times10^{1}$ & 0.9996
                     & 1.49 & 0.101  & 0.9992
                     & 1.43 & 6.36   & 0.9987\\
    \colrule
    0.01 & original & 3.43 & $3.80\times10^{1}$ & 0.9965
                     & 1.78 & 0.584  & 0.9954
                     & 1.23 & 15.8   & 0.9954\\
         & UWC       & 2.77 & $1.07\times10^{1}$ & 0.9992
                     & 1.47 & 0.192  & 0.9974
                     & 1.33 & 4.97   & 0.9990\\
  \end{tabular}
  \end{ruledtabular}
\end{table*}
 
The exponent of the gate count is reduced by UWC at every value of $\xi$, from $3.43$--$3.69$ for the original representation to $2.77$--$2.84$ after optimization. The improvement is therefore not confined to the prefactor: the advantage grows with system size, from roughly one order of magnitude at $N=10$ to two orders of magnitude at $N\gtrsim60$.
 
Figs.~\ref{fig:scaling}(b) and (c) identify where the improvement originates. Since the randomized cost coefficient $g^{(\mathrm{rand})}_{M}$ is independent of the representation (Appendix~\ref{append:prnd}), the randomized term of Eq.~\eqref{eq:abstract_cost_model} alone would give $\beta_{G}=2\beta_{\lambda_{R}}$, and the fitted values are close to this relation: $2\beta_{\lambda_{R}}$ exceeds $\beta_{G}$ by $0.1$--$0.2$ for every data set, the small deficit being expected since $G_{M}$ is a sum of two terms with different exponents and its effective single power law is therefore slightly shallower. 
More tellingly, the improvement in the gate-count exponent is accounted for almost entirely by the improvement in $\lambda_{R}$: the reduction $\Delta\beta_{G}=0.86$, $0.75$ and $0.66$ at $\xi=0.10$, $0.05$ and $0.01$ is matched by $2\Delta\beta_{\lambda_{R}}=0.98$, $0.82$ and $0.62$. The deterministic segment contributes little by comparison: its exponent changes by $\Delta\beta_{L_{D}}=0.22$, $0.13$ and $-0.10$, with no consistent sign, and it enters Eq.~\eqref{eq:abstract_cost_model} linearly rather than squared, so even the largest scaling improvement of these is a small correction; the improvement there appears instead in the prefactor, which falls by a factor of three to five (Table~\ref{tab:exponents}). The scaling advantage of UWC thus arises predominantly from a faster suppression of the weight carried by the randomized segment.

One limitation should be kept in mind. Hydrogen chains are one-dimensional systems treated in a minimal basis, and the numerical values of the exponents should not be extrapolated to the three-dimensional, multi-shell active spaces of Table~\ref{tab:mol_info_representative}; what the series establishes is that the improvement is not restricted to the prefactor. As a consistency check, the reduction factors obtained for the molecular models of Table~\ref{tab:mol_info_representative} fall in the same range as those of the hydrogen chains of comparable active-space size.

\section{Resource estimation for early fault-tolerant quantum hardware \label{sec:resource_estimation}}

Having established that partially randomized QPE combined with UWC can reduce algorithmic gate counts by up to three orders of magnitude relative to deterministic Trotterization, we now assess whether these improvements translate into practical feasibility on realistic quantum hardware. In this section, we present end-to-end resource estimates for executing the partially randomized RPE algorithm on early-FTQC hardware, focusing on devices with physical qubit budgets on the order of $10^5$.

\subsection{Partially fault-tolerant quantum computing architecture based on STAR-magic mutation \label{subsec:STAR}}

We begin by outlining the early-FTQC framework used for resource estimation throughout this work. To translate algorithmic gate counts into physically meaningful resource estimates, we adopt a resource estimation framework based on an improved variant of the STAR architecture~\cite{Akahoshi2024-hj,Toshio2025-nn,Akahoshi2025-ra}. This improved variant is detailed in a separate work~\cite{Toshio2026}. 
Here, we summarize only the features directly relevant to our resource estimation, and defer implementation details to Ref.~\cite{Toshio2026}.

The defining feature of the STAR architecture is its direct support for analog Pauli rotation gates $\hat{R}_P(\theta_L):=e^{i\theta_L\hat{P}}$, where $\hat{P}$ is a Pauli string operator and $\theta_L \in \mathbb{R}$ is a logical rotation angle. Such analog rotation gates form the dominant cost component in Trotter-based quantum simulation and single-ancilla QPE. Rather than decomposing these rotation gates into long sequences of $T$-gates~\cite{Bocharov2015-de}, which typically requires a costly magic-state distillation protocol~\cite{Bravyi2005-df,Fowler2012-le,Gidney2019-ch}, the STAR architecture enables their execution as native logical operations. This capability is particularly well-suited to early-FTQC regimes, where large-scale magic-state factories may be impractical due to the physical-qubit overhead. 

Specifically, the STAR architecture implements analog rotation gates by preparing a special non-Clifford ancilla state $\ket{m_{\theta_L}}_L := e^{i\theta_L\hat{Z}_L}\ket{+}_L$ in a non-fault-tolerant manner, and then uses this state to perform a Pauli-$Z$ rotation $\hat{R}_{z,L}(\theta_L)\coloneqq e^{i\theta_L\hat{Z}_L}$ via gate teleportation~\cite{Akahoshi2024-hj}. This gate-teleportation protocol is stochastically implemented as a \emph{repeat-until-success} (RUS) process: a failure event, occurring with a probability of $1/2$, is corrected by an inverse rotation, thereby doubling the logical rotation angle until teleportation succeeds. On average, this RUS process succeeds in two trials. 
The residual logical error arising from this RUS process is addressed using error mitigation techniques such as probabilistic error cancellation~\cite{Temme2017-hs,Endo2018-zs}. In contrast, logical Clifford gates are implemented fault-tolerantly based on lattice surgery~\cite{Litinski2019-qf}. In this sense, the STAR architecture provides a means to achieve partial fault tolerance on limited-scale, near-term quantum hardware. 

In the improved version of the STAR architecture considered throughout this work, logical analog rotations are executed using a protocol called \emph{STAR-magic mutation} (SMM)~\cite{Toshio2026}. SMM combines the STAR-based analog rotation protocol with \emph{magic state cultivation}~\cite{Gidney2024-lq,Hirano2025-mz,Hetenyi2026-ga}, a technique to prepare a clean magic state using only a single surface code patch. In the RUS process, failure events cause a sequential doubling of the rotation angle, leading to an increase from $\theta_L$ to $\theta_{\rm RUS} (>\theta_L)$. In the SMM protocol, once the post-RUS angle $\theta_{\rm RUS}$ exceeds a certain threshold $\theta_{\rm th}$, the target rotation gate $\hat{R}_{P}(\theta_{\rm RUS})$ is decomposed into a sequence of $T$-gates, each executed using the magic state cultivation protocol. This significantly increases the efficiency of the RUS process and results in an effective logical error rate for analog rotation gates of $\mathcal{O}(\theta_L^{2(1-\Theta(1/d))} p_{\rm ph})$~\cite{Toshio2026}, where $d$ is the code distance of the underlying surface code. This error scaling indicates that the SMM protocol is particularly beneficial for small-angle rotation gates, which are the main components of product-formula circuits.

Under the SMM protocol, the effective logical error rate for executing an analog rotation gate $\hat{R}_P(\theta_L)$ is formally described as
\begin{align}
    P_L(\theta_L) = \alpha_{\rm RUS}(\theta_L) \cdot \theta_L \cdot p_{\rm ph},  
    \label{eq:P_logical_star}
\end{align}
where $p_{\rm ph}$ is the physical error rate, and $\alpha_{\rm RUS}(\theta_L)$ is defined as the RUS factor~\cite{Toshio2026}.
The angular dependence of the RUS factor $\alpha_{\rm RUS}(\theta_L)$ accounts for parameter choices such as the threshold $\theta_{\rm th}$ and physical error rate $p_{\rm ph}$, as demonstrated in Ref.~\cite{Toshio2026} and Appendix~\ref{append:resource_assumptions}. In the SMM protocol, this logical error arising from the $\hat{R}_P(\theta_L)$ gate is mitigated using the probabilistic error cancellation technique~\cite{Temme2017-hs,Endo2018-zs}. This error mitigation process is accompanied by an increased sampling overhead $\gamma_{\theta_L}^2$, which is approximately given by $\gamma_{\theta_L}^2 \simeq 1 + 4P_L(\theta_L)$~\cite{Toshio2026}.

Now, consider a quantum circuit composed of $N_{\rm rot}$ analog Pauli rotation gates with rotation angles $\{ \theta_j \}_{j=1,2,\ldots,N_{\rm rot}}$. The total logical error rate for this circuit is described as
\begin{align}
    P_{\rm total} = \sum_{j=1}^{N_{\rm rot}}P_L(\theta_j) = \sum_{j=1}^{N_{\rm rot}} \alpha_{\rm RUS}(\theta_j) \cdot |\theta_j| \cdot p_{\rm ph} ,
    \label{eq:P_total_star}
\end{align}
and the corresponding total sampling overhead is given by
\begin{align}
    \gamma_{\rm total}^2 = \prod_{j=1}^{N_{\rm rot}} \gamma_{\theta_j}^2 \simeq e^{4 P_{\rm total}} .
    \label{eq:pec_gamma_total}
\end{align}
In the following resource estimation, we estimate the total time-to-solution by considering the error mitigation overhead based on Eq.~\eqref{eq:pec_gamma_total}.

\subsection{Resource estimation framework \label{subsec:resource_estimation_setup}}

We now describe the detailed resource estimation framework based on the SMM protocol. Our goal is not to provide asymptotic bounds, but rather to obtain concrete, end-to-end estimates for QPE execution under early-FTQC constraints. Accordingly, we focus on physically meaningful metrics that directly capture early-FTQC hardware limitations, including physical qubit counts, circuit execution time, total time-to-solution, and achievable parallelism.

To obtain these physical resource estimates, we first determine the algorithmic gate cost for partially randomized RPE, as described in Sec.~\ref{subsec:alg_cost_model}. Specifically, for product-formula circuits $\{ \mathcal{C}_m \}_{m=0,1,\ldots,M}$ approximating the time evolution $e^{-it_m\hat{H}}$ ($t_m=2^m$), we estimate the number of non-Clifford Pauli rotation gates, denoted $\{ G_{m} \}_{m=0,1,\cdots,M}$, and the number of shots for each circuit, $\{ N_{m} \}_{m=0,1,\cdots,M}$, required to execute the RPE algorithm. Throughout this paper, algorithmic costs are estimated for a target precision corresponding to chemical accuracy, $\epsilon=1.6$ mHa.

Next, we translate the algorithmic cost into physical resource estimates using the SMM-based partially fault-tolerant quantum computing architecture. In this scheme, time-evolution circuits are decomposed into a sequence of Clifford gates and non-Clifford Pauli-rotation gates, which are then implemented via sequential Pauli-based computation~\cite{Litinski2019-qf}, combined with SMM. 
To estimate physical resources, we first determine the code distance $d$ of the underlying surface code, which is chosen to suppress logical errors from Clifford operations. Specifically, we adopt prior results for a rotated surface code, which state that the logical error rate per code cycle is approximately given by~\cite{Fowler2018-or}
\begin{align}
    p_L(p_{\rm ph}, d ) &= 0.1 \times (100p_{\rm ph})^{(d+1)/2}. 
    \label{eq:logical_error_clifford}
\end{align}
The optimal code distance $d$ is determined such that the probability of a single logical error on any of the logical patches is less than 1\%, following prior studies~\cite{Babbush2018-bs,Litinski2019-qf,Toshio2025-nn}. For a quantum circuit composed of $N_{\rm rot}$ Pauli rotations with rotation angles $\{ \theta_j \}_{j=1,\ldots,N_{\rm rot}}$, this condition is formally expressed as
\begin{align}
    p_L(p_{\rm ph}, d )^{-1} \geq 100 \times d \, N_{\rm patch} C_{\mathrm{total}} , 
    \label{eq:code_dist_condition}
\end{align}
where $N_{\rm patch}$ is the number of logical code patches and $C_{\mathrm{total}}$ denotes the total number of ``clocks"—corresponding to $d$ code cycles—required to execute the circuit, and is defined as
\begin{align}
    C_{\mathrm{total}} := \sum_{j=1}^{N_{\rm rot}} C_{\rm smm}(\theta_j). 
    \label{eq:total_clock_star}
\end{align}
Here, $C_{\rm smm}(\theta_L)$ denotes the number of clocks required to execute a single analog rotation with angle $\theta_L$ using SMM. 
The remaining Clifford errors, kept to at most 1\%, can then be mitigated with almost negligible overhead using probabilistic error cancellation~\cite{Suzuki2022-fz,Huggins2025-bo}. We determine the optimal code distance $d_m$ for each circuit $\mathcal{C}_m$ according to Eq.~\eqref{eq:code_dist_condition}. Note that the code distance can be chosen separately for each circuit $\mathcal{C}_m$, as each is executed independently in the single-ancilla QPE scheme. 

Once the code distance is determined, the physical space-time costs for the RPE algorithm can be estimated. The number of physical qubits required per quantum processing unit (QPU) is estimated as $\mathcal{Q}_{\rm QPU} = N_{\rm patch} \times 2d_{\rm max}^2$, where $d_{\mathrm{max}}\coloneqq d_M$ is the maximum code distance across the circuit set $\{ \mathcal{C}_m \}_{m=0,1,\ldots,M}$. Throughout this paper, we assume that lattice surgery operations are executed using the \emph{fast block layout}~\cite{Litinski2019-qf}, which requires $2N_{L} + \sqrt{8N_{L}} + 1$ patches to store $N_{L}$ logical data qubits. In the SMM protocol, an additional 10 patches are introduced to ensure a sufficient supply of non-Clifford ancilla states, resulting in a code layout with $N_{\rm patch}=2N_{L} + \sqrt{8N_{L}} + 11$~\cite{Toshio2026}. The time cost is estimated by specifying a physical timescale for a single code cycle. Current superconducting qubit chips achieve single-round syndrome measurements in less than 1 \si{\micro\second}~\cite{Arute2019-re,Google_Quantum_AI2023-rq}. Therefore, we assume a 1 \si{\micro\second} duration for a single code cycle, consistent with previous works~\cite{Babbush2018-bs,Kivlichan2020-hf,Yoshioka2024-sw,Toshio2025-nn}. The circuit runtime per single shot of the RPE algorithm is therefore evaluated as 
\begin{align}
    \mathcal{T}_{m} = C_{\mathrm{total},m} d_m \quad [\si{\micro\second}], 
    \label{eq:circuit_runtime}
\end{align}
where $C_{\mathrm{total},m}:= \sum_{j=1}^{G_m} C_{\mathrm{smm}}(\theta_j)$ denotes the total number of clocks required for executing circuit $\mathcal{C}_m$. Since the last round ($m=M$) requires the deepest circuit, the maximum per-shot circuit runtime is given by $\mathcal{T}_{\rm max}\coloneqq\mathcal{T}_{M}$. Multiplying the sampling cost by each circuit runtime $\mathcal{T}_{m}$, the total time-to-solution is formally expressed as
\begin{align}
    \mathcal{T}_{\rm total} = \sum_{m=0}^{M}e \gamma_{m}^2 N_{m} \mathcal{T}_{m}  \quad [\si{\micro\second}], 
    \label{eq:total_runtime}
\end{align}
where $\gamma_{m}^2 \simeq e^{4P_{\mathrm{total},m}}$ with $P_{\mathrm{total},m}=\sum_{j=1}^{G_{m}}\alpha_{\rm RUS}(\theta_j)\cdot |\theta_j| \cdot p_{\rm ph}$ is the sampling overhead for probabilistic error cancellation, as given by Eq.~\eqref{eq:pec_gamma_total}. The prefactor $e\simeq 2.72$ arises from the sampling overhead associated with the randomized product formula.

To estimate physical space-time resources according to this procedure, we also need to specify the actual rotation-angle dependence of $C_{\rm smm}(\theta_L)$ and $\alpha_{\rm RUS}(\theta_L)$. This dependence is determined by implementation details of SMM, such as the threshold $\theta_{\rm th}$ and the physical error rate $p_{\rm ph}$. We remark here that the SMM protocol exhibits a trade-off between per-gate execution time $C_{\rm smm}(\theta_L)$ and the logical error factor $\alpha_{\rm RUS}(\theta_L)$, especially when the target logical rotation angle $\theta_L$ is relatively large~\cite{Toshio2026}. In the SMM protocol, large-angle rotation gates are primarily executed through the magic state cultivation process~\cite{Gidney2024-lq}, which is accurate but slow due to $T$-gate synthesis. This process can be accelerated by directly executing such large-angle rotations without $T$-gate synthesis, by choosing a sufficiently large threshold ratio $\theta_{\rm th}/\theta_L$. However, this simultaneously increases the failure probability of the RUS process and the logical error rate.

Maximizing the performance of SMM therefore requires careful adjustment of the threshold angle $\theta_{\rm th}$. In this paper, for simplicity, we determine $C_{\rm smm}(\theta_L)$ and $\alpha_{\rm RUS}(\theta_L)$ for two distinct patterns:
\begin{enumerate}
    \item In the ``accuracy-prioritized" setting, the threshold $\theta_{\rm th}$ is optimized for each target angle $\theta_L$ such that the resulting logical error rate and RUS factor $\alpha_{\rm RUS}(\theta_L)$ are minimized.
    \item In the ``speed-prioritized" setting, the maximum threshold ratio $\theta_{\rm th}/\theta_L$ is fixed to a sufficiently large value, leading to low accuracy but fast execution of large-angle analog rotations.
\end{enumerate}
Further technical details and the actual $\theta_L$-dependence of $C_{\rm smm}(\theta_L)$ and $\alpha_{\rm RUS}(\theta_L)$ are presented in Appendix~\ref{append:resource_assumptions}. It is worth noting that a more granular adjustment of the threshold ratio $\theta_{\rm th}/\theta_L$ might further reduce space-time resources. Such optimization of SMM performance is left for future study.

Finally, we account for the possibility of parallel execution across multiple QPUs operating concurrently. For a given physical-qubit budget $\mathcal{Q}_{\rm budget}$, we define the maximum parallelism factor $k^\ast=\lfloor \mathcal{Q}_{\rm budget}/\mathcal{Q}_{\rm QPU} \rfloor$ as the largest number of identical QPUs that can be instantiated simultaneously without exceeding the total qubit budget. The corresponding parallelized time-to-solution is obtained by dividing the single-QPU runtime by $k^\ast$. 
Here, we emphasize that these QPUs are assumed to operate independently without coherent interconnection, in contrast to distributed quantum computing architectures~\cite{Caleffi2024-kw,Mohseni2024-ji}.
The distinction between single-QPU and multi-QPU execution is essential for assessing the feasibility of single-ancilla QPE schemes (such as RPE), where the total time-to-solution can be substantially shortened by executing individual circuits in parallel within a fixed hardware budget. For the following resource estimation, we set $\mathcal{Q}_{\rm budget}=5 \times 10^5$. 
This budget is chosen to match the computational requirements. As shown in Sec.~\ref{subsec:resource_results}, each QPU requires approximately $8\times10^{4}$ to $4\times10^{5}$ physical qubits across the 20--50-orbital regime considered here, allowing modest parallelization for the smaller instances. The budget also represents a meaningful hardware scale, as $\mathcal{O}(10^{5})$ physical qubits is expected to fit within a single dilution refrigerator in superconducting architectures~\cite{Mohseni2024-ji}. These qubit counts, however, reflect our error-correction assumptions rather than an intrinsic requirement of the algorithm: they assume a rotated surface code with $p_{\rm ph}=10^{-3}$ and would decrease with a lower physical error rate or a more efficient code. For comparison, Appendix~\ref{append:ftqc} presents estimates for qubitized QPE in a full-fledged FTQC architecture, which requires millions of physical qubits for the molecular models considered here under the same rotated-surface-code assumptions~\cite{Reiher2017-bd,Berry2019-sy,Von_Burg2021-du,Lee2021-tz,Rocca2024-yx,Caesura2025-tc,Low2025-tb,Goings2022-so}.

It is also useful to define two spacetime volumes built from the quantities above:
\begin{align}
  \mathcal{V}^{\max}_{\rm coh} = \mathcal{Q}_{\rm QPU}\cdot\mathcal{T}_{\max},
  \quad
  \mathcal{V}_{\rm total} = \mathcal{Q}_{\rm QPU}\cdot\mathcal{T}_{\rm total} .
  \label{eq:volumes}
\end{align}
The former measures the footprint of a single uninterrupted circuit execution and thus the minimum indivisible demand on the hardware, whereas the latter measures the total computational cost. For single-ancilla QPE schemes such as RPE, these volumes differ by the number of circuit repetitions and can therefore span several orders of magnitude. By contrast, they nearly coincide for algorithms that execute a single long circuit, such as qubitized QPE. Importantly, $\mathcal{V}_{\rm total}$ is invariant under the parallelization described above: using $k$ QPUs increases the total qubit count by a factor of $k$ while reducing the runtime by the same factor. It is therefore a suitable metric for comparing architectures with different degrees of parallelism. Both volumes are reported for all benchmark models in Appendix~\ref{append:full_resource_estimates}. The corresponding logical volumes, $N_LG_M$ and $N_LG_{\rm total}$, are independent of error-correction assumptions and can be computed directly from the logical resource tables provided there.

\subsection{End-to-end resource estimates for molecules \label{subsec:resource_results}}

\begin{figure*}[tbp]
  \centering
  \includegraphics[width=1.0\textwidth]{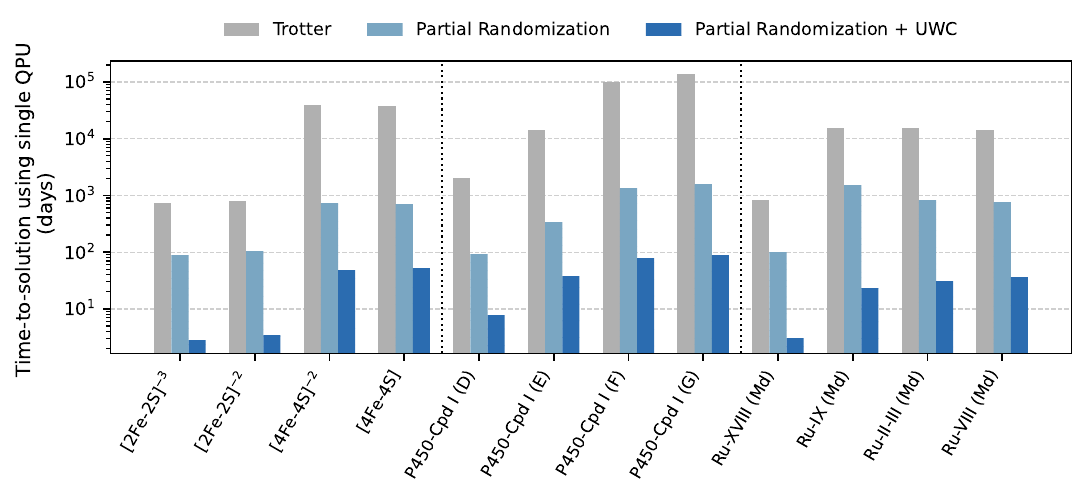}
  \caption{
    Time-to-solution for representative molecular active-space models, utilizing a single QPU. The figure compares three schemes: second-order Trotterization (gray bars), partially randomized time evolution (light-blue cross-hatched bars), and partially randomized time evolution combined with UWC (blue hatched bars), all calculated for an initial-state parameter of $\xi=0.01$. All values are presented in days. The target accuracy $\epsilon$ and physical error rate $p_{\rm ph}$ are assumed to be $\epsilon=1.6$ mHa and $p_{\rm ph}=10^{-3}$, respectively. 
  }
  \label{fig:uwc_runtime_reduction}
\end{figure*}

We now present end-to-end resource estimates for executing QPE on early-FTQC hardware, following the framework described in Sec.~\ref{subsec:resource_estimation_setup}. Our analysis combines the algorithmic gate counts obtained from partially randomized time evolution with UWC-optimized Hamiltonians (Sec.~\ref{sec:hamiltonian}) and the SMM-based resource estimation framework to yield physical space-time costs for a broad set of molecular active-space models. 
The following resource estimations primarily focus on a currently realistic physical error rate of $p_{\rm ph}=10^{-3}$. Resource estimates for more optimistic physical error rates, reflecting anticipated improvements in future hardware performance (specifically, $p_{\rm ph}=5\times10^{-4}$ and $p_{\rm ph}=10^{-4}$), are presented in Sec.~\ref{subsubsec:reduced_error_rate}.

\subsubsection{Resource reduction by partial randomization and UWC}

We first demonstrate the impact of partial randomization and UWC on physical resource estimates. 
Figure~\ref{fig:uwc_runtime_reduction} compares the single-QPU time-to-solution $\mathcal{T}_{\rm total}$ for the initial-state parameter $\xi=0.01$ across three schemes: second-order Trotterized evolution, partially randomized product-formula evolution, and partially randomized product-formula evolution combined with UWC. 
For the second-order Trotterized evolution estimates, we truncate the smallest coefficients of the Hamiltonian $\hat{H}$ such that the total weight of truncated terms is at most a small threshold $10^{-3}$~\cite{Gunther2025-zk}. 
The partially randomized product formula is composed of deterministic second-order Trotterization and randomized qDRIFT parts, as described in Sec.~\ref{subsec:uwc_coefficients}. 
Across all benchmark models, partial randomization alone achieves a one-to-two orders of magnitude reduction in time-to-solution from fully deterministic Trotterization. Combining UWC with partial randomization further yields an additional one-to-two orders of magnitude reduction in time-to-solution compared to the partial randomization alone. In total, partial randomization combined with UWC achieves a two-to-three orders of magnitude reduction in time-to-solution, reflecting the gate-count reduction shown in Fig.~\ref{fig:aggregate_gates}. 

The absolute values of the time-to-solution in Fig.~\ref{fig:uwc_runtime_reduction} are also noteworthy. 
For Trotterized evolution of the original Hamiltonian representation, the time-to-solution lies in the range of $10^3$ to $10^5$ days, or equivalently, several years to centuries, rendering it impractical. 
Partial randomization and UWC reduce this unrealistic time-to-solution to a realistic range of several days to weeks.  

It is also worth noting that the code distance and the resulting physical qubit count per QPU ($\mathcal{Q}_{\rm QPU}$) are reduced due to the gate count reduction enabled by partial randomization and UWC.
For instance, the physical qubit count required to simulate the P450-Cpd I (E) model for $\xi=0.01$ is $\mathcal{Q}_{\rm QPU}=2.32 \times 10^5$ for Trotterization, while $\mathcal{Q}_{\rm QPU}=1.41\times 10^5$ for partial randomization combined with UWC, representing an approximately 40\% reduction in physical qubit overhead.
This physical qubit reduction also influences the time-to-solution when using multiple QPUs, since the available QPU parallelism $k^\ast$ increases under a fixed qubit budget.

\subsubsection{Physical resource estimates for molecular active-space models }

We now present physical resource estimates for a range of chemically relevant molecular active-space models employing partial randomization combined with UWC. 

\begin{table*}[tbp]
\caption{ 
    Physical resource estimates for the RPE algorithm applied to representative molecular active-space models. These estimates are derived from partially randomized product-formula circuits and UWC-optimized Pauli-LCU Hamiltonian representations. The initial-state infidelity parameter is set to $\xi=0.01$, corresponding to an initial-state overlap $p_0=|\braket{\psi_0|\psi}|^2 > 0.99$. The target accuracy $\epsilon$ and physical error rate $p_{\rm ph}$ are assumed to be $\epsilon=1.6$ mHa and $p_{\rm ph}=10^{-3}$, respectively. 
}
\label{tab:resource_summary}
\begin{ruledtabular}
\begin{tabular}{lccccc}
    Molecule ID & \makecell{Physical qubits\\per QPU} & \makecell{Maximum per-shot\\runtime [\si{\second}]} &  \makecell{Time-to-solution\\(single QPU) [days]} & \makecell{QPU parallelism $k^*$\\($\mathcal{Q}_{\rm budget}=5\times10^5$)} & \makecell{Time-to-solution\\($k^\ast$ QPUs) [days]} \\
    \midrule
    \multicolumn{6}{c}{(a) Iron-sulfur clusters }\\
    \midrule
$\mathrm{[2Fe\mathchar`-2S]}^{-3}$ & $8.02\times 10^{4}$ & $2.51\times 10^{0}$ & $2.81\times 10^{0}$ &      6 & $4.68\times 10^{-1}$ \\
$\mathrm{[2Fe\mathchar`-2S]}^{-2}$ & $8.02\times 10^{4}$ & $3.21\times 10^{0}$ & $3.37\times 10^{0}$ &      6 & $5.62\times 10^{-1}$ \\
$\mathrm{[4Fe\mathchar`-4S]}^{-2}$ & $1.60\times 10^{5}$ & $5.07\times 10^{1}$ & $4.84\times 10^{1}$ &      3 &  $1.61\times 10^{1}$ \\
     $\mathrm{[4Fe\mathchar`-4S]}$ & $1.60\times 10^{5}$ & $5.12\times 10^{1}$ & $5.15\times 10^{1}$ &      3 &  $1.72\times 10^{1}$ \\
    \midrule
    \multicolumn{6}{c}{(b) Cytochrome P450 }\\ 
    \midrule
    P450-Cpd I (D) & $8.98\times 10^{4}$ &  $6.55\times 10^{0}$ &  $7.74\times 10^{0}$ &      5 &  $1.55\times 10^{0}$ \\
    P450-Cpd I (E) & $1.41\times 10^{5}$ &  $3.39\times 10^{1}$ &  $3.74\times 10^{1}$ &      3 &  $1.25\times 10^{1}$ \\
    P450-Cpd I (F) & $2.15\times 10^{5}$ &  $7.08\times 10^{1}$ &  $7.64\times 10^{1}$ &      2 &  $3.82\times 10^{1}$ \\
    P450-Cpd I (G) & $2.24\times 10^{5}$ &  $8.11\times 10^{1}$ &  $8.74\times 10^{1}$ &      2 &  $4.37\times 10^{1}$ \\
    \midrule
    \multicolumn{6}{c}{(c) Ruthenium-based catalysts for \ce{CO2} utilization }\\
    \midrule
    Ru-XVIII (Md) & $8.02\times 10^{4}$ &  $2.87\times 10^{0}$ &  $3.05\times 10^{0}$ &      6 & $5.09\times 10^{-1}$ \\
     Ru-IX (Md) & $1.21\times 10^{5}$ &  $1.98\times 10^{1}$ &  $2.31\times 10^{1}$ &      4 &  $5.77\times 10^{0}$ \\
     Ru-II-III (Md) & $1.33\times 10^{5}$ &  $2.08\times 10^{1}$ &  $3.05\times 10^{1}$ &      3 &  $1.02\times 10^{1}$ \\
   Ru-VIII (Md) & $1.33\times 10^{5}$ &  $2.17\times 10^{1}$ &  $3.61\times 10^{1}$ &      3 &  $1.20\times 10^{1}$ \\
\end{tabular}
\end{ruledtabular}
\end{table*}
\begin{figure*}[htbp]
  \centering
  \includegraphics[width=1.0\textwidth]{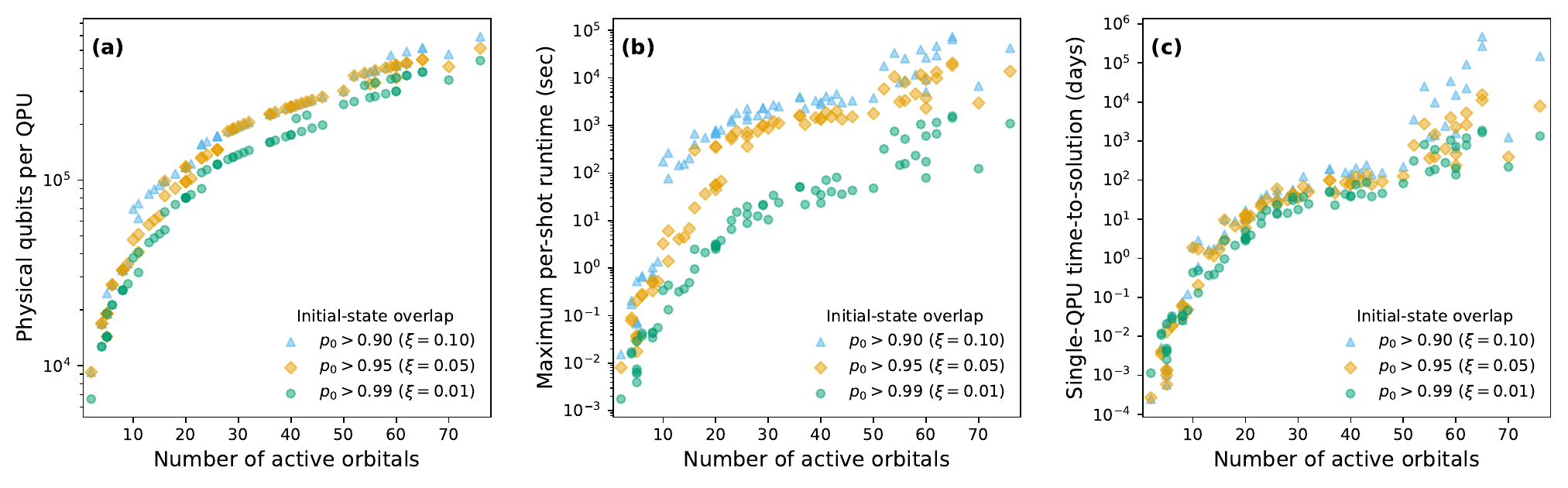}
  \caption{
    Scaling of key resource components for partially randomized RPE on SMM-based partially fault-tolerant quantum computing architectures. (a) Number of physical qubits required per QPU ($\mathcal{Q}_{\rm QPU}$), (b) Maximum per-shot runtime ($\mathcal{T}_{\rm max}$), and (c) time-to-solution for single QPU execution ($\mathcal{T}_{\rm total}$) as a function of the number of active orbitals. The marker shape in all panels indicates the initial-state infidelity parameter $\xi$, along with the corresponding lower bound for the initial-state overlap $p_0=|\braket{\psi_0|\psi}|^2$. Panels (b) and (c) use different time units to emphasize the distinct physical meanings of per-circuit execution cost and overall computational effort. All data points are derived from UWC-optimized Hamiltonian representations of benchmark molecular active-space models, detailed in Appendix~\ref{append:models_detail}, encompassing hydrogen chains, iron-sulfur clusters, P450 active sites, and ruthenium-based catalysts. The target accuracy $\epsilon$ and physical error rate $p_{\rm ph}$ are assumed to be $\epsilon=1.6$ mHa and $p_{\rm ph}=10^{-3}$, respectively. 
    }
  \label{fig:resource_scaling}
\end{figure*}
Table~\ref{tab:resource_summary} summarizes the physical resource estimates for representative molecular models, first listed in Table~\ref{tab:mol_info_representative}, for $\xi=0.01$. For each molecular instance, we list the number of physical qubits required per QPU ($\mathcal{Q}_{\rm QPU}$), the maximum per-shot circuit execution time ($\mathcal{T}_{\rm max}=\mathcal{T}_M$), the total time-to-solution assuming a single QPU ($\mathcal{T}_{\rm total}$), and the degree of QPU parallelism ($k^\ast=\lfloor \mathcal{Q}_{\rm budget}/\mathcal{Q}_{\rm QPU} \rfloor$) achievable given a budget of $\mathcal{Q}_{\rm budget}=5\times 10^5$ physical qubits. The logical resources used to derive these values, such as the analog rotation gate count and code distance, are provided for all molecular models in Appendix~\ref{append:logical_resource}.

For instance, ground-state energy estimation for the 31-orbital P450-Cpd I (E) model can be executed within 37 days using a QPU with $1.41 \times 10^5$ physical qubits, assuming $p_{\rm ph}=10^{-3}$. 
Assuming an available physical-qubit budget of $\mathcal{Q}_{\rm budget}=5\times 10^5$, parallel execution using $k^\ast=3$ QPUs is possible, reducing the time-to-solution to within 13 days. It is noteworthy that the maximum per-shot runtime $\mathcal{T}_{\rm max}$ is less than 34 seconds, significantly smaller than the total time-to-solution. 
This reflects the characteristic of single-ancilla QPE that makes it suited to early-FTQC hardware: the computation is carried out as a large number of repetitions of low-depth circuits, so that what the device must sustain in one uninterrupted run is far smaller than the total cost.

To complement these representative values, Fig.~\ref{fig:resource_scaling} presents the corresponding resource estimates for the full benchmark set as a function of the number of active orbitals for initial-state infidelity parameters $\xi \in \{0.1, 0.05, 0.01 \}$, corresponding to initial-state overlaps $p_0 > 0.9$. Specifically, the physical qubit requirement per QPU, the maximum per-shot circuit execution time, and the single QPU time-to-solution are plotted for all molecular instances considered in this work. See Appendix~\ref{append:mol_data_full} for model details and actual resource estimate values. 

Across the benchmark set, Fig.~\ref{fig:resource_scaling}(a) demonstrates that the physical qubit requirement per QPU grows almost linearly with active-space size and remains well below $\sim10^6$, a typical requirement for a full-fledged FTQC setting. This linear behavior arises from the fact that the logical qubit count $N_L$ is proportional to the number of orbitals $N$ as $N_L=2N+1$, including $2N$ data qubits encoding spin-orbitals and a single ancilla qubit for the Hadamard test.

In contrast, the maximum per-shot runtime $\mathcal{T}_{\rm max}$ (Fig.~\ref{fig:resource_scaling}(b)) and the single QPU time-to-solution $\mathcal{T}_{\rm total}$ (Fig.~\ref{fig:resource_scaling}(c)) exhibit a much stronger and complex dependence on the number of active orbitals. This primarily reflects the combined effects of the increased gate count and sampling overhead inherent in the SMM protocol. Consequently, molecular models with more than 50 orbitals become impractical to execute on a single QPU due to the prohibitive time-to-solution.

We note that the absolute values of these resource estimates strongly depend on assumptions regarding the initial-state overlap, which are quantified by the parameter $\xi$. When the initial-state overlap is increased (or equivalently, $\xi$ is decreased), the maximum per-shot runtime $\mathcal{T}_{\rm max}$ can be significantly reduced, largely in accordance with the algorithmic scaling shown in Table~\ref{tab:part_rand}. The behavior of the total time-to-solution $\mathcal{T}_{\rm total}$ is primarily determined by the combined effects of the algorithmic scaling and the error mitigation cost of SMM. Specifically, $\mathcal{T}_{\rm total}$ increases exponentially, as shown in Eq.~\eqref{eq:total_runtime}, when both the maximum circuit depth and the resulting total logical error rate $P_{\rm total}$ are relatively large. Such behavior is observed in Fig.~\ref{fig:resource_scaling}(c) for models with more than 50 orbitals, where the value of $\mathcal{T}_{\rm total}$ increases exponentially for larger $\xi$.

\subsubsection{Feasibility map \label{subsec:feasibility}}

\begin{figure*}[tbp]
  \centering
  \includegraphics[width=1.0\textwidth]{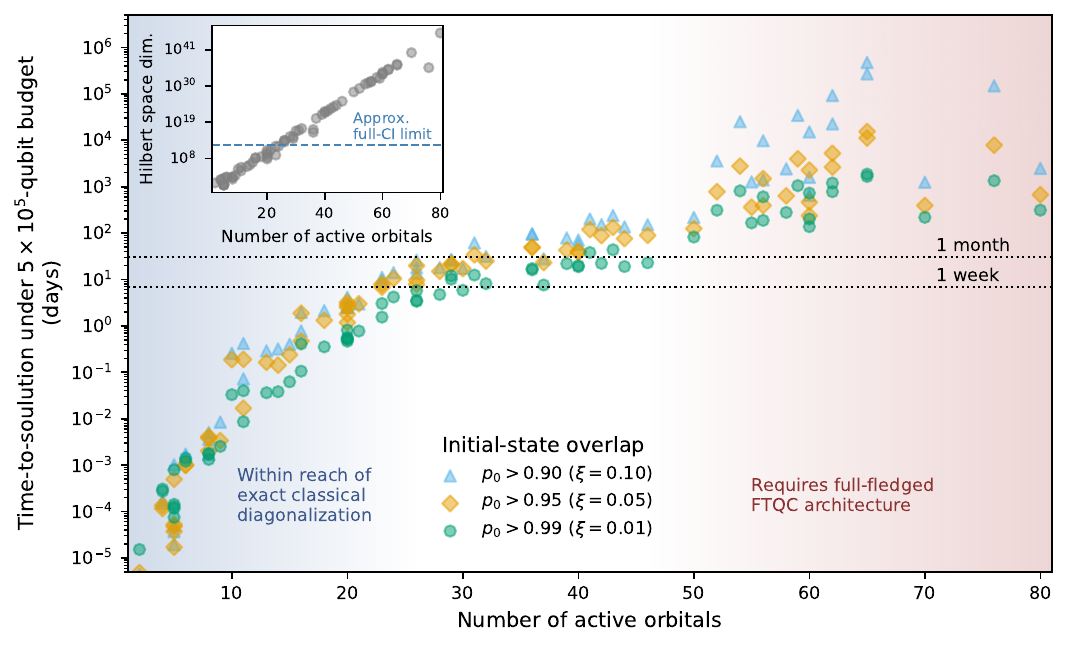}
  \caption{
    Feasibility map for ground-state energy estimation via partially randomized RPE under a fixed physical-qubit budget of $\mathcal{Q}_{\rm budget}=5 \times 10^5$. The figure shows the time-to-solution within the SMM-based partially fault-tolerant quantum computing architecture as a function of the number of active orbitals.
    For each system, the degree of parallelism is chosen as the maximum allowed by the qubit budget, enabling the independent execution of Hadamard test circuits.
    The marker shape indicates the initial-state infidelity parameter $\xi$, along with the corresponding lower bound for the initial-state overlap $p_0=|\braket{\psi_0|\psi}|^2$.
    The horizontal dotted lines mark representative timescales of 1 week and 1 month, indicating practical targets for sustained quantum-chemical calculations. 
    Shaded regions denote the regime within reach of exact classical diagonalization (blue-gray) and that requiring a full-fledged FTQC architecture (red-gray), while the intermediate region highlights the operating window of early-FTQC processors. The blue-gray boundary is that of exact diagonalization (full-CI) alone; quasi-exact classical methods such as DMRG and HCI remain applicable well beyond it.
    The inset shows the Hilbert space dimension as a function of the number of active orbitals; the approximate full-CI limit of $\sim10^{12}$ is indicated by the dashed line.
    Together, this map identifies a range of chemically relevant systems beyond the reach of classical full-CI methods for which accurate ground-state energy estimation may be achievable on early-FTQC hardware.
    For each data point, the reported time-to-solution corresponds to the value obtained for UWC-optimized Hamiltonian representations of benchmark molecular active-space models, detailed in Appendix~\ref{append:models_detail}, encompassing hydrogen chains, iron-sulfur clusters, P450 active sites, and ruthenium-based catalysts. The target accuracy $\epsilon$ and physical error rate $p_{\rm ph}$ are assumed to be $\epsilon=1.6$ mHa and $p_{\rm ph}=10^{-3}$, respectively.
  }
  \label{fig:feasibility_map}
\end{figure*}
To synthesize the resource estimates presented previously and to clarify the practical scope of early-FTQC for chemistry, we construct a feasibility map that summarizes the execution requirements across system size and runtime constraints. This map provides a compact, architecture-aware view of which molecular active-space models can realistically be addressed within a fixed physical-qubit budget and execution-time horizon.

\begin{table*}[tbp]
\centering
\caption{
    Sensitivity of resource estimates to the assumed physical error rate. This table compares the physical space-time cost for representative benchmark systems under two different physical error rates, $p_{\rm ph}=10^{-3}$, $p_{\rm ph}=5\times 10^{-4}$, and $p_{\rm ph}=10^{-4}$, for $\xi=0.01$. Specifically, it lists the physical qubits per QPU and single-QPU time-to-solution for selected models.
    The data for $p_{\rm ph}=10^{-4}$ is derived using the SMM protocol's performance calibrated at $p_{\rm ph}=5\times 10^{-4}$, as further explained in the text.
}
\label{tab:error_rate_sensitivity}
\begin{ruledtabular}
    \begin{tabular}{lccccccc}
    &
    &
    \multicolumn{3}{c}{Physical qubits per QPU} &
    \multicolumn{3}{c}{Time-to-solution (single QPU) [days]} \\
    \cmidrule(lr){3-5} \cmidrule(lr){6-8}
    Molecule ID & Orbitals & 
    $p_{\rm ph}=10^{-3}$ & $p_{\rm ph}=5\times10^{-4}$ & $p_{\rm ph}=10^{-4}$ &
    $p_{\rm ph}=10^{-3}$ & $p_{\rm ph}=5\times10^{-4}$ & $p_{\rm ph}=10^{-4}$ \\
    \midrule
    \multicolumn{8}{c}{(a) Iron-sulfur clusters }\\
    \midrule
$\mathrm{[2Fe\mathchar`-2S]}^{-3}$ & 20 & $8.02\times 10^{4}$ & $5.00\times 10^{4}$ & $1.80\times 10^{4}$ & $2.81\times 10^{0}$ & $1.55\times 10^{0}$ & $6.40\times 10^{-1}$ \\
$\mathrm{[2Fe\mathchar`-2S]}^{-2}$ & 20 & $8.02\times 10^{4}$ & $5.00\times 10^{4}$ & $1.80\times 10^{4}$ & $3.37\times 10^{0}$  & $1.90\times 10^{0}$ & $8.07\times 10^{-1}$\\
$\mathrm{[4Fe\mathchar`-4S]}^{-2}$ & 36 & $1.60\times 10^{5}$ & $1.05\times 10^{5}$ & $4.38\times 10^{4}$ & $4.84\times 10^{1}$ & $2.79\times 10^{1}$ & $1.34\times 10^{1}$ \\
$\mathrm{[4Fe\mathchar`-4S]}$ & 36 & $1.60\times 10^{5}$ & $1.05\times 10^{5}$ & $4.38\times 10^{4}$ & $5.15\times 10^{1}$ & $2.96\times 10^{1}$ & $1.37\times 10^{1}$ \\
    \midrule
    \multicolumn{8}{c}{(b) Cytochrome P450 }\\ 
    \midrule
    P450-Cpd I (D) & 23 & $8.98\times 10^{4}$ &  $5.60\times 10^{4}$ & $2.02\times 10^{4}$ &  $7.74\times 10^{0}$ & $4.02\times 10^{0}$ & $1.64\times 10^{0}$ \\
    P450-Cpd I (E) & 31 & $1.41\times 10^{5}$ & $9.22\times 10^{4}$ & $3.86\times 10^{4}$ & $3.74\times 10^{1}$ & $2.14\times 10^{1}$ & $9.23\times 10^{0}$ \\
    P450-Cpd I (F) & 41 & $2.15\times 10^{5}$ & $1.17\times 10^{5}$ & $4.91\times 10^{4}$ &  $7.64\times 10^{1}$ & $3.51\times 10^{1}$ & $1.65\times 10^{1}$ \\
    P450-Cpd I (G) & 43 & $2.24\times 10^{5}$ & $1.22\times 10^{5}$ & $5.12\times 10^{4}$ & $8.74\times 10^{1}$ & $4.13\times 10^{1}$ &  $1.92\times 10^{1}$ \\
    \midrule
    \multicolumn{8}{c}{(c) Ruthenium-based catalysts for \ce{CO2} utilization }\\
    \midrule
   Ru-XVIII (Md) & 20 & $8.02\times 10^{4}$ & $5.00\times 10^{4}$ & $1.80\times 10^{4}$ & $3.05\times 10^{0}$ & $1.65\times 10^{0}$ & $7.20\times 10^{-1}$ \\
     Ru-IX (Md) & 26 & $1.21\times 10^{5}$ & $6.19\times 10^{4}$ & $2.23\times 10^{4}$ & $2.31\times 10^{1}$ & $1.18\times 10^{1}$ & $4.50\times 10^{0}$ \\
 Ru-II-III (Md) & 29 & $1.33\times 10^{5}$ & $6.78\times 10^{4}$ & $3.65\times 10^{4}$ & $3.05\times 10^{1}$ & $1.37\times 10^{1}$ & $5.89\times 10^{0}$  \\
   Ru-VIII (Md) & 29 & $1.33\times 10^{5}$ & $6.78\times 10^{4}$ & $3.65\times 10^{4}$ & $3.61\times 10^{1}$ & $1.45\times 10^{1}$ & $6.17\times 10^{0}$ \\
    \end{tabular}
\end{ruledtabular}
\end{table*}

Figure~\ref{fig:feasibility_map} plots the estimated time-to-solution for executing the partially randomized RPE algorithm under a fixed total physical-qubit budget of $\mathcal{Q}_{\rm budget}=5 \times 10^5$ as a function of the number of active orbitals $N$. For each molecular instance, the plotted time-to-solution is derived assuming parallel execution using $k^\ast$ QPUs, as described in Sec.~\ref{subsec:resource_estimation_setup}. This representation explicitly incorporates the trade-off among circuit depth, repeated sampling, and available parallelism, which is central to single-ancilla QPE algorithms in the early-FTQC regime.

The feasibility map delineates three qualitatively distinct regions. 
At small orbital counts, specifically $N \lesssim 20$, the problem lies within the reach of exact classical diagonalization (full-CI)~\cite{Vogiatzis2017-fc,Gao2024-ac}, as also shown in the inset of Fig.~\ref{fig:feasibility_map}, implying that quantum computation offers no clear advantage. 
We stress that this boundary marks the regime accessible to direct full-CI calculations. DMRG and selected configuration interaction can provide quasi-exact estimates well beyond it, and their practical reach is not determined by orbital count alone; we return to the comparison with classical methods in Sec.~\ref{sec:conclusion}.
Conversely, for very large orbital counts, $N \gtrsim 50$, the required time-to-solution far exceeds practical limits even when full parallelism under the fixed qubit budget is exploited, placing these systems beyond the reach of our early-FTQC framework. Between these extremes, the map identifies an intermediate regime in which ground-state energy estimation via RPE can be executed within days to weeks, utilizing $\sim10^5$ physical qubits.
Importantly, this qubit scale remains below the regime typically associated with multi-cryostat distributed architectures~\cite{Mohseni2024-ji}, reinforcing the practical relevance of this intermediate window within the early-FTQC regime.

Notably, this intermediate region encompasses active-space models with approximately 20--50 orbitals, which are already well beyond the capabilities of exact classical full-CI computation~\cite{Vogiatzis2017-fc,Gao2024-ac}. 
Although these systems remain smaller than those with more than 50 orbitals—typically targeted in proposals for full-fledged fault-tolerant, millions-of-qubit quantum computers~\cite{Babbush2025-id,Reiher2017-bd,Berry2019-sy,Von_Burg2021-du,Lee2021-tz,Rocca2024-yx,Caesura2025-tc,Low2025-tb,Goings2022-so}—they represent a scientifically meaningful and practically relevant class of problems for early fault-tolerant quantum chemistry simulations.

We emphasize that the boundaries shown in Fig.~\ref{fig:feasibility_map} are not sharp transitions but rather reflect smooth crossovers determined by time-to-solution thresholds, qubit-budget constraints, and assumptions regarding initial-state overlap and error tolerance. Nevertheless, the feasibility map provides a concrete and experimentally relevant target for early-FTQC hardware, clarifying how algorithmic optimizations, such as partially randomized time evolution and UWC, translate into practical capabilities at the system level.

\subsubsection{Resource estimates at reduced physical error rates \label{subsubsec:reduced_error_rate}}

To assess the sensitivity of resource requirements to hardware performance, we present resource estimation data for reduced physical error rates of $p_{\rm ph}=5\times10^{-4}$ and $p_{\rm ph}=10^{-4}$. By providing these estimates and comparing them with those for $p_{\rm ph}=10^{-3}$, we demonstrate the significant impact of improved hardware fidelity on required quantum resources.

Table~\ref{tab:error_rate_sensitivity} summarizes the physical space-time cost, physical qubits per QPU and time-to-solution under single-QPU execution, for representative molecular models (as detailed in Table~\ref{tab:mol_info_representative}) at both $p_{\rm ph}=10^{-3}$ and the more optimistic $p_{\rm ph}=5\times10^{-4}$ and $p_{\rm ph}=10^{-4}$. 
While this table highlights the reductions for representative molecules, a comprehensive analysis of the resource scaling trends across the full benchmark set at $p_{\rm ph}=5\times10^{-4}$ and $p_{\rm ph}=10^{-4}$ is provided in Appendix~\ref{append:error_rate_sensitivity}. 
For the $p_{\rm ph}=10^{-4}$ case, the data are obtained by leveraging the SMM protocol's performance calibrated at $p_{\rm ph}=5\times 10^{-4}$ (see Fig.~\ref{fig:star_magic_mutation} in Appendix~\ref{append:resource_assumptions}). This approach is necessitated because the performance data for $p_{\rm ph}=10^{-4}$ is not directly available in Ref.~\cite{Gidney2024-lq} for magic state cultivation protocols, which form the basis for SMM~\cite{Toshio2026}. Crucially, since the SMM protocol's performance at $p_{\rm ph}=10^{-4}$ is guaranteed to be no worse than at $p_{\rm ph}=5\times 10^{-4}$ (due to higher fidelity), our resource estimates for $p_{\rm ph}=10^{-4}$ provide a conservative upper bound.

As quantitatively demonstrated in Table~\ref{tab:error_rate_sensitivity}, reducing the physical error rate has significant impacts on resource requirements. Halving the physical error rate from the baseline $p_{\rm ph}=10^{-3}$ to $p_{\rm ph}=5\times10^{-4}$ reduces the number of physical qubits by approximately 34--49\% and the single-QPU time-to-solution by about 42--60\%. This effect is even more pronounced as the error rate improves to $p_{\rm ph}=10^{-4}$, leading to a substantial reduction of approximately 73--82\% in physical qubits and 72--83\% in the time-to-solution, compared to the $p_{\rm ph}=10^{-3}$ case. 
These findings underscore that lowering the physical error rate directly expands the scope of solvable problems under practical time and qubit constraints. This improvement translates to an effective expansion of the feasibility window shown in Fig.~\ref{fig:feasibility_map}, critically advancing the timeline for achieving practical chemistry simulations on quantum computers. 

\subsubsection{Comparison with fully fault-tolerant estimates}
\label{subsec:ftqc_comparison}
 
To clarify the trade-off between coherent-circuit requirements and total computational cost, we compare our partially fault-tolerant estimates with full-fledged FTQC estimates based on qubitized QPE. Four molecular models are common to both analyses: [2Fe-2S]$^{-2}$ ($N=20$), [4Fe-4S]$^{-2}$ ($N=36$), FeMoco (Sm) ($N=54$)~\cite{Reiher2017-bd}, and FeMoco (Lg) ($N=76$)~\cite{Li2019-ah}. The full-fledged FTQC estimates, presented in Appendix~\ref{append:ftqc}, are obtained for qubitized QPE with a double-factorized tensor-hypercontraction (DFTHC) block encoding~\cite{Low2025-tb}. This shared benchmark set enables a side-by-side comparison of the two routes, subject to the qualifications discussed below.

\begin{figure*}[t]
  \includegraphics[width=\textwidth]{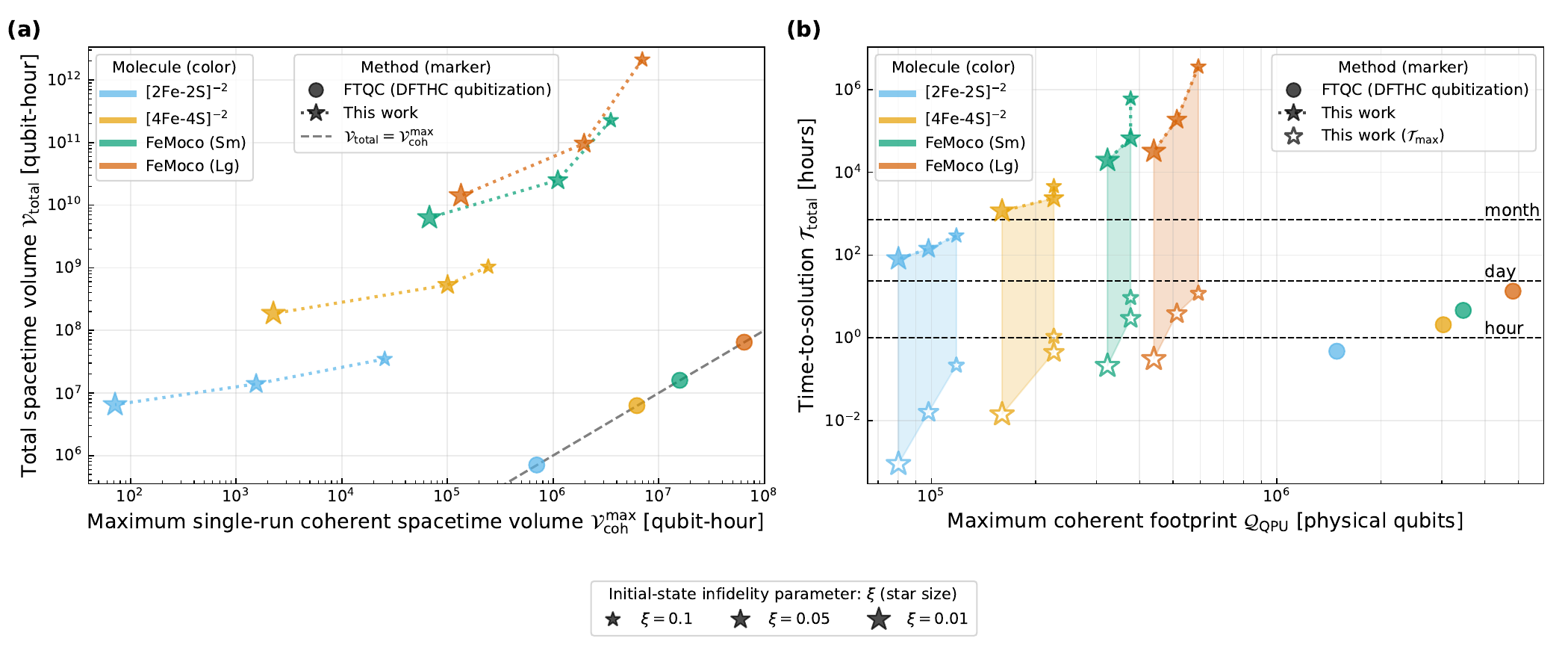}
  \caption{Comparison of the partially and fully fault-tolerant routes for the four shared molecular models. (a) Total spacetime volume $\mathcal{V}_{\rm total}$ versus maximum single-run coherent volume $\mathcal{V}^{\max}_{\rm coh}$ [Eq.~\eqref{eq:volumes}]. The dashed diagonal, $\mathcal{V}_{\rm total}=\mathcal{V}^{\max}_{\rm coh}$, corresponds to a computation executed as a single circuit; the fully fault-tolerant estimates lie on this line. (b) Time to solution $\mathcal{T}_{\rm total}$ versus maximum coherent footprint. Open stars indicate the per-circuit execution time $\mathcal{T}_{\max}$ rather than the time to solution, and the interval between $\mathcal{T}_{\max}$ and $\mathcal{T}_{\rm total}$ represents the cost of repetition. The shaded region shows the time to solution achievable through parallel execution on multiple QPUs. Star size encodes the initial-state infidelity parameter $\xi$, with larger stars corresponding to smaller $\xi$. Both sets of estimates assume a rotated surface code with $p_{\rm ph}=10^{-3}$.
  }
  \label{fig:ftqc_comparison}
\end{figure*}

Figure~\ref{fig:ftqc_comparison}(a) plots $\mathcal{V}_{\rm total}$ against $\mathcal{V}^{\max}_{\rm coh}$. Since a single-circuit computation satisfies $\mathcal{V}^{\max}_{\rm coh}=\mathcal{V}_{\rm total}$, the fully fault-tolerant estimates, obtained assuming an input state with nearly unit ground-state overlap, lie on the diagonal. The partially fault-tolerant estimates lie above and to its left: the horizontal displacement quantifies the reduction in coherent demand, whereas the vertical displacement quantifies the increase in total cost. Across the four models, the coherent volume of a single circuit is two to four orders of magnitude smaller than that of a qubitized-QPE run, whereas the total volume is one to three orders of magnitude larger. The latter difference grows from approximately one order of magnitude near the lower end of the 20--50-orbital regime to two or three orders beyond it. Figure~\ref{fig:ftqc_comparison}(b) presents the same instances in terms of time-to-solution and maximum coherent footprint. The separation between $\mathcal{T}_{\max}$ and $\mathcal{T}_{\rm total}$ represents the cost of circuit repetition.

This comparison makes the trade-off explicit. The partially fault-tolerant route does not reduce the total spacetime volume; instead, it decomposes the computation into many short, independent circuits. Shorter circuits generally permit a higher logical error rate per operation and hence lower error-correction overhead. Their independence also enables parallel execution: using $k$ QPUs reduces the time to solution by approximately a factor of $k$ while leaving $\mathcal{V}_{\rm total}$ unchanged. The practical choice therefore depends not only on the available qubit count but also on whether the device can sustain coherent execution for many hours. Under the assumptions considered here, qubitized QPE becomes more spacetime-efficient once such long coherent computations are feasible. The increasing total-volume penalty with system size is also the quantitative origin of the upper boundary of the feasibility window discussed in Sec.~\ref{subsec:feasibility}.

Two caveats qualify this comparison. First, it involves different algorithms as well as different architectures, so the observed ratios cannot be attributed to algorithmic scaling or architectural overhead alone. The logical volumes reported in Appendices~\ref{append:full_resource_estimates} and~\ref{append:ftqc} help separate these contributions. Second, although both estimates assume a rotated surface code at the same physical error rate, their compilation strategies differ. In particular, magic-state cultivation is used in the SMM protocol but not in the qubitization estimates of Appendix~\ref{append:ftqc} obtained by using \texttt{Qualtran}~\cite{harrigan2024qualtran}. The resulting ratios should therefore be interpreted as indicative rather than as a fully controlled comparison.

\section{Conclusion \label{sec:conclusion}}

In this work, we investigated the feasibility of QPE-based ground-state energy estimation for chemically relevant molecular systems on early-FTQC hardware. 
The key constraint in this regime is not only the number of available physical qubits but also the maximum spacetime volume of a single coherent execution. Since non-Clifford rotations are error-mitigated rather than implemented using distilled magic states, individual circuits must remain short, and the computation must be divided among many independent repetitions. Within this constraint, we combined single-ancilla QPE with a partially randomized product formula and introduced UWC, a novel Hamiltonian-optimization strategy that reduces the algorithmic cost.

We demonstrated that the partially randomized product formula alone yields substantial reductions in gate count relative to deterministic Trotterization. Furthermore, the additional application of UWC systematically enhances these gains by concentrating the Hamiltonian coefficient distribution. Across a diverse benchmark set of molecular active-space models, including iron-sulfur clusters, cytochrome P450 active sites, and ruthenium-based catalysts, the combined approach achieves gate-count reductions of approximately three orders of magnitude relative to deterministic Trotter baselines. These algorithmic improvements translate directly into reduced circuit depth and non-Clifford gate overheads, which are critical bottlenecks in the early-FTQC regime.

Building on these algorithmic improvements, we performed an end-to-end resource estimation for the STAR architecture using a framework that incorporates the SMM improvement~\cite{Toshio2026}. 
Our analysis accounts for physical qubit overhead, per-shot circuit runtime, and total time-to-solution, including the effects of parallel execution under a fixed qubit budget.
Our results identify a concrete feasibility window in which active-space models with approximately 20--50 orbitals—well beyond the reach of classical full-CI computations—can be addressed within days to weeks of time-to-solution using at most $5 \times 10^5$ physical qubits. 
Comparison with fully fault-tolerant, qubitization-based estimates for the same molecular models reveals an asymmetric trade-off: the spacetime volume of a single coherent run is two to four orders of magnitude smaller, whereas the total spacetime volume is one to three orders of magnitude larger. Thus, the partially fault-tolerant route is not more economical overall; its value lies in enabling useful computations before devices can sustain coherent execution for many hours.
The resulting feasibility map delineates a practically meaningful intermediate regime between systems accessible to exact diagonalization and the large-scale molecular targets envisioned for full-fledged FTQC. 
Notably, this feasibility window lies below the scale at which distributed, multi-cryostat architectures become necessary for fault-tolerant operation~\cite{Mohseni2024-ji}, underscoring its relevance as a realistic intermediate milestone.

Two limitations bound the applicability of this approach, and both deserve to be stated plainly. The first concerns the classical reference point. The 20--50 orbital window lies beyond direct full-CI calculation, but DMRG and selected configuration-interaction methods such as HCI can provide quasi-exact estimates in this regime~\cite{chan2011density,baiardi2020density,Holmes2016-jd,Li2020-ih}. 
The practical cost of these methods is not determined by the orbital count alone. For DMRG, the bond dimension required to reach a given accuracy depends on the entanglement structure of the state, as well as on the orbital basis and ordering~\cite{chan2011density,baiardi2020density}. 
For HCI, the number of selected determinants required depends on how compactly the wavefunction can be represented in the determinant basis generated by the chosen orbitals~\cite{Holmes2016-jd,Li2020-ih}. Consequently, for the chemically complex transition-metal active spaces considered here, convergence of either method would have to be established case by case.
What distinguishes QPE is not that a result cannot be obtained classically, but that its error enters as an algorithmic parameter: given a valid lower bound on the overlap of the initial state with the ground state, and under the stated models for Hamiltonian-simulation and logical errors, the ground-state eigenvalue of the active-space Hamiltonian can be estimated to within a prescribed $\epsilon$ with a quantifiable failure probability, and additional resources can be used systematically to reduce either quantity~\cite{Gunther2025-zk,Ni2023-it}. 
At finite bond dimension or determinant-selection threshold, the variational energies obtained with DMRG and HCI provide upper bounds on the ground-state energy. The remaining errors are generally assessed through convergence studies or extrapolation rather than rigorously bounded, so that having reached chemical accuracy is inferred rather than certified~\cite{chan2011density,baiardi2020density,Holmes2016-jd,Sharma2017-sg}. Moreover, the perturbatively corrected HCI energies commonly reported are not themselves variational~\cite{Holmes2016-jd,Sharma2017-sg}.
Two qualifications should be added. The guarantee applies to the eigenvalue of the chosen active-space Hamiltonian and not to the exact molecular energy; basis-set incompleteness and the choice of active space lie outside it, and are common to both routes. It is conditional on the actual initial-state overlap being at least as large as the assumed lower bound~\cite{Gunther2025-zk,Ni2023-it}, which is an external input to the resource model rather than something the model predicts; this is one reason our estimates are reported as a function of $\xi$.
 
The second is a genuine limitation intrinsic to partial fault tolerance. Because errors in analog rotations are mitigated rather than corrected, their rate scales linearly with the physical error rate $p_{\rm ph}$ rather than being exponentially suppressed with code distance, as for Clifford operations. Consequently, the sampling overhead $\gamma_{\rm total}^{2}\simeq e^{4P_{\rm total}}$ cannot be reduced by adding qubits. Across the 20--50-orbital regime, $P_{\rm total}$ remains below $0.33$ for all instances and initial-state assumptions considered, limiting the overhead to about a factor of four. Beyond 50 orbitals, however, $P_{\rm total}$ approaches and then exceeds unity, causing the overhead to grow by up to two orders of magnitude. This mechanism closes the feasibility window; its boundary should therefore be interpreted as a limitation of the approach, rather than of the systems examined. Lowering $p_{\rm ph}$ shifts this boundary outward, as shown by the estimates for $p_{\rm ph}=5\times10^{-4}$ and $10^{-4}$ in Sec.~\ref{subsubsec:reduced_error_rate} and Appendix~\ref{append:error_rate_sensitivity}, but does not change the exponential scaling.  Possible routes to substantially extending the feasibility window include algorithmic advances that reduce the number or magnitude of analog rotations and architectural improvements that suppress rotation errors beyond their current linear dependence on $p_{\rm ph}$. 

Several important algorithmic directions remain for future work. First, the resource estimates presented here assume a relatively high overlap between the prepared initial state and the true ground state. While algorithms capable of achieving such overlaps are known~\cite{Tubman2018-lg,Lee2023-rd,Fomichev2024-us,Ollitrault2024-ko,Morchen2024-ts,Erakovic2025-cl,Berry2025-dq}, their resource requirements can be substantial for strongly correlated molecules and, in some cases, may be comparable or exceed the cost of the time-evolution circuits used in QPE itself~\cite{Berry2025-dq,Thomas2026-tz}. Achieving practical quantum advantage in quantum chemistry will therefore necessitate the development of state-preparation techniques that can reliably produce high-overlap initial states at a cost comparable to—or lower than—that of the subsequent time-evolution operations.
In addition, while this work focused on Pauli-LCU representations, extending partially randomized time evolution and UWC to alternative LCU forms, such as double factorized LCU representation~\cite{Motta2021-kh} and other fermionic LCU representations~\cite{Patel2025-np,Loaiza2025-mf}, is an interesting avenue for future research. If the effective weight concentration in such LCU representations can be enhanced, even larger gate-count reductions may become possible.
Finally, the UWC framework itself is flexible and could incorporate alternative cost metrics beyond gate count. These could include rotation angles that directly affect execution time and logical error rates in STAR-based architectures~\cite{Toshio2025-nn,Toshio2026}.

From an architectural perspective, our resource estimates considered two representative operating modes of the STAR architecture, corresponding to accuracy-prioritized and speed-prioritized configurations. In practice, these choices could be optimized in a problem-dependent manner, and further runtime reductions may be achievable through architecture-aware compilation. In particular, the STAR architecture admits parallel execution of rotation gates utilizing fermionic swap operations~\cite{Akahoshi2025-ra}. This suggests additional opportunities for acceleration by developing appropriate scheduling and compilation strategies, potentially utilizing a generalized fermionic swap protocol for molecular electronic structure Hamiltonians~\cite{O-Gorman2019-qn}.
Moreover, the STAR architecture itself admits further improvement. Integrating recently proposed techniques~\cite{Zhang2025-rt,Pei2025-gg} may help reduce effective logical error rates, potentially increasing the tolerable circuit depth and relaxing initial-state overlap requirements for reliable QPE. 
While our analysis focused on superconducting platforms, extending this framework to other hardware modalities, such as neutral-atom arrays~\cite{Ismail2025-qp} and trapped-ion systems~\cite{Dasu2026-bl}, is an important future direction. Differences in native gates, physical error rates, and execution times may substantially reshape the resulting feasibility landscape~\cite{Leone2025-ft,Sunami2025-rf,Gratsea2025-tj}.
Finally, the implementation cost of FTQC operations continues to improve, as exemplified by recent advances such as magic state cultivation~\cite{Gidney2024-lq,Hirano2025-mz,Hetenyi2026-ga} and high-rate quantum error correction codes~\cite{Yoder2025-ff,Paul2026-sj,Cain2026-zp,Low2026-fr,Ismail2026-fk}. For instance, Ref.~\cite{Low2026-fr} reported a substantial reduction in the physical-qubit overhead of qubitization-based QPE by using a denser planar surface code and a cultivation-based CCZ factory. These improvements are orthogonal to the algorithmic contributions of the present work and are expected to benefit both partially and fully fault-tolerant approaches, although not necessarily to the same extent. Ongoing benchmarking of early-FTQC approaches, such as the one presented here, against the evolving landscape of full-fledged FTQC constitutes an important direction for future work.

Taken together, our results demonstrate that while full-fledged fault-tolerant quantum computers are envisioned for simulations of even larger molecular models, chemically meaningful quantum chemistry problems can already become accessible in an experimentally relevant early fault-tolerant regime through co-designed algorithms, Hamiltonian representations, and error correction architecture optimized for early-FTQC constraints.

\begin{acknowledgments}
We are grateful to Yutaro Akahoshi, Norifumi Matsumoto, Shinichiro Yamano, Koki Chinzei, and Quoc Hoan Tran for fruitful discussions. We are indebted to Jun Fujisaki, Shintaro Sato, and Keisuke Fujii for their support in fostering a conducive research environment. We also acknowledge Shintaro Sato and Keisuke Fujii for reviewing the manuscript. We acknowledge Masatoshi Ishii for technical support in performing numerical simulations. 
\end{acknowledgments} 

\section*{Data availability}
The data that support the findings of this article are openly available~\cite{kanasugi_github}. 

\section*{Author contributions}
S. K. was responsible for the detailed theoretical framework conceptualization, methodology development, and performing numerical simulations and resource estimation for all figures and tables. S. K. also led the primary investigation, performed formal data analysis, and prepared the original draft of the manuscript. R. T. contributed to the technical discussion on resource estimation in Sec.~\ref {sec:resource_estimation} and provided numerical data for the performance of the STAR-magic mutation protocol in Fig.~\ref{fig:star_magic_mutation}. K. M. and H. O. were instrumental in securing necessary resources and overseeing project administration. All authors participated in the review and editing of the manuscript.

\appendix

\section{Robust phase estimation \label{append:rpe}}
In this Appendix, we detail the RPE algorithm~\cite{Ni2023-it,Gunther2025-zk}, which is briefly reviewed in Sec.~\ref{sec:single_ancilla_prpe} and serves as the cost model throughout this work. 
The analysis presented here follows the results detailed in Refs.~\cite{Ni2023-it,Gunther2025-zk}. 

The core idea of RPE is to estimate the signal $g(t):=\bra{\psi}e^{-it\hat{H}}\ket{\psi}$ for a given state $\ket{\psi}$ at timesteps $t=2^m$ for $m=0,1,\cdots,M$. 
The integer $M$ is set to $M=\lceil\log_2{\epsilon^{-1}}\rceil$ for estimating the ground state energy $E_0$ with target precision $\epsilon$. In the $m$-th round, the Hadamard test shown in Fig.~\ref{fig:measurement_circuit} with $t=2^m$ is repeated $N_m/2$ times each for both the real part ($W=I$) and imaginary part ($W=S^\dagger$) to obtain the sampling mean $\bar{\bm{Z}}_m$, which is an estimate of $g(2^m)=\bra{\psi}e^{-i2^m\hat{H}}\ket{\psi}$. The angle $\arg{(\bar{\bm{Z}}_m)}$ provides an estimate of $2^mE_0$ modulo $2\pi$~\cite{Ni2023-it}. 
Then, the candidate set for $E_0$ at the $m$-th round is obtained as 
\begin{align}
    \mathcal{S}_m := \Set{ \frac{\arg{(\bar{\bm{Z}}_m)} + 2\pi k}{2^m} }_{k=0,1,\cdots,2^{m}-1}. 
\end{align}
Given an estimate $\vartheta_{m-1}$ for $E_0$ from the previous $m-1$ rounds, an updated estimate of the ground-state energy, $\vartheta_m$, is selected from the set $\mathcal{S}_m$ as
\begin{align}
    \vartheta_m = \underset{\vartheta \in \mathcal{S}_m } {\rm argmin} \abs{\vartheta - \vartheta_{m-1}}_{2\pi}, 
\end{align}
where $\abs{\vartheta_1-\vartheta_2}_{2\pi}:=\min_{n\in\mathbb{Z}}\abs{\vartheta_1-\vartheta_2+2n\pi}$ denotes the angular distance between $\theta_1$ and $\theta_2$. 
If the input state $\ket{\psi}$ satisfies $p_0:=|\braket{\psi|\psi_0}|^2>\eta>4-2\sqrt{3}\simeq 0.54$, this procedure achieves the Heisenberg-limited scaling in total runtime~\cite{Ni2023-it}. 

The RPE algorithm can be further refined for the regime where the initial-state overlap $p_0$ approaches 1 (i.e., $\eta\to 1$). In such a scenario, the maximum circuit depth can be reduced to $T_{\rm max}=\mathcal{O}((1-p_0)\epsilon^{-1})$~\cite{Ni2023-it}. 
Specifically, the maximum number of iterations $M$ is modified to $M= \lceil \log_2{(\xi\epsilon^{-1})} \rceil$ with a prefactor $\xi$ satisfying $1 > \xi > \frac{3}{\pi}\arcsin{(\frac{1-\eta}{\eta})}$. 
The modified RPE algorithm then proceeds as summarized in Algorithm~\ref{alg:RPE}. 
\begin{figure}
\begin{algorithm}[H] 
    \caption{RPE algorithm for the large initial-state overlap regime~\cite{Ni2023-it,Gunther2025-zk}.}
    \label{alg:RPE}
    \begin{algorithmic}[1] 
    \State \textbf{Input:} $\epsilon$: target precision, $\eta$: lower bound for the squared overlap $p_0=|\braket{\psi_0|\psi}|^2$, $\xi$: prefactor for the maximal timestep satisfying $1 > \xi > \frac{3}{\pi}\arcsin{(\frac{1-\eta}{\eta})}$, $M=\lceil\log_2{(\frac{\xi}{\epsilon})}\rceil$: number of iterations, $\set{ N_m }^{M}_{m=0}$: schedule of sample numbers per iteration. 
    \State $\vartheta_{-1}\leftarrow 0$ 
    \For{$m=0,1,\ldots,M$}
    \State Construct the estimator $\bar{\bm{Z}}_m$ by performing the Hadamard test (Fig.~\ref{fig:measurement_circuit}) with $t=2^m$ for $N_m/2$ shots each for the real and imaginary parts.
    \State Define a candidate set:
    \[
     \mathcal{S}_m = \Set{ \frac{\arg{(\bar{\bm{Z}}_m)} + 2\pi k}{2^m} }_{k=0,1,\cdots,2^{m}-1}. 
    \]
    \State Update the estimate by selecting the candidate closest to the previous estimate:
  \[
      \vartheta_m \leftarrow \arg\min_{\vartheta \in \mathcal{S}_m}\abs{\vartheta - \vartheta_{m-1}}_{2\pi} \, .
  \] 
  \EndFor
  \State $\vartheta^{\ast}\leftarrow \vartheta_M$
  \State \textbf{Output:} $\vartheta^\ast$ as an estimate of $E_0$. 
  \end{algorithmic}
\end{algorithm}
\end{figure}

In Algorithm~\ref{alg:RPE}, the number of samples $\{ N_m \}$ is determined following Ref.~\cite{Gunther2025-zk}. 
For $m<M$, $N_m$ is chosen as 
\begin{align}
    N_{m<M} &= \frac{2}{\beta^{2}}(\log{(\xi^{-1})}+\log{2}(\alpha(M-m)+1)) , 
    \label{eq:number_of_samples}
\end{align}
where $\beta\coloneqq\eta(1+\sin(\frac{\pi}{3}))-1$ and $\alpha>1$ is a constant. 
Based on Hoeffding's inequality, this choice of $N_m$ ensures that~\cite{Gunther2025-zk}
\begin{align}
    \mathbb{P}\left( \abs{\arg{(\bar{\bm{Z}}_m)} - 2^m E_0}_{2\pi} \geq \frac{\pi}{3} \right) \leq \xi^2 4^{-\alpha(M-m)} . 
    \label{eq:rpe_condition_1}
\end{align}
The final round ($m=M$) requires $N_M=2\xi^{-2}$ samples, leading to the following bound~\cite{Gunther2025-zk}:
\begin{align}
    \mathbb{E}\left( \abs{\arg{(\bar{\bm{Z}}_M)}-2^ME_0}_{2\pi} \right) \lesssim \xi^2. 
    \label{eq:rpe_condition_2}
\end{align}
Under the conditions in Eqs.~\eqref{eq:rpe_condition_1} and \eqref{eq:rpe_condition_2}, Ref.~\cite{Gunther2025-zk} shows that Algorithm~\ref{alg:RPE} yields an estimate $\vartheta^\ast$ of $E_0$ such that
\begin{align}
    \mathbb{E}\left( \abs{\vartheta^\ast - E_0}_{2\pi}^{2} \right) \leq (1+\rho ) \xi^2 4^{-M} ,
    \label{eq:rpe_bound}
\end{align}
where $\rho=\frac{1}{4^{\alpha-1}-1}(\frac{16\pi}{3})^2$ is a constant determined by the choice of $\alpha>1$. 
In this work, we set $\alpha \simeq 10$ such that $\rho \simeq10^{-3}$, which ensures that the prefactor $(1+\rho)$ on the right-hand side of Eq.~\eqref{eq:rpe_bound} is close to unity.
Consequently, the total evolution time $T_{\rm total}$ scales as $T_{\rm total}\simeq N_M T_{\rm max} =\mathcal{O}(\xi^{-1}\epsilon^{-1})$, exhibiting the trade-off between $T_{\rm max}$ and $T_{\rm total}$, governed by $\xi$~\cite{Ni2023-it,Gunther2025-zk}.   

\section{Gate counts for partially randomized robust phase estimation \label{append:prnd}}
In this Appendix, we outline the gate cost model for the RPE algorithm when implemented with partially randomized product formulas. The analysis presented here follows the framework detailed in Ref.~\cite{Gunther2025-zk}. 

As described in Appendix~\ref{append:rpe}, the RPE algorithm proceeds by performing the Hadamard test on the time-evolution unitary $e^{-it\hat{H}}$ for $t=2^m$, where $m=0,1,\cdots,M$. 
We approximate this unitary using a quantum circuit based on the partially randomized product formulas introduced in Sec.~\ref{subsec:pr_time_evolution}. 
The Trotter step size $\delta$ is chosen to suppress this error to a negligible level. 
In this approach, only a part of the Trotter circuit $\hat{S}_{p}(\delta)^r \approx e^{-it\hat{H}}$ is stochastically implemented by randomized product formulas. Consequently, the overall circuit is subject to the inherent Trotter discretization error. 
The step size $\delta$ is determined to suppress this Trotter error to a negligible level. 
For QPE, the Trotter error is naturally quantified by the resulting bias in the ground-state energy, $\epsilon_{\rm trot}\coloneqq|E_0-E_{\mathrm{eff},0}|\leq C_{\rm gs}\delta^{p}$, as given in Eq.~\eqref{eq:trotter_error_energy}. 
The total estimation error arises from two independent sources: this Trotter error and the algorithmic error from the QPE protocol itself, $\epsilon_{\rm qpe}$. To achieve a total target precision $\epsilon$ for the true ground-state energy $E_0$, these errors are typically combined in quadrature, requiring:
\begin{align}
    \epsilon^2 = \epsilon_{\rm qpe}^2 + \epsilon_{\rm trot}^2 \leq \epsilon_{\rm qpe}^2 + C_{\rm gs}^2 \delta^{2p}. 
    \label{eq:qpe_trot_error}
\end{align}
Applying the RPE algorithm to the Trotter unitary $\hat{S}_p(\delta)$ to estimate its eigenenergy $E_{\mathrm{eff},0}$ with precision $\epsilon_{\rm qpe}$ requires a maximum evolution time of $\mathcal{O}(\epsilon_{\rm qpe}^{-1})$. This corresponds to $T_{\rm max}/\delta \simeq \xi(\epsilon_{\rm qpe}\delta)^{-1}$ applications of the Trotter step unitary $\hat{S}_p(\delta)$. Thus, the number of RPE rounds becomes $M=\lceil \log_2(\xi\epsilon_{\rm qpe}^{-1}\delta^{-1}) \rceil$.
The total number of Trotter steps, and therefore the gate count, scales as $\mathcal{O}(\epsilon_{\rm qpe}^{-1}\delta^{-1})=\mathcal{O}(\epsilon_{\rm qpe}^{-1}(C_{\rm gs}/\epsilon_{\rm trot})^{-1/p})$. Under the constraint of Eq.~\eqref{eq:qpe_trot_error}, this gate count is minimized when $\epsilon_{\rm qpe} = \epsilon \sqrt{\frac{p}{1+p}}$, which corresponds to the optimal Trotter step size:
\begin{align}
    \delta = \left( \frac{\epsilon}{C_{\rm gs}} \right)^{\frac{1}{p}} \left( \frac{1}{1+p} \right)^{\frac{1}{2p}}. 
    \label{eq:trotter_step_optimal}
\end{align}
As mentioned in Sec.~\ref{subsubsec:det}, we determine the Trotter constant $C_{\rm gs}=C_{\rm gs}(p,\{\hat{H}_\ell\})$ using the heuristic estimate from Ref.~\cite{Gunther2025-zk}, which shows a strong correlation between $C_{\rm gs}$ and the Hamiltonian's $\ell_1$-norm, $\lambda$. Specifically, for second-order Trotterization ($p=2$), numerical results in Ref.~\cite{Gunther2025-zk} indicate an approximate scaling of $C_{\rm gs}\simeq a\lambda^b$ with $a\simeq 3.41\times 10^{-5}$ and $b\simeq 2.09$. 

In the partially randomized RPE framework, the Hamiltonian $\hat{H}$ is decomposed into a deterministic part $\hat{H}_D$ and a randomized part $\hat{H}_R$, as in Eq.~\eqref{eq:ham_partially_randomized}. In the $m$-th round of RPE, the evolution corresponding to $\hat{H}_R$ within the full Trotterized unitary $(\hat{S}_p(\delta))^{2^m}$ is implemented using a randomized product formula. 
The resulting gate count in the $m$-th round, $G_m$, is given by~\cite{Gunther2025-zk}: 
\begin{align}
    G_m &= G_m^{(\rm det)} + G_m^{(\rm rand)}, \\
    G_m^{(\rm det)} &= \frac{1}{2}C_{\rm gate}N_{\rm stage} 2^{m} L_D \eqqcolon g_m^{(\rm det)}L_D, \label{eq:gate_det} \\
    G_m^{(\rm rand)} &= C_{\rm gate} \gamma\delta^2 2^{2m} \lambda_R^2 \eqqcolon g_m^{(\rm rand)}\lambda_R^2,  \label{eq:gate_rand}
\end{align}
where $G_m^{(\rm det)}$ and $G_m^{(\rm rand)}$ are the gate counts for the deterministic and randomized contributions, respectively. 
Here, $N_{\rm stage}$ is the number of stages in the symmetric decomposition, e.g., $N_{\rm stage}=2\cdot5^{k-1}$ for a $2k$-th order Suzuki-Trotter product formula. The factor of $1/2$ in Eq.~\eqref{eq:gate_det} arises from the ``halving trick" applicable to the Hadamard test for symmetric Suzuki-Trotter product formulas~\cite{Reiher2017-bd,Gunther2025-zk}.  
The constant $\gamma$ depends on the chosen randomized protocol: $\gamma=1$ for qDRIFT and $\gamma=2$ for RTE~\cite{Gunther2025-zk}, and $\gamma=2$ for TE-PAI~\cite{Kiumi2025-qm}.
In Eqs.~\eqref{eq:gate_det} and~\eqref{eq:gate_rand}, $C_{\rm gate}$ is a factor that converts the number of Pauli rotation gates into the number of elementary gates for a specific hardware architecture. 
For the STAR architecture considered in this work, we set $C_{\rm gate}=1$, as single-qubit Pauli rotations are native elementary gates~\cite{Akahoshi2024-hj,Toshio2025-nn}.

\section{Spin-extended BLISS transformation \label{append:BLISS_spin}}

In this Appendix, we detail the extension of the BLISS transformation to incorporate spin symmetries. As explained in Sec.~\ref{subsubsec:BLISS}, the BLISS transformation is conventionally formulated for particle-number symmetry only~\cite{loaiza2013_bliss,Patel2025-np}. However, the framework can be extended to other symmetries, such as spin and spatial symmetries, a possibility mentioned in the original proposal~\cite{loaiza2013_bliss}.
If the target eigenstate $\ket{\psi_k}$ is known to be an eigenstate of a set of symmetry operators $\{\hat{Q}_i\}$ with eigenvalues $\{Q_i\}$, such that $\hat{Q}_i\ket{\psi_k}=Q_i\ket{\psi_k}$, a corresponding BLISS transformation can be systematically constructed as $\hat{H}' = \hat{H}-\hat{O}_i(\hat{Q}_i - Q_i)$, where $\hat{O}_i$ is an arbitrary Hermitian operator that commutes with $\hat{Q}_i$. A key constraint for this construction, imposed to simplify subsequent optimization, is that the BLISS operator $\hat{O}_i(\hat{Q}_i - Q_i)$ must have the same polynomial degree in its fermionic operator expansion as the original Hamiltonian $\hat{H}$~\cite{loaiza2013_bliss}. 

For molecular electronic structure Hamiltonians, the relevant symmetries are the conserved total electron number $\hat{N}_e = \sum_{p\sigma}\hat{a}_{p\sigma}^{\dagger}\hat{a}_{p\sigma}$, the spin projection $\hat{S}_z$, and the total spin $\hat{S}^2$. Note that the chemically complex molecules considered in this work generally do not exhibit spatial point-group symmetries.
The total spin operator $\hat{S}^2$, however, presents a challenge. These spin operators are defined as
\begin{align}
    \hat{S}_z & \coloneqq \frac{1}{2}\sum_{p,\sigma}\sigma\,\hat{a}_{p\sigma}^{\dagger}\hat{a}_{p\sigma} \, , \\
    \hat{S}^2 & \coloneqq \hat{S}_{+}\hat{S}_{-} + \hat{S}_z(\hat{S}_z-1) \, , 
\end{align}
where $\hat{S}_{+}\coloneqq \sum_{p}\hat{a}_{p\uparrow}^{\dagger}\hat{a}_{p\downarrow}$ and $\hat{S}_{-}\coloneqq \sum_{p}\hat{a}_{p\downarrow}^{\dagger}\hat{a}_{p\uparrow}$, involves spin-flipping terms. 
We define a spin variable $\sigma$ such that an up-spin ($\uparrow$) corresponds to $\sigma=+1$ and a down-spin ($\downarrow$) corresponds to $\sigma = -1$. In this context, $\sigma$ effectively acts as a sign function for the spin state.
The off-diagonal part $\hat{S}_{+}\hat{S}_{-}=\sum_{pq}\hat{a}_{p\uparrow}^{\dagger}\hat{a}_{p\downarrow}\hat{a}_{q\downarrow}^{\dagger}\hat{a}_{q\uparrow}$ introduces four-fermion operators with mixed spin indices that are absent in the original electronic structure Hamiltonian~\eqref{eq:elec_ham}. Consequently, a BLISS transformation based on $\hat{S}^2$ would introduce new Pauli strings not present in $\hat{H}$, increasing the Hamiltonian's $\ell_1$-norm and, therefore, the simulation cost. To maintain or reduce the gate count, as is the goal of UWC, we thus exclude the $\hat{S}^2$ symmetry from our BLISS construction. 

To incorporate the $\hat{S}_z$ symmetry into the BLISS framework, we first express the electronic structure Hamiltonian in a spin-explicit representation:
\begin{align}
    \hat{H} &= \sum_{pq}^{N}\sum_{\sigma} h_{pq}^{(\sigma)}\hat{a}_{p\sigma}^{\dagger}\hat{a}_{q\sigma} + \frac{1}{2}\sum_{pqrs}^{N}\sum_{\sigma\tau} g_{pqrs}^{(\sigma\tau)}\hat{a}_{p\sigma}^{\dagger}\hat{a}_{r\tau}^{\dagger}\hat{a}_{s\tau}\hat{a}_{q\sigma} \nonumber\\
    &\eqqcolon \sum_{pq}^{N}\sum_{\sigma} k_{pq}^{(\sigma)}\hat{f}_{pq}^{(\sigma)} + \frac{1}{2}\sum_{pqrs}^{N}\sum_{\sigma\tau} g_{pqrs}^{(\sigma\tau)}\hat{f}_{pq}^{(\sigma)} \hat{f}_{rs}^{(\tau)},
    \label{eq:elec_ham_spin}
\end{align}
where $\hat{f}_{pq}^{(\sigma)} \coloneqq \hat{a}_{p\sigma}^{\dagger}\hat{a}_{q\sigma}$ and $k_{pq}^{(\sigma)} \coloneqq h_{pq}^{(\sigma)} - \frac{1}{2}\sum_r g_{prrq}^{(\sigma\sigma)}$. This representation reduces to the conventional form in Eq.~\eqref{eq:elec_ham} under the assumption that $h_{pq}=h_{pq}^{(\uparrow)}=h_{pq}^{(\downarrow)}$ and $g_{pqrs}=g_{pqrs}^{(\uparrow\uparrow)}=g_{pqrs}^{(\downarrow\downarrow)}=g_{pqrs}^{(\uparrow\downarrow)}=g_{pqrs}^{(\downarrow\uparrow)}$. 
Analogous to the derivation leading to Eq.~\eqref{eq:majorana_ham}, the corresponding Majorana representation is obtained as
\begin{widetext}
\begin{align}
    \hat{H} &= \frac{i}{2}\sum_{pq,\sigma}\left( k_{pq}^{(\sigma)} + \frac{1}{2}\sum_{r,\tau} g_{pqrr}^{(\sigma\tau)}\right) \hat{\gamma}_{p\sigma}\hat{\bar{\gamma}}_{q\sigma} 
    + \frac{1}{4} \sum_{p>r,s>q}\sum_\sigma (g_{pqrs}^{(\sigma\sigma)} - g_{psrq}^{(\sigma\sigma)} )\hat{\gamma}_{p\sigma}\hat{\gamma}_{r\sigma}\hat{\bar{\gamma}}_{q\sigma}\hat{\bar{\gamma}}_{s\sigma} 
    + \frac{1}{4}\sum_{pqrs}g_{pqrs}^{(\uparrow\downarrow)}\hat{\gamma}_{p\uparrow}\hat{\gamma}_{r\downarrow}\hat{\bar{\gamma}}_{q\uparrow}\hat{\bar{\gamma}}_{s\downarrow} + \mathrm{const} . 
    \label{eq:majorana_spin}
\end{align}

Using this spin-explicit representation, we can now define a BLISS operator that incorporates both particle-number and spin-projection symmetries. For a target eigenstate $\ket{\psi_k}$ with well-defined quantum numbers $N_e$ and $S_z$, such that $\hat{N}_e\ket{\psi_k}=N_e\ket{\psi_k}$ and $\hat{S}_z\ket{\psi_k}=S_z\ket{\psi_k}$, the spin-extended BLISS operator is defined as
\begin{align}
    \hat{T}_S(\vec{\nu}, \vec{\zeta})
    &\coloneqq \nu_1(\hat{S}_z-S_z) + \nu_2\left(\left(\hat{S}_z^2 - \frac{1}{4}\hat{N}_e \right) - \left(S_z^2 - \frac{1}{4}N_e \right) \right) + \nu_3\left( \hat{S}_z(\hat{N}_e - 1) - S_z(N_e-1)\right) \nonumber\\
    &+ \sum_{pq}\zeta_{pq}^{(1)}\hat{F}_{pq}(\hat{S}_z-S_z) 
    + \sum_{pq}\zeta_{pq}^{(2)}\hat{F}_{pq}^{(z)}(\hat{N}_e-N_e) 
    + \sum_{pq}\zeta_{pq}^{(3)}\hat{F}_{pq}^{(z)}(\hat{S}_z-S_z), 
    \label{eq:BLISS_op_spin}
\end{align}
where $\nu_{i},\,\zeta_{pq}^{(j)} \in \mathbb{R}$ with $\zeta_{pq}^{(j)}=\zeta_{qp}^{(j)}$, and $\hat{F}_{pq}\coloneqq \sum_{\sigma}\hat{f}_{pq}^{(\sigma)}$ and $\hat{F}_{pq}^{(z)}\coloneqq \sum_{\sigma}\sigma\,\hat{f}_{pq}^{(\sigma)}$. The parameters are denoted as $\vec{\nu}=\{\nu_1, \nu_2, \nu_3 \}$ and $\vec{\zeta}= \set{\zeta_{pq}^{(j)} \mid p\leq q; j=1,2,3}$, leading to $3+\frac{3}{2}N(N-1)$ independent parameters in total.
Each term in Eq.~\eqref{eq:BLISS_op_spin} is constructed to be a quadratic or quartic fermionic operator, thereby matching the polynomial degree of the electronic Hamiltonian. Specifically, the terms associated with $\nu_2$ and $\nu_3$ are quartic operators:
\begin{align}
    \hat{S}_z^2 - \frac{1}{4}\hat{N}_e &= \frac{1}{4}\sum_{pqrs}\sum_{\sigma\tau}\sigma\tau\, \delta_{pq}\delta_{rs}\hat{a}_{p\sigma}^{\dagger}\hat{a}_{r\tau}^{\dagger}\hat{a}_{s\tau}\hat{a}_{q\sigma} ,\\
    \hat{N}_e\hat{S}_z - \hat{S}_z &=  \frac{1}{4}\sum_{pqrs}\sum_{\sigma\tau}(\sigma+\tau)\delta_{pq}\delta_{rs}\hat{a}_{p\sigma}^{\dagger}\hat{a}_{r\tau}^{\dagger}\hat{a}_{s\tau}\hat{a}_{q\sigma} ,
\end{align}
These expressions are consistent with the fermionic representation in Eq.~\eqref{eq:elec_ham_spin}. 
The spin-extended BLISS transformation is then defined by $\hat{H}'(\vec{\nu}, \vec{\zeta}) \coloneqq \hat{H} - \hat{T}_S(\vec{\nu}, \vec{\zeta})$, yielding a transformed Hamiltonian
\begin{align}
    \hat{H}'(\vec{\nu}, \vec{\zeta}) &\coloneqq \hat{H} - \hat{T}_S(\vec{\nu}, \vec{\zeta}) 
    = \sum_{pq}^{N}\sum_{\sigma} k_{pq}^{(\sigma)'}(\vec{\nu}, \vec{\zeta})\hat{f}_{pq}^{(\sigma)} + \frac{1}{2}\sum_{pqrs}^{N}\sum_{\sigma\tau} g_{pqrs}^{(\sigma\tau)'}(\vec{\nu}, \vec{\zeta})\hat{f}_{pq}^{(\sigma)} \hat{f}_{rs}^{(\tau)}, 
    \label{eq:spin_BLISS_transform}
\end{align}
with the transformed one- and two-electron integrals 
\begin{align}
    h_{pq}^{(\sigma)'}(\vec{\nu}, \vec{\zeta}) &= h_{pq}^{(\sigma)} - \frac{1}{2}\nu_1\sigma\delta_{pq}
    + \left(S_z-\frac{1}{2}\sigma \right)\zeta_{pq}^{(1)} 
    + \sigma\left(N_e - 1\right)\zeta_{pq}^{(2)} 
    + \left(\sigma S_z-\frac{1}{2}\right)\zeta_{pq}^{(3)}, \\
    g_{pqrs}^{(\sigma\tau)'}(\vec{\nu}, \vec{\zeta}) &= g_{pqrs}^{(\sigma\tau)} - \frac{1}{2}[\nu_2\sigma\tau + \nu_3(\sigma+\tau)]\delta_{pq}\delta_{rs} - \frac{1}{2}[\tau\zeta_{pq}^{(1)}\delta_{rs}+\sigma\zeta_{rs}^{(1)}\delta_{pq}] \nonumber\\
    &- [\sigma\zeta_{pq}^{(2)}\delta_{rs}+\tau\zeta_{rs}^{(2)}\delta_{pq}] 
     - \frac{1}{2}\sigma\tau[\zeta_{pq}^{(3)}\delta_{rs}+\zeta_{rs}^{(3)}\delta_{pq}] ,
\end{align}
\end{widetext}
and $k_{pq}^{(\sigma)'}(\vec{\nu}, \vec{\zeta}) \coloneqq h_{pq}^{(\sigma)'}(\vec{\nu}, \vec{\zeta}) - \frac{1}{2}\sum_r g_{prrq}^{(\sigma\sigma)'}(\vec{\nu}, \vec{\zeta})$. 

In the UWC optimization detailed in Appendix~\ref{append:uwc_details}, we implemented this spin-extended BLISS transformation alongside both orbital optimization and the conventional particle-number BLISS. For the systems studied, the additional resource reduction from spin-extended BLISS was typically negligible compared to that from conventional BLISS alone, a finding consistent with previous works~\cite{loaiza2013_bliss,Patel2025-np}. Nevertheless, we included it in our analysis for all molecules to ensure our resource estimates are as accurate as possible, capturing even minor constant-factor improvements.

\section{Details of the UWC optimization \label{append:uwc_details}}
This appendix collects the details of the UWC optimization. Appendix~\ref{append:uwc_algorithm} specifies the iterative procedure, including the orbital initialization and the termination criteria, and shows that the representation it returns never carries a larger gate count than the one supplied as input. Appendix~\ref{append:uwc_l1_comparison} examines the relation between the $\ell_{1}$-norm and the partially randomized gate count, identifies the regime in which the former is a faithful surrogate for the latter, and compares the two objectives numerically. Appendix~\ref{append:soft_cost_params} then shows that the two hyperparameters of the softened gate-cost function are fixed by the cost model rather than tuned, and
Appendix~\ref{append:uwc_comp_details} lists the remaining computational settings.

\subsection{Iterative optimization procedure\label{append:uwc_algorithm}}

First, we detail the iterative UWC optimization procedure employed in this work. As outlined in Sec.~\ref{subsec:uwc}, our approach combines OO and BLISS transformations within a sequential, iterative framework. The precise procedure is summarized in Algorithm~\ref{alg:UWC}.
\begin{figure}
\begin{algorithm}[H]
    \caption{Iterative UWC optimization procedure.}
    \label{alg:UWC}
    \begin{algorithmic}[1] 
    \State \textbf{Input:} Initial Hamiltonian $\hat{H}_{\rm init}$, target electron number $N_e$, target spin projection $S_z$.
    \State \textbf{Parameters:} Convergence threshold $\delta_{\rm th}$, max iterations $N_{\rm iter}$.
    \Statex \textbf{Orbital Initialization}
    \State Generate a set of Cholesky orbital basis representations for $\hat{H}_{\rm init}$.
    \State Select, from $\hat{H}_{\rm init}$ and these representations, the one with the minimum $G_M$ as the starting Hamiltonian, $\hat{H}$.
    \State $G_{\rm prev} \leftarrow G_M \,\, \text{for} \,\, \hat{H}$
    \State $\hat{H}_{\rm best} \leftarrow \hat{H}$
    \Statex \textbf{Main Optimization Loop}
    \For{$n=1,2,\ldots,N_{\rm iter}$}
        \State Determine the optimal partition $L_D^\ast$ and weight parameter $w_{\rm soft}$ for $\hat{H}$, according to Eqs.~\eqref{eq:LD_opt} and \eqref{eq:w_soft}.
        \State $\vec{\kappa}^\ast \leftarrow \arg\min_{\vec{\kappa}} \, G_{\rm soft}(\hat{H}(\vec{\kappa}))$ 
        \State $\hat{H}_{\rm OO} \leftarrow \hat{H}(\vec{\kappa}^\ast)$
        \State $(\vec{\mu}^\ast,\,\vec{\xi}^\ast) \leftarrow \arg\min_{(\vec{\mu},\,\vec{\xi})} \, G_{\rm soft}(\hat{H}_{\rm OO}(\vec{\mu},\vec{\xi}))$
        \State $\hat{H}_{\rm BLISS} \leftarrow \hat{H}_{\rm OO}(\vec{\mu}^\ast,\vec{\xi}^\ast)$
        \State $(\vec{\nu}^\ast,\,\vec{\zeta}^\ast) \leftarrow \arg\min_{(\vec{\nu},\,\vec{\zeta})} \, G_{\rm soft}(\hat{H}_{\rm BLISS}(\vec{\nu},\vec{\zeta}))$ 
        \State $\hat{H}_{\rm next} \leftarrow \hat{H}_{\rm BLISS}(\vec{\nu}^\ast,\vec{\zeta}^\ast)$
        \Statex \textbf{Convergence and Degradation Check}
        \State Calculate the discrete gate cost $G_{\rm next}$ for $\hat{H}_{\rm next}$.
        \If{$G_{\rm next} > G_{\rm prev}$}
            \State \textbf{terminate loop} \Comment{Cost increased, discard result and terminate}
        \ElsIf{$(G_{\rm prev}-G_{\rm next})/G_{\rm prev} < \delta_{\rm th}$}
            \State $\hat{H}_{\rm best} \leftarrow \hat{H}_{\rm next}$ 
            \State \textbf{terminate loop} \Comment{Converged}
        \EndIf
        \Statex \textbf{Update for Next Iteration}
        \State $\hat{H} \leftarrow \hat{H}_{\rm next}$
        \State $G_{\rm prev} \leftarrow G_{\rm next}$
        \State $\hat{H}_{\rm best} \leftarrow \hat{H}_{\rm next}$
  \EndFor
  \State \textbf{Output:} The UWC-optimized Hamiltonian representation $\hat{H}_{\rm best}$.
  \end{algorithmic}
\end{algorithm}
\end{figure}

The optimization begins with an \emph{orbital initialization} step, in which the starting representation is chosen among several candidates by the cost $G_{M}$ of Eq.~\eqref{eq:abstract_cost_model} itself. 
Specifically, we compare the initial canonical molecular orbital basis with a set of alternative bases derived from a \emph{double factorization} procedure, which we term the \emph{Cholesky orbital bases}~\cite{Motta2021-kh}. 
This procedure involves first performing a Cholesky decomposition of the two-electron integral tensor $g_{pqrs}$, viewed as an $N^2 \times N^2$ positive-semidefinite matrix:
\begin{align}
    g_{pqrs} = \sum_{t=1}^{N_{\rm DF}}L_{pq}^{(t)} (L_{rs}^{(t)})^{\dagger}, 
    \label{eq:cholesky}
\end{align}
where $N_{\rm DF}\leq N^2$ and each $L^{(t)}$ is a real symmetric $N\times N$ matrix. A second factorization is then performed via eigenvalue decomposition of each Cholesky vector $L^{(t)}$:
\begin{align}
    L_{pq}^{(t)}=\sum_{i=1}^{N}U_{pi}^{(t)} W_{i}^{(t)} U_{qi}^{(t)}, 
    \label{eq:second_factorization}
\end{align}
where $W_{i}^{(t)}$ are the real-valued eigenvalues and $U^{(t)}$ are the $N\times N$ unitary matrices of eigenvectors. 
Among the canonical and in each of these Cholesky bases, we take the representation with the smallest $G_M$ as the starting point, $\hat{H}$, for the main optimization loop. 
This choice was motivated by our observation that rotating to a Cholesky orbital basis often reduces the cost $G_{M}$, consistent with the results in Ref.~\cite{Gunther2025-zk}, which show a reduced $\ell_1$-norm in the Cholesky orbital basis. 

Following initialization, we enter the main iterative loop.
In each iteration, we first determine the optimal partitioning boundary $L_D^\ast$ (Eq.~\eqref{eq:LD_opt}) and the corresponding softened gate-cost hyperparameter $w_{\rm soft}$ (Eq.~\eqref{eq:w_soft}) for the current Hamiltonian representation $\hat{H}$. We then minimize the softened gate-cost function $G_{\rm soft}$ (Eq.~\eqref{eq:uwc_def}) by applying a sequence of transformations: first OO, then the standard particle-number BLISS, and finally the spin-extended BLISS (Appendix~\ref{append:BLISS_spin}). This specific order was chosen based on empirical tests showing that performing OO first consistently yielded the best results.

After completing this sequence, we obtain the optimized Hamiltonian for the current iteration, $\hat{H}_{\rm next}$, and evaluate its true cost by calculating the discrete gate count, $G_{\rm next}$. This cost is then compared to the cost from the previous iteration, $G_{\rm prev}$. 
The iterative process is terminated under two conditions, tested in this order: (1) if the cost has increased, $G_{\rm next}>G_{\rm prev}$, the result of the current iteration is discarded and the optimization stops; (2) if the relative improvement in cost falls below a threshold $\delta_{\rm th}$, the current iteration is accepted and the loop stops (convergence). If neither condition is met, the procedure continues to the next iteration using $\hat{H}_{\rm next}$ as the starting point. The final output is the best representation encountered.
Because this test is applied to the exact discrete cost $G_{M}$ rather than to the softened objective $G_{\rm soft}$, and because $\hat{H}_{\rm init}$ is itself among the candidates compared in the initialization step, the returned representation never carries a larger $G_{M}$ than the input one---for any Hamiltonian, and irrespective of how closely $G_{\rm soft}$ approximates $G_{M}$ or how well the inner minimizations converge. Since the end-to-end cost $G_{\rm total}$ of Eq.~\eqref{eq:total_cost} is dominated by its final round for the systems considered here (Sec.~\ref{subsec:alg_cost_model}), the same holds for $G_{\rm total}$ to that accuracy.

\subsection{Relation to $\ell_1$-norm optimization
\label{append:uwc_l1_comparison}}


Both OO and BLISS were originally introduced as means of minimizing the $\ell_1$-norm $\lambda$ of the Hamiltonian~\cite{loaiza2013_bliss,Koridon2021-xm}, and for the algorithms they were developed for this is a well-founded choice: the query complexity of qubitization scales as $\mathcal{O}(\lambda)$~\cite{Lee2021-tz,Von_Burg2021-du,Low2025-tb,Low2019-op} and the gate count of fully randomized product formulas as $\mathcal{O}(\lambda^{2})$ as described in Sec.~\ref{subsec:pr_time_evolution}, so in both cases the cost is a strictly increasing function of $\lambda$ alone and minimizing $\lambda$ is \emph{equivalent} to minimizing the cost. For partially randomized simulation, this equivalence fails. We make precise below in what sense it fails, and then compare the two optimizations numerically.

\subsubsection{Theoretical comparison}

To theoretically compare the $\ell_1$-norm optimization and UWC, we first introduce the \emph{tail-sum profile} of a Hamiltonian representation,
\begin{align}
  \lambda_{R}(k) := \sum_{\ell>k}\lvert c_{\ell}\rvert ,
  \qquad k = 0,1,\dots,L ,
  \label{eq:tailprofile}
\end{align}
where, as in Sec.~\ref{subsec:alg_cost_model}, the coefficients are sorted in descending order of magnitude. The sum is empty at $k=L$, so $\lambda_{R}(0)=\lambda$ and $\lambda_{R}(L)=0$, the endpoints corresponding to the fully randomized and fully deterministic partitions. For any positive constants $g_{\rm d}$ and $g_{\rm r}$, we define
\begin{align}
    \Phi(k) &:= g_{\rm d} k + g_{\rm r}\lambda_{R}(k)^{2}, \quad
    G := \min_{0\le k\le L}\Phi(k),
  \label{eq:Phi}
\end{align}
with $L_{D}^{*}=\arg\min_{k}\Phi(k)$. Equations~\eqref{eq:abstract_cost_model} and~\eqref{eq:LD_opt} are the case with $g_{\rm d}=g^{(\mathrm{det})}_{M}$ and $g_{\rm r}=g^{(\mathrm{rand})}_{M}$, for which $G=G_{M}$. By Eqs.~\eqref{eq:gate_det} and~\eqref{eq:gate_rand}, every other round $G_{m}$ has the same form. Written this way, $G$ is a functional of the \emph{entire} profile $\lambda_{R}(\cdot)$, whereas $\lambda$ is the value of that profile at the single point $k=0$. The three statements below quantify the resulting discrepancy.
They follow from Eqs.~\eqref{eq:tailprofile}--\eqref{eq:Phi} and hold for arbitrary positive $g_{\rm d},g_{\rm r}$ and arbitrary coefficient distributions, independently of the transformation family, of the optimizer, and of the system considered.

\paragraph*{(i) A two-sided bound in terms of $\lambda$.}
The $\ell_{1}$-norm constrains $G$ only within a range that widens linearly in $\lambda$. 
The upper bound is immediate: since a minimum is at most the value of the function at any single point, evaluating $\Phi$ at $k=0$, where $\lambda_{R}(0)=\lambda$, gives $G\le g_{\rm r}\lambda^{2}$ without any further assumption. 
For a lower bound, no coefficient exceeds $\lvert c_{1}\rvert$, so $\lambda_{R}(k)\ge\max\{0,\,\lambda-k\lvert c_{1}\rvert\}$ and
\begin{align}
    \Phi(k) &\ge f\bigl(k\lvert c_{1}\rvert\bigr), \nonumber\\
    f(u) &:= \frac{g_{\rm d}}{|c_{1}|} u
          + g_{\rm r}(\max\{0,\lambda-u\})^{2},
  \label{eq:relax}
\end{align}
the truncation at zero being required for the squared bound to remain valid. Relaxing $u$ to a non-negative real variable, $f$ is convex with stationary point $\lambda-u=g_{\rm d}/(2g_{\rm r}\lvert c_{1}\rvert)$, and substituting it back gives
\begin{align}
  \frac{g_{\rm d}}{\lvert c_{1}\rvert}\,\lambda
  -\frac{g_{\rm d}^{2}}{4g_{\rm r}\lvert c_{1}\rvert^{2}}
  \le G \le 
  g_{\rm r}\lambda^{2}.
  \label{eq:twosided}
\end{align}
The upper bound is quadratic in $\lambda$ and the lower bound linear, so knowledge of $\lambda$ does not determine $G$ even up to a constant factor.
If the stationary point falls below $u=0$, the constrained minimum of $f$ on $u\ge0$ is attained at $u=0$, where $\lambda_{R}(0)=\lambda$ exactly, so that $G=g_{\rm r}\lambda^{2}$ and the two bounds coincide; this is the case treated in statement~(ii). The lower bound displayed in Eq.~\eqref{eq:twosided} is the unconstrained stationary value of $f$, and is therefore negative in this regime; it remains valid but carries no information.

Both bounds are essentially attained, so the range cannot be narrowed. At fixed $\lambda$, and assuming $\lambda>\sqrt{g_{\rm d}/g_{\rm r}}$ so that retaining a single term deterministically is cheaper than randomizing the whole Hamiltonian, a representation whose weight is carried by a single coefficient has $\lvert c_{1}\rvert=\lambda$ and $\lambda_{R}(1)=0$, hence $G=\min\{g_{\rm r}\lambda^{2},g_{\rm d}\}=g_{\rm d}$ in the regime of interest, saturating the lower bound up to its additive term. A uniform distribution over all $L$ coefficients has $\lambda_{R}(k)=\lambda(1-k/L)$, and since $(1-k/L)^{2}\ge 1-2k/L$,
\begin{align}
  \Phi(k)-\Phi(0) \ge
  k\bigl[g_{\rm d}-2g_{\rm r}\lambda^{2}/L\bigr],
  \label{eq:uniformcase}
\end{align}
which is non-negative once $L\ge 2g_{\rm r}\lambda^{2}/g_{\rm d}$; then $L_{D}^{*}=0$ and $G=g_{\rm r}\lambda^{2}$, attaining the upper bound exactly. The gate count therefore varies over the full range at fixed $\ell_{1}$-norm, and no function $F$ with $G=F(\lambda)$ exists.

\paragraph*{(ii) When $\lambda$ is a faithful surrogate.}
The $\ell_{1}$-norm is a faithful surrogate precisely when the optimal partition is trivial. If $L_{D}^{*}=0$ then $G=g_{\rm r}\lambda^{2}$ exactly, and minimizing $\lambda$ over any family of transformations is \emph{equivalent} to minimizing $G$. By definition, this requires $\Phi(0)\le\Phi(k)$ for all $k$, but the $L$ conditions collapse to a single inequality. Using $\lambda_{R}(k-1)=\lambda_{R}(k)+\lvert c_{k}\rvert$, the forward difference of $\Phi$ is
\begin{align}
    \Delta(k) &:= \Phi(k)-\Phi(k-1) \nonumber \\
    &= g_{\rm d}-g_{\rm r}\lvert c_{k}\rvert
      \bigl[\lambda_{R}(k-1)+\lambda_{R}(k)\bigr],
  \label{eq:Phidiff}
\end{align}
which is non-decreasing in $k$ because $\lvert c_{k}\rvert$ and $\lambda_{R}(k)$ are both non-negative and non-increasing; hence $\Phi$ is convex on the integers, has no spurious local minima, and changes from decreasing to increasing at most once. Taking $L_{D}^{*}$ to be the smallest minimizer, the two ends of its range are then fixed by $\Delta(1)$ and $\Delta(L)$. At the lower end,
\begin{align}
    L_{D}^{*}>0
      &\quad\Longleftrightarrow\quad \Delta(1)\le 0 \nonumber\\
      &\quad\Longleftrightarrow\quad
        g_{\rm r}\lvert c_{1}\rvert
        \bigl(2\lambda-\lvert c_{1}\rvert\bigr)\ge g_{\rm d},
  \label{eq:nontrivialpartition}
\end{align}
i.e., the partition is nontrivial as soon as treating the single largest coefficient deterministically already pays for itself. At the upper end, $\Delta(L)=g_{\rm d}-g_{\rm r}\lvert c_{L}\rvert^{2}$, so $L_{D}^{*}<L$ whenever $g_{\rm r}\lvert c_{L}\rvert^{2}<g_{\rm d}$, which excludes the fully deterministic partition. The regime in which $\ell_{1}$-norm minimization is the correct objective is thus exactly the fully randomized one; every instance studied in this work has $0<L_{D}^{*}<L$ (see Table~\ref{tab:h_chain_logical}--\ref{tab:co2-cat_2_logical}) and lies outside it, so that $G$ is fixed by the profile $\lambda_{R}(\cdot)$ at $k>0$, on which $\lambda$ imposes no constraint beyond Eq.~\eqref{eq:twosided}.

\paragraph*{(iii) Monotonicity in the tail-sum profile.}
Consider two representations and suppose that the tail-sum profile of one lies pointwise below that of the other,
\begin{align}
  \lambda'_{R}(k) \le \lambda_{R}(k)
  \quad\text{for } \forall k .
  \label{eq:profileorder}
\end{align}
Then $\Phi'(k)\le\Phi(k)$ for every $k$. Writing $k^{*}:=\arg\min_{0\le k\le L}\Phi(k)$, so that $\Phi(k^{*})=G$, and evaluating the primed cost at that same $k^{*}$,
\begin{align}
  G'=\min_{0\le k\le L}\Phi'(k)
  \le\Phi'(k^{*})\le\Phi(k^{*})=G ,
  \label{eq:profilemonotone}
\end{align}
the first inequality holding for any choice of evaluation point. The two minimizers therefore need not coincide.

At fixed $\lambda$, Eq.~\eqref{eq:profileorder} may be rewritten in terms of the partial sums $\Sigma(k):=\sum_{\ell\le k}\lvert c_{\ell}\rvert=\lambda-\lambda_{R}(k)$ as $\Sigma'(k)\ge \Sigma(k)$ for every $k$, with $\Sigma'(L)=\Sigma(L)=\lambda$: the same total weight is carried more heavily by the leading coefficients. This is the majorization order, and Eq.~\eqref{eq:profilemonotone} states that $G$ is non-increasing along it, i.e., Schur-concave. The mechanism is visible in Eq.~\eqref{eq:Phi}: the deterministic term is linear in the \emph{number} of retained terms while the randomized term is quadratic in their total \emph{weight}, so a concentrated distribution buys a given amount of weight with fewer deterministic rotations. This gives the notion of unitary weight concentration a precise meaning---the object sought by UWC is a representation whose tail-sum profile lies below that of the reference---and locates the $\ell_{1}$-norm within that ordering, as the $k=0$ component of the $L+1$ conditions in Eq.~\eqref{eq:profileorder}.

One caveat is worth recording. Equation~\eqref{eq:profilemonotone} was derived at fixed $g_{\rm d},g_{\rm r}$, whereas in the cost model of Appendix~\ref{append:prnd}, these are not universal constants: by Eqs.~\eqref{eq:gate_det} and~\eqref{eq:gate_rand}, $g^{(\mathrm{rand})}_{M}$ is independent of the representation, but $g^{(\mathrm{det})}_{M}\propto\delta^{-1}\propto C_{\rm gs}^{1/p}$ through Eq.~\eqref{eq:trotter_step_optimal}. The empirical relation used here, $C_{\rm gs}\simeq a\lambda^{b}$ with $b\simeq2.09$~\cite{Gunther2025-zk}, is increasing in $\lambda$, and Eq.~\eqref{eq:profileorder} already implies $\lambda'\le\lambda$ through its $k=0$ component; the primed representation therefore carries the smaller deterministic coefficient, the pointwise inequality $\Phi'(k)\le\Phi(k)$ survives, and Eq.~\eqref{eq:profilemonotone} is unaffected.

The above statements (i)--(iii) concern the cost functional itself. They show that the $\ell_{1}$-norm cannot serve as a faithful surrogate for $G$ outside the fully randomized regime. They do not imply that UWC outperforms $\ell_{1}$ minimization on any given instance: both are non-convex optimizations over the same transformation family, neither is guaranteed to reach a global minimum, and no ordering between two such local searches can be established in general. What is guaranteed is that the output never carries a higher cost than the representation the optimization started from (Appendix~\ref{append:uwc_algorithm}).

Nor should an unconditional \emph{strict} advantage be expected for any method of this class: the OO and BLISS families together carry $\mathcal{O}(N^{2})$ parameters and cannot arbitrarily reshape a coefficient distribution with $\mathcal{O}(N^{4})$ entries, so the achievable concentration is limited by the transformation family itself. The magnitude of the achievable reduction is therefore an empirical question, which we address next.

\subsubsection{Numerical comparison}

\begin{figure}[tbp]
  \centering
  \includegraphics[width=0.48\textwidth]{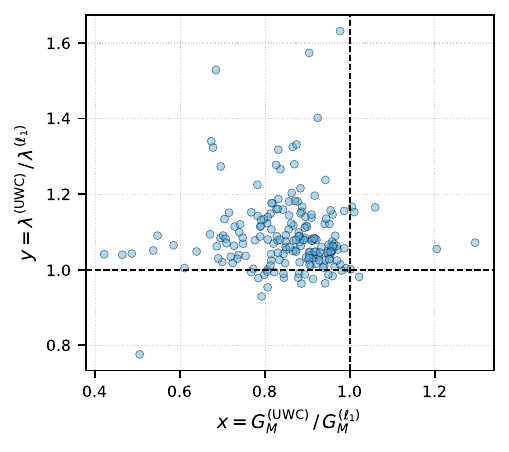}
  \caption{Performance comparison of UWC and $\ell_1$-norm optimization for all benchmark molecules. The plot compares the final $\ell_1$-norm ($\lambda$) and maximum gate count ($G_M$) from the two methods. The vertical axis shows the ratio $\lambda^{(\text{UWC})} / \lambda^{(\ell_1)}$, while the horizontal axis shows the ratio $G_{M}^{(\text{UWC})} / G_{M}^{(\ell_1)}$. Points to the left of the vertical dashed line ($x=1$) indicate that UWC achieved a lower gate count, the primary optimization target. }
    \label{fig:l1_vs_uwc}
\end{figure}
To compare the two objectives, we applied the $\ell_{1}$-norm optimization procedure---incorporating OO, conventional BLISS, and spin-extended BLISS---to all benchmark systems (see Appendix~\ref{append:models_detail}). The difference between the two runs is the function driving the optimization: the UWC method minimizes the softened gate-cost $G_{\rm soft}$, while the $\ell_{1}$-norm optimization minimizes $\lambda$ directly. All other settings, including the transformation sequence, the optimizer, and the initialization, are identical.

The results are summarized in Fig.~\ref{fig:l1_vs_uwc}. 
As expected, optimizing $\lambda$ directly yields the smaller $\ell_{1}$-norm for most instances (points above $y=1$), while UWC yields the smaller gate count for almost all of them (points left of $x=1$), with reductions of up to $60\%$. Within this transformation family, the two objectives are strongly correlated in practice: we find no instance for which $\ell_{1}$-norm optimization returns a higher gate count than the reference representation, so the failure of $\lambda$ as a surrogate does not manifest itself here as an outright degradation. It manifests itself instead as a systematic shortfall, the representation with the smaller $\ell_{1}$-norm consistently not being the one with the smaller gate count. This is the behaviour anticipated by statements~(i)--(iii): a smaller $\lambda$ constrains the tail-sum profile only at $k=0$, and the profiles reached by the two optimizations
differ where it matters, near $L_{D}^{*}$.

We therefore do not claim that UWC is superior to $\ell_{1}$-norm optimization as a matter of principle, nor that the latter is harmful. The claims are that (a) the $\ell_{1}$-norm is not a faithful surrogate for the partially randomized gate count, by Eq.~\eqref{eq:twosided} and statement~(ii); (b) UWC never returns a representation worse than the one it started from, by the non-degradation property of Algorithm~\ref{alg:UWC} (Appendix~\ref{append:uwc_algorithm}); and (c) empirically, the additional reduction obtained by targeting the gate count directly rather than a proxy is substantial, reaching $60\%$ for the instances studied here.

\subsection{The softened gate-cost function and its hyperparameters}\label{append:soft_cost_params}
The softened gate-cost of Eq.~\eqref{eq:uwc_def} involves two hyperparameters, the partition threshold $w_{\rm soft}$ and the sigmoid width $\epsilon_{\rm soft}$. Neither is free: the first is fixed by the cost coefficients, and the second is bounded above by the requirement that the softened partition still resolve the true one.

The two hyperparameters entering Eq.~\eqref{eq:uwc_def}, the partition threshold $w_{\rm soft}$ and the sigmoid width $\epsilon_{\rm soft}$, are fixed as follows. The threshold is not free. As shown in Appendix~\ref{append:uwc_l1_comparison}, the discrete cost is convex on the integers and $L_{D}^{*}$ is fixed by the change of sign of the forward difference of Eq.~\eqref{eq:Phidiff}, $\Delta(L_{D}^{*})\le0<\Delta(L_{D}^{*}+1)$. Neglecting the difference between $\lambda_{R}(k-1)$ and $\lambda_{R}(k)$, which is legitimate whenever $\lvert c_{L_{D}^{*}}\rvert\ll\lambda_{R}$, these two inequalities read
\begin{align}
 \lvert c_{L_{D}^{*}+1}\rvert
 < w^{*} \le
 \lvert c_{L_{D}^{*}}\rvert ,
 \quad
 w^{*} := \frac{g_{\rm d}}{2 g_{\rm r} \lambda_{R}} ,
 \label{eq:wstar}
\end{align}
where $\lambda_{R}\equiv\lambda_{R}(L_{D}^{*})$. The optimal partition is therefore the one induced by thresholding at $w^{*}$, a level fixed by the algorithmic cost coefficients themselves. Equation~\eqref{eq:wstar} locates $w^{*}$ only within the gap between two consecutive sorted coefficients, and the midpoint of that gap---which is what Eq.~\eqref{eq:w_soft} prescribes---is the choice minimizing the worst-case offset $\lvert w^{*}-w_{\rm soft}\rvert$. The threshold is thus determined by the cost model rather than tuned. Consistent with Eq.~\eqref{eq:wstar} and with $g_{\rm r}/g_{\rm d}\propto\xi$, the observed thresholds decrease with $\xi$ across the benchmark set, lying between $2\times10^{-3}$ and $6\times10^{-2}$ for all systems and all values of $\xi$ considered.

Equation~\eqref{eq:wstar} presumes that $\lvert c_{L_{D}^{*}}\rvert$ and $\lvert c_{L_{D}^{*}+1}\rvert$ are distinct, and that presumption is not vacuous: no threshold can separate coefficients of exactly equal magnitude, and such degeneracies do occur, since the fermion-to-qubit mapping distributes a single electronic integral over several Pauli strings with commensurate magnitudes. Where the partition boundary falls inside a degenerate block, the description of the optimal partition as a thresholding breaks down, although the optimal partition itself remains well defined. As noted below, this affects how much reduction the search finds, not the validity of the result.

\begin{figure}[t]
 \includegraphics[width=\columnwidth]{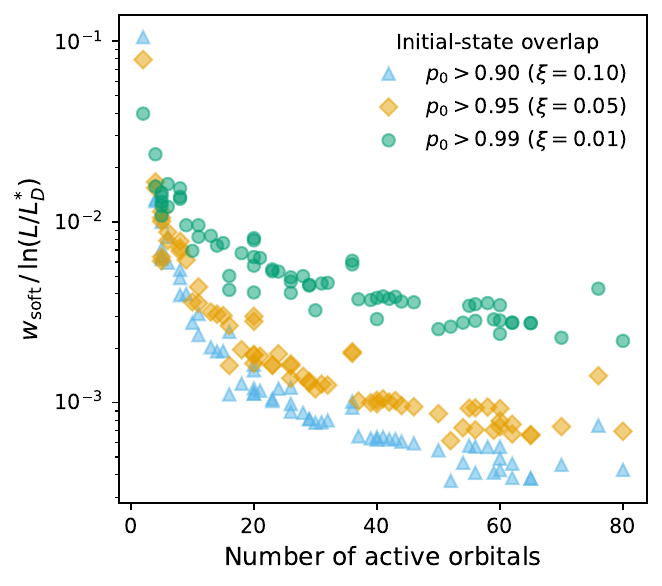}
 \caption{Upper bound on the sigmoid width, Eq.~\eqref{eq:epsupper}, evaluated on the reference representation of each benchmark system for three values of $\xi$. The bound decreases with system size and is smallest at the largest $\xi$, the most restrictive case being approximately $4\times10^{-4}$. The value $\epsilon_{\rm soft}=10^{-4}$ used throughout this work lies below the bound for every instance.}
 \label{fig:epsceiling}
\end{figure}

The sigmoid width is bounded from above, and the bound can be made quantitative. 
A coefficient of magnitude $\lvert c_{\ell}\rvert$ below the threshold
contributes $\varsigma(-d/\epsilon_{\rm soft})\simeq
e^{-d/\epsilon_{\rm soft}}$ to the softened count of deterministic terms,
where $d=w_{\rm soft}-\lvert c_{\ell}\rvert$ is its distance from the
threshold. Because that term counts \emph{terms} rather than weight, the
contribution is not suppressed by the smallness of the coefficient, and
the vast majority of the $L$ coefficients have
$\lvert c_{\ell}\rvert\ll w_{\rm soft}$ and hence $d\simeq w_{\rm soft}$.
Requiring their aggregate spurious contribution to remain below the true
count, $L\,e^{-w_{\rm soft}/\epsilon_{\rm soft}}\ll L_{D}^{*}$, gives
\begin{align}
 \epsilon_{\rm soft} \lesssim 
 \frac{w_{\rm soft}}{\ln\!\left(L/L_{D}^{*}\right)} .
 \label{eq:epsupper}
\end{align}
This is a condition on $\epsilon_{\rm soft}$ relative to the threshold rather than on its absolute value, and it involves only quantities computed in the course of the optimization. Its right-hand side is shown in Fig.~\ref{fig:epsceiling} for the reference representation of every benchmark system. It falls with system size and levels off near $4\times10^{-4}$ for the largest active spaces at $\xi=0.1$, the most restrictive case. The value $\epsilon_{\rm soft}=10^{-4}$ used throughout therefore lies below the ceiling for every instance, by a factor of three or more, whereas $10^{-3}$ would exceed it for most of the larger systems.

What lies beyond the ceiling is not merely a blurred objective. Taking the limit $\epsilon_{\rm soft}\rightarrow\infty$ in Eq.~\eqref{eq:uwc_def}, every $\varsigma(\cdot)\rightarrow\tfrac12$ irrespective of the coefficient, and
\begin{align}
 G_{\rm soft} \xrightarrow[\epsilon_{\rm soft} \to \infty]{}  
 \frac{g_{\rm d}L}{2} + \frac{g_{\rm r}}{4} \lambda^{2}.
 \label{eq:largeepslimit}
\end{align}
The first term is a constant, independent of the representation, and the second is a monotone function of the $\ell_{1}$-norm alone: in this limit the objective retains no information about how the weight is distributed among the coefficients, and minimizing it reduces exactly to $\ell_{1}$-norm minimization. The sigmoid width thus interpolates between the two objectives compared in Appendix~\ref{append:uwc_l1_comparison}---the exact partially randomized gate count as $\epsilon_{\rm soft}\rightarrow0$, and the $\ell_{1}$-norm surrogate as $\epsilon_{\rm soft}\rightarrow\infty$---and Eq.~\eqref{eq:epsupper} marks where the deterministic term begins to lose its ability to resolve the partition. Since the $\ell_{1}$-norm is not a faithful surrogate outside the fully randomized regime, exceeding the ceiling drives the optimization towards a different and, for the present purpose, inappropriate target.


One further remark clarifies the role of the outer loop. For a fixed threshold $w$, define the cost incurred by the partition that $w$ induces,
\begin{align}
 \tilde{G}(w) &:= g_{\rm d}\sum_{\ell}\theta(\lvert c_{\ell}\rvert-w)
 \nonumber\\
 & + g_{\rm r}\left(\sum_{\ell}\lvert c_{\ell}\rvert
   \bigl[1-\theta(\lvert c_{\ell}\rvert-w)\bigr]\right)^{2},
 \label{eq:Gtilde}
\end{align}
with $\theta$ the Heaviside step function; Eq.~\eqref{eq:uwc_def} is obtained from Eq.~\eqref{eq:Gtilde} by replacing $\theta$ with the sigmoid at $w=w_{\rm soft}$. 
Since $G_{M}$ minimizes over all partitions, $\tilde{G}(w)\ge G_{M}$ for every $w$. Moreover, provided the two coefficients bracketing the partition in Eq.~\eqref{eq:wstar} are distinct, Eq.~\eqref{eq:w_soft} places $w_{\rm soft}$ strictly between them, the induced partition is exactly the optimal one, and $\tilde{G}(w_{\rm soft})=G_{M}$ with equality. The function minimized in each iteration is then an upper bound on the true cost that is tight at the current iterate, and the outer loop of Algorithm~\ref{alg:UWC}, which recomputes $L_{D}^{*}$ and $w_{\rm soft}$ before each inner optimization, is a majorization--minimization scheme rather than a heuristic update of a hyperparameter. The sigmoid serves only to render this bound differentiable, with the fidelity controlled by Eq.~\eqref{eq:epsupper}.

\subsection{Computational details}\label{append:uwc_comp_details}

Here, we specify the computational settings used to obtain the UWC-optimized Hamiltonians for the numerical results in Sec.~\ref{sec:numerical} and~\ref{sec:resource_estimation}.

To generate the Cholesky orbital basis representations, we employed the double factorization functionality implemented in the \texttt{ffsim} library~\cite{ffsim}. The subsequent OO and BLISS optimizations were performed using the sequential least squares programming (SLSQP) optimizer available in the \texttt{SciPy} package~\cite{2020SciPy-NMeth}. The choice of SLSQP was motivated by the nature of our softened gate-cost function, $G_{\rm soft}$ (Eq.~\eqref{eq:uwc_def}). This function includes an absolute value term $|c_{L_D}|$, which makes its gradient discontinuous. We empirically observed that optimizers reliant on continuous gradients, such as L-BFGS-B, often failed to minimize $G_{\rm soft}$ effectively. To accelerate the optimization, all gradients were computed using automatic differentiation via the \texttt{JAX} library~\cite{jax2018github}. We also found that the raw values of $G_{\rm soft}$ and its gradients could become very large, leading to numerical instability in the optimization process. This issue was mitigated by rescaling the cost function by the initial gate cost $G_M$ for the starting Hamiltonian $\hat{H}$. 
The sigmoid is evaluated with \texttt{jax.nn.sigmoid}. A naive implementation as $1/(1+e^{-x})$ overflows in double precision once $\lvert x\rvert\gtrsim710$, a regime the argument $(\lvert c_{\ell}\rvert-w_{\rm soft})/\epsilon_{\rm soft}$ readily reaches; the value remains correct, but the reverse-mode derivative evaluates $0\times\infty$ and returns NaN.


In practice, we set $\epsilon_{\rm soft}=10^{-4}$ throughout, this being the largest round value admissible under Eq.~\eqref{eq:epsupper} for all systems studied. Spot checks at other values are consistent with the picture above. Reducing $\epsilon_{\rm soft}$ makes the surrogate more faithful and lowers the gate count marginally---by a few per cent, against an overall reduction of about an order of magnitude---at the cost of slower inner minimizations. Raising it above the ceiling of Eq.~\eqref{eq:epsupper} instead returns a representation with a \emph{smaller} $\ell_{1}$-norm but a larger gate count, as anticipated by Eq.~\eqref{eq:largeepslimit}. The results reported here are therefore insensitive to $\epsilon_{\rm soft}$ at the level of a few per cent within the admissible range, and are conservative in the sense that a smaller value would improve them slightly.

For all numerical results presented, the UWC optimization (Algorithm~\ref{alg:UWC}) was executed with a maximum of $N_{\rm iter}=10$ iterations and a convergence threshold of $\delta_{\rm th}=0.1\%$. While these strict criteria ensure thorough optimization, we note for practical purposes that they could likely be relaxed. Empirically, we consistently observed that the majority of the cost reduction occurred during the first iteration, with subsequent iterations providing only minor refinements.

\section{Detailed resource estimation data for molecular models \label{append:mol_data_full}}

In this Appendix, we provide comprehensive resource estimation data for all molecular models investigated in this work. The results presented in Sec.~\ref{sec:numerical} and Sec.~\ref{sec:resource_estimation} are derived from and substantiated by the detailed information compiled within this Appendix.

\subsection{Details of molecular models \label{append:models_detail}}

This section provides essential information for the molecular active-space models investigated in this work. We focus on four distinct families: hydrogen chains, iron-sulfur clusters (including FeMoco), cytochrome P450 active sites, and ruthenium-based catalysts for \ce{CO2} utilization. All Hamiltonian data were sourced from publicly available datasets accompanying the relevant literature~\cite{Beinert1997-ok,Lee2023-rd,Ollitrault2024-ko,Reiher2017-bd,Li2019-ah,Goings2022-so,Von_Burg2021-du}.

The pertinent details for each molecular family, including the number of orbitals, number of electrons, and Hilbert space dimension, are summarized in Tables~\ref{tab:h_chain} (hydrogen chain models), \ref{tab:fe-s} (iron-sulfur clusters), \ref{tab:p450} (cytochrome P450 active-site models), and \ref{tab:co2_cat} (ruthenium-based catalyst series). 
For consistency, each table includes columns for
\begin{itemize}
    \item \textbf{``Molecule ID"}: Identifier for specific molecular active-space models. 
    \item \textbf{``Orbitals''}: Number of spatial orbitals, $N$. 
    \item \textbf{``Electrons"}: Number of spin-up and spin-down electrons, ($N_\uparrow,N_\downarrow$).
    \item \textbf{``Hilbert space dimension"}: Dimension of the Hilbert space calculated as $\binom{N}{N_\uparrow}\binom{N}{N_\downarrow}$. 
\end{itemize}
Specific naming conventions and data sources for each model are elaborated within the respective table captions.

It is important to note that the number of orbitals in a model can vary even for the same molecule, depending on the chosen active space. Such variations are distinguished by identifiers in parentheses, for example, FeMoco (Sm) and FeMoco (Lg) denote smaller and larger active-space models for FeMoco, respectively.

\begin{table}[tbp]
\caption{Summary of molecular active-space models for hydrogen chain models. H$_N$ identifies a linear hydrogen chain model with $N$ hydrogen atoms. The Hamiltonian data for these models were obtained from a publicly available Zenodo repository~\cite{joonho_lee_2020_4248322} associated with Ref.~\cite{Lee2021-tz}. }
\label{tab:h_chain}
\begin{ruledtabular}
\begin{tabular}{lccc}
    Molecule ID &  Orbitals &  \makecell{Electrons\\($N_\uparrow, N_\downarrow$)} & \makecell{Hilbert space\\dimension} \\
    \colrule
\ce{H{10}} & 10 & (5, 5) & $6.35\times 10^{4}$ \\
\ce{H{20}} & 20 & (10, 10) & $3.41\times 10^{10}$ \\
\ce{H{30}} & 30 & (15, 15) & $2.41\times 10^{16}$ \\
\ce{H{40}} & 40 & (20, 20) & $1.90\times 10^{22}$ \\
\ce{H{50}} & 50 & (25, 25) & $1.60\times 10^{28}$ \\
\ce{H{60}} & 60 & (30, 30) & $1.40\times 10^{34}$ \\
\ce{H{70}} & 70 & (35, 35) & $1.26\times 10^{40}$ \\
\ce{H{80}} & 80 & (40, 40) & $1.16\times 10^{46}$ \\
    \end{tabular}
\end{ruledtabular}
\end{table}
\begin{table}[tbp]
\caption{Summary of molecular active-space models for iron-sulfur clusters.
The Hamiltonian data for $\mathrm{[2Fe\mathchar`-2S]}^{-2}$, $\mathrm{[2Fe\mathchar`-2S]}^{-3}$, $\mathrm{[4Fe\mathchar`-4S]}$, and $\mathrm{[4Fe\mathchar`-4S]}^{-2}$ were retrieved from a publicly available Zenodo repository~\cite{ollitrault_2024_10944099} associated with Ref.~\cite{Ollitrault2024-ko}.
FeMoco (Sm) and FeMoco (Lg) denote the smaller 54-orbital and larger 76-orbital models for FeMoco, respectively, derived in Refs.~\cite{Reiher2017-bd} and \cite{Li2019-ah}. The Hamiltonian data for these FeMoco models were obtained from a separate publicly available Zenodo repository~\cite{joonho_lee_2020_4248322} linked to Ref.~\cite{Lee2021-tz}.
}
\label{tab:fe-s}
\begin{ruledtabular}
\begin{tabular}{lccc}
    Molecule ID  &  Orbitals &  \makecell{Electrons\\($N_\uparrow, N_\downarrow$)} & \makecell{Hilbert space\\dimension} \\
    \colrule
    $\mathrm{[2Fe\mathchar`-2S]}^{-3}$ &  20 &  (16, 15) &  $7.51\times 10^{7}$ \\
    $\mathrm{[2Fe\mathchar`-2S]}^{-2}$ &  20 &  (15, 15) &  $2.40\times 10^{8}$ \\
         $\mathrm{[4Fe\mathchar`-4S]}^{-2}$ &  36 &  (27, 27) &  $8.86\times 10^{15}$ \\
    $\mathrm{[4Fe\mathchar`-4S]}$ &  36 &  (26, 26) &  $6.46\times 10^{16}$ \\
                           FeMoco (Sm) &  54 &  (27, 27) &  $3.79\times 10^{30}$ \\
                           FeMoco (Lg) &  76 &  (58, 55) &  $3.64\times 10^{35}$ \\
    \end{tabular}
\end{ruledtabular}
\end{table}
\begin{table}[tbp]
\caption{Summary of molecular active-space models for the cytochrome P450 active site. The Hamiltonian data for these models were obtained from a publicly available Zenodo repository~\cite{goings_2022_5941130} associated with Ref.~\cite{Goings2022-so}.
The Molecule IDs ``Cpd I", ``water", ``inhibited", and ``empty" denote the Compound I intermediate and various ligand-bound or empty states in the P450 catalytic cycle, as described in Ref.~\cite{Goings2022-so}.
The active space labels A-G-X follow a hierarchical construction introduced in Ref.~\cite{Goings2022-so}, where orbitals are progressively added based on their relevance to spin character, ligand binding, and spatial proximity to the iron center:
`A': Five open-shell orbitals crucial for spin character;
`B': Adds iron 3d/4s and iron-nitrogen antibonding orbitals;
`C': Further includes iron 4p/d' and iron-axial ligand bonding/antibonding orbitals;
`D': Incorporates spatially closest iron-nitrogen $\sigma$ orbitals;
`E': Adds heme $\pi$ orbitals on nitrogen atoms;
`F': Adds heme $\pi$ orbitals on carbon atoms;
`G': Includes distal carbon-sulfur bonding/antibonding orbitals;
`X': Finally adds less-important heme carbon-nitrogen $\sigma$ orbitals.
}
\label{tab:p450}
\begin{ruledtabular}
\begin{tabular}{lccc}
    Molecule ID &  Orbitals & \makecell{Electrons\\($N_\uparrow, N_\downarrow$)} & \makecell{Hilbert space\\dimension} \\
    \midrule
    \multicolumn{4}{c}{(a) Compound I (Cpd I) }\\
    \midrule
        P450-Cpd I (A) &         5 &    (5, 0) &                 $1.00\times 10^{0}$  \\
    P450-Cpd I (B) &         8 &    (8, 3) &                 $5.60\times 10^{1}$  \\
    P450-Cpd I (C) &        15 &   (11, 6) &                 $6.83\times 10^{6}$  \\
    P450-Cpd I (D) &        23 &  (15, 10) &                $5.61\times 10^{11}$ \\
    P450-Cpd I (E) &        31 &  (19, 14) &                $3.74\times 10^{16}$ \\
    P450-Cpd I (F) &        41 &  (25, 20) &                $2.77\times 10^{22}$ \\
    P450-Cpd I (G) &        43 &  (26, 21) &                $4.43\times 10^{23}$ \\
    P450-Cpd I (X) &        58 &  (34, 29) &                $3.86\times 10^{32}$ \\
    \midrule
    \multicolumn{4}{c}{(b) Resting state with water bound to heme }\\
    \midrule
     P450-rest (A) &         5 &    (5, 0) &                 $1.00\times 10^{0}$   \\
     P450-rest (B) &         8 &    (7, 2) &                 $2.24\times 10^{2}$  \\
     P450-rest (C) &        13 &   (10, 5) &                 $3.68\times 10^{5}$ \\
     P450-rest (D) &        20 &   (14, 9) &                 $6.51\times 10^{9}$ \\
     P450-rest (E) &        28 &  (18, 13) &                $4.91\times 10^{14}$ \\
     P450-rest (F) &        40 &  (24, 19) &                $8.25\times 10^{21}$ \\
     P450-rest (G) &        42 &  (25, 20) &                $1.31\times 10^{23}$ \\
     P450-rest (X) &        56 &  (33, 28) &                $2.42\times 10^{31}$ \\
    \midrule
    \multicolumn{4}{c}{(c) Pyridine inhibitor bound model complex }\\
    \midrule
    P450-inhibited (A) &         5 &    (5, 0) &                 $1.00\times 10^{0}$ \\
    P450-inhibited (B) &         9 &    (8, 3) &                 $7.56\times 10^{2}$ \\
    P450-inhibited (C) &        14 &   (11, 6) &                 $1.09\times 10^{6}$ \\
    P450-inhibited (D) &        21 &  (15, 10) &                $1.91\times 10^{10}$ \\
    P450-inhibited (E) &        32 &  (20, 15) &                $1.28\times 10^{17}$ \\
    P450-inhibited (F) &        44 &  (27, 22) &                $1.44\times 10^{24}$ \\
    P450-inhibited (G) &        46 &  (28, 23) &                $2.32\times 10^{25}$ \\
    P450-inhibited (X) &        60 &  (36, 31) &                $4.13\times 10^{33}$\\
    \midrule
    \multicolumn{4}{c}{(d) Pentacoordinated empty state with no ligands }\\
    \midrule
    P450-empty (A) &         5 &    (5, 0) &                 $1.00\times 10^{0}$ \\
    P450-empty (B) &         8 &    (7, 2) &                 $2.24\times 10^{2}$ \\
    P450-empty (C) &        11 &    (9, 4) &                 $1.82\times 10^{4}$ \\
    P450-empty (D) &        18 &   (13, 8) &                 $3.75\times 10^{8}$ \\
    P450-empty (E) &        26 &  (17, 12) &                $3.02\times 10^{13}$ \\
    P450-empty (F) &        37 &  (22, 17) &                $1.49\times 10^{20}$ \\
    P450-empty (G) &        39 &  (23, 18) &                $2.35\times 10^{21}$ \\
    P450-empty (X) &        55 &  (31, 26) &                $8.86\times 10^{30}$ \\
    \end{tabular}
\end{ruledtabular}
\end{table}
\begin{table}[tbp]
\caption{Summary of molecular active-space models for ruthenium-based catalysts used in \ce{CO2} utilization.
The Hamiltonian data for these models were retrieved from a publicly available Zenodo repository~\cite{vera_von_burg_2021_4769113}, associated with Ref.~\cite{Von_Burg2021-du}.
The numerical labels in the Molecule IDs (i.e., I, II, II-III, V, VIII, VIII-IX, IX, and XVIII) correspond to specific intermediate and transition state structures within the catalytic cycle, as detailed in Ref.~\cite{Von_Burg2021-du}.
The active space labels `Sm', `Md', and `Lg' denote a hierarchical construction introduced in Ref.~\cite{Von_Burg2021-du}, where orbitals are progressively added based on their electronic correlation and chemical relevance:
`Sm' represents a small active space constructed from orbitals selected via an entanglement criterion;
`Md' designates a middle-sized active space, adding valence orbitals of ruthenium/ligands and ligand-metal bonding orbitals;
`Lg' refers to a large active space, further incorporating additional $\pi$/$\pi^*$ orbitals on the triphos ligand.}
\label{tab:co2_cat}
\begin{ruledtabular}
\begin{tabular}{lccc}
    Molecule ID &  Orbitals & \makecell{Electrons\\($N_\uparrow, N_\downarrow$)} & \makecell{Hilbert space\\dimension} \\
    \midrule
    \multicolumn{4}{c}{(a) Stable intermediate I }\\
    \midrule
      Ru-I (Sm) &         5 &    (2, 2) &                 $1.00\times 10^{2}$   \\
      Ru-I (Md) &        16 &    (7, 7) &                 $1.31\times 10^{8}$  \\
      Ru-I (Lg) &        52 &  (24, 24) &                $1.82\times 10^{29}$  \\
    \midrule
    \multicolumn{4}{c}{(b) Stable intermediate II }\\
    \midrule
     Ru-II (Sm) &         6 &    (4, 4) &                 $2.25\times 10^{2}$  \\
     Ru-II (Md) &        26 &  (17, 17) &                $9.76\times 10^{12}$ \\
     Ru-II (Lg) &        62 &  (35, 35) &                $7.82\times 10^{34}$  \\
    \midrule
    \multicolumn{4}{c}{(c) Transition state II-III }\\
    \midrule
 Ru-II-III (Sm) &         6 &    (4, 4) &                 $2.25\times 10^{2}$  \\
 Ru-II-III (Md) &        29 &  (19, 19) &                $4.01\times 10^{14}$  \\
 Ru-II-III (Lg) &        65 &  (37, 37) &                $3.86\times 10^{36}$   \\
     \midrule
    \multicolumn{4}{c}{(e) Stable intermediate V }\\
    \midrule
      Ru-V (Sm) &        11 &    (6, 6) &                 $2.13\times 10^{5}$  \\
      Ru-V (Md) &        24 &  (16, 16) &                $5.41\times 10^{11}$  \\
      Ru-V (Lg) &        60 &  (34, 34) &                $4.88\times 10^{33}$  \\
    \midrule
    \multicolumn{4}{c}{(f) Stable intermediate VIII }\\
    \midrule
   Ru-VIII (Sm) &         2 &    (1, 1) &                 $4.00\times 10^{0}$  \\
   Ru-VIII (Md) &        29 &  (20, 20) &                $1.00\times 10^{14}$  \\
   Ru-VIII (Lg) &        65 &  (38, 38) &                $2.10\times 10^{36}$  \\
    \midrule
    \multicolumn{4}{c}{(g) Transition state VIII-IX }\\
    \midrule
Ru-VIII-IX (Sm) &         4 &    (2, 2) &                 $3.60\times 10^{1}$  \\
Ru-VIII-IX (Md) &        23 &  (18, 18) &                 $1.13\times 10^{9}$ \\
Ru-VIII-IX (Lg) &        59 &  (36, 36) &                $2.08\times 10^{32}$  \\
    \midrule
    \multicolumn{4}{c}{(d) Stable intermediate IX }\\
    \midrule
     Ru-IX (Sm) &        16 &    (8, 8) &                 $1.66\times 10^{8}$  \\
     Ru-IX (Md) &        26 &  (16, 16) &                $2.82\times 10^{13}$  \\
     Ru-IX (Lg) &        62 &  (34, 34) &                $1.22\times 10^{35}$   \\
    \midrule
    \multicolumn{4}{c}{(h) Stable intermediate XVIII }\\
    \midrule
  Ru-XVIII (Sm) &         4 &    (2, 2) &                 $3.60\times 10^{1}$  \\
  Ru-XVIII (Md) &        20 &  (14, 14) &                 $1.50\times 10^{9}$  \\
  Ru-XVIII (Lg) &        56 &  (32, 32) &                $1.90\times 10^{31}$ \\
    \end{tabular}
\end{ruledtabular}
\end{table}

\subsection{Full list of resource estimates \label{append:full_resource_estimates}}

This section summarizes the logical and physical resource estimates for all molecular active-space models introduced previously. These estimates specifically pertain to partially randomized RPE under the UWC-optimized Hamiltonian representation. The calculations are performed for three different control parameters: $\xi=0.1$, $\xi=0.05$, and $\xi=0.01$. It is important to note that the UWC-optimized Hamiltonian representation is distinct for each $\xi$ value. Furthermore, the partially randomized circuits are assumed to comprise deterministic second-order Trotterization and randomized qDRIFT parts. For all resource estimates, a realistic physical error rate of $p_{\rm ph}=10^{-3}$ and the chemical accuracy target $\epsilon=1.6\times 10^{-3}$ are adopted.

\subsubsection{Logical resource estimates \label{append:logical_resource}}
Logical resource estimates, encompassing information on the UWC-optimized Hamiltonian representations, logical circuits for RPE, and partial fault tolerance based on the SMM protocol, are detailed in Tables~\ref{tab:h_chain_logical} (hydrogen chain models), \ref{tab:fe-s_logical} (iron-sulfur clusters), \ref{tab:p450_1_logical}, \ref{tab:p450_2_logical} (cytochrome P450 active-site models), and \ref{tab:co2-cat_1_logical}, \ref{tab:co2-cat_2_logical} (ruthenium-based catalyst series). For consistency, each table features ``Hamiltonian", ``Logical circuit", and ``Error correction" sections.
\begin{itemize}
    \item The \textbf{``Hamiltonian"} columns report the number of deterministically treated terms ($L_D^*$), the total weight for the randomized part ($\lambda_R$), and the $\ell_1$-norm ($\lambda$). $L_D^*$ and $\lambda_R$ represent the optimal values for the last round $m=M$, chosen to minimize the gate count as defined in Eq.~\eqref{eq:abstract_cost_model}.
    \item The \textbf{``Logical circuit"} columns provide the number of logical data qubits ($N_L=2N+1$), the maximum gate count ($G_M$) at the last round (Eq.~\eqref{eq:abstract_cost_model}), and the total logical gate cost $G_{\rm total}$ (Eq.~\eqref{eq:total_cost}).
    \item The \textbf{``Error correction"} columns detail the partially fault-tolerant setup for the last round $m=M$, including the selected SMM priority (``accuracy" or ``speed" prioritized setting in Appendix~\ref{append:resource_assumptions}), the code distance ($d$), the average rotation angle ($\bar{\theta}_L$) within the partially randomized circuit, and the total logical error rate ($P_{\rm total}$).
\end{itemize}
The partially randomized circuit consists of second-order Trotterization and qDRIFT. The deterministic component involves Pauli rotations with angles $\{ c_{\ell}\frac{\delta}{2} \}_{\ell=1,\cdots,L_D}$, where $\delta$ is the Trotterization step size (Eq.~\eqref{eq:trotter_step_optimal}). The randomized part uses Pauli rotation gates with a fixed angle $\arctan(1/(2\lambda_R\delta 2^{m}))$ for the $m$-th round~\cite{Gunther2025-zk,Campbell2019-of}. Using this information alongside the angular dependence of $\alpha_{\rm RUS}(\theta_L)$ and $C_{\rm smm}(\theta_L)$, determined from Fig.~\ref{fig:star_magic_mutation}, the values of $d$ and $P_{\rm total}$ are derived according to the procedure outlined in Sec.~\ref{subsec:resource_estimation_setup}.

\begin{table*}[htbp]
\caption{
    Logical resource requirements for the hydrogen chain models across three initial-state infidelity parameter values: (a) $\xi=0.1$, (b) $\xi=0.05$, and (c) $\xi=0.01$.
    Columns are grouped into ``Hamiltonian" ($L_D^*$, $\lambda_R$, $\lambda$), ``Logical circuit" ($N_L$, $G_M$, $G_{\rm total}$), and ``Error correction" (SMM priority, $d$, $\bar{\theta}_L$, $P_{\rm total}$).
    Detailed definitions for these parameters and the underlying assumptions can be found in Appendix~\ref{append:logical_resource}.
}
\label{tab:h_chain_logical}
\begin{ruledtabular}
\begin{tabular}{l c c c c c c c c c c}
    & \multicolumn{3}{c}{Hamiltonian} & \multicolumn{3}{c}{Logical circuit} & \multicolumn{4}{c}{Error correction} \\
    \cmidrule(lr){2-4} \cmidrule(lr){5-7} \cmidrule(lr){8-11}
    Molecule ID & $L_D^*$ & $\lambda_R$ & $\lambda$ & \makecell{$N_{L}$} & \makecell{$G_M$} & $G_{\rm total}$ & SMM priority & \makecell{$d$} & \makecell{$\bar{\theta}_L$} & \makecell{$P_{\rm total}$} \\
    \midrule
    \multicolumn{11}{c}{(a) $\xi=0.1$}\\
    \midrule
\ce{H10} & 217 & 2.66 & 25.52 & 21 & $1.36\times 10^{5}$ & $1.49\times 10^{8}$ & accuracy & 21 & $7.19\times 10^{-3}$ & $3.98\times 10^{-2}$ \\
\ce{H20} & 552 & 7.59 & 62.54 & 41 & $9.53\times 10^{5}$ & $1.44\times 10^{9}$ & accuracy & 23 & $2.51\times 10^{-3}$ & $1.04\times 10^{-1}$ \\
\ce{H30} & 1028 & 13.59 & 106.02 & 61 & $3.07\times 10^{6}$ & $3.06\times 10^{9}$ & accuracy & 25 & $1.32\times 10^{-3}$ & $1.85\times 10^{-1}$ \\
\ce{H40} & 1602 & 19.99 & 150.75 & 81 & $6.81\times 10^{6}$ & $8.43\times 10^{9}$ & accuracy & 25 & $8.47\times 10^{-4}$ & $2.82\times 10^{-1}$ \\
\ce{H50} & 2340 & 30.19 & 202.62 & 101 & $1.42\times 10^{7}$ & $1.39\times 10^{10}$ & accuracy & 25 & $5.45\times 10^{-4}$ & $4.08\times 10^{-1}$ \\
\ce{H60} & 2772 & 34.68 & 244.14 & 121 & $1.99\times 10^{7}$ & $2.84\times 10^{10}$ & accuracy & 27 & $4.70\times 10^{-4}$ & $5.09\times 10^{-1}$ \\
\ce{H70} & 3580 & 48.27 & 299.75 & 141 & $3.42\times 10^{7}$ & $4.03\times 10^{10}$ & accuracy & 27 & $3.36\times 10^{-4}$ & $6.52\times 10^{-1}$ \\
\ce{H80} & 4378 & 56.68 & 350.23 & 161 & $4.83\times 10^{7}$ & $5.05\times 10^{10}$ & accuracy & 27 & $2.77\times 10^{-4}$ & $7.81\times 10^{-1}$ \\
    \midrule
    \multicolumn{11}{c}{(b) $\xi=0.05$}\\
    \midrule
\ce{H10} & 165 & 3.14 & 24.41 & 21 & $4.88\times 10^{4}$ & $1.30\times 10^{8}$ & speed & 19 & $9.58\times 10^{-3}$ & $6.86\times 10^{-1}$ \\
\ce{H20} & 466 & 8.93 & 62.45 & 41 & $3.76\times 10^{5}$ & $1.12\times 10^{9}$ & accuracy & 23 & $3.18\times 10^{-3}$ & $5.08\times 10^{-2}$ \\
\ce{H30} & 871 & 16.01 & 105.12 & 61 & $1.21\times 10^{6}$ & $3.09\times 10^{9}$ & accuracy & 23 & $1.66\times 10^{-3}$ & $9.03\times 10^{-2}$ \\
\ce{H40} & 1296 & 24.70 & 150.87 & 81 & $2.70\times 10^{6}$ & $7.43\times 10^{9}$ & accuracy & 25 & $1.07\times 10^{-3}$ & $1.40\times 10^{-1}$ \\
\ce{H50} & 1766 & 33.62 & 198.31 & 101 & $4.94\times 10^{6}$ & $1.26\times 10^{10}$ & accuracy & 25 & $7.68\times 10^{-4}$ & $1.97\times 10^{-1}$ \\
\ce{H60} & 2206 & 43.08 & 243.89 & 121 & $7.81\times 10^{6}$ & $2.27\times 10^{10}$ & accuracy & 25 & $5.97\times 10^{-4}$ & $2.53\times 10^{-1}$ \\
\ce{H70} & 2774 & 58.95 & 298.71 & 141 & $1.30\times 10^{7}$ & $3.51\times 10^{10}$ & accuracy & 25 & $4.39\times 10^{-4}$ & $3.24\times 10^{-1}$ \\
\ce{H80} & 3202 & 72.05 & 348.72 & 161 & $1.84\times 10^{7}$ & $4.75\times 10^{10}$ & accuracy & 27 & $3.64\times 10^{-4}$ & $3.88\times 10^{-1}$ \\
    \midrule
    \multicolumn{11}{c}{(c) $\xi=0.01$}\\
    \midrule
\ce{H10} & 103 & 5.39 & 23.51 & 21 & $5.83\times 10^{3}$ & $3.20\times 10^{8}$ & speed & 17 & $1.55\times 10^{-2}$ & $1.50\times 10^{-1}$ \\
\ce{H20} & 267 & 16.37 & 61.86 & 41 & $4.51\times 10^{4}$ & $2.47\times 10^{9}$ & speed & 19 & $5.25\times 10^{-3}$ & $2.65\times 10^{-1}$ \\
\ce{H30} & 472 & 28.17 & 103.49 & 61 & $1.36\times 10^{5}$ & $7.46\times 10^{9}$ & speed & 21 & $2.92\times 10^{-3}$ & $3.26\times 10^{-1}$ \\
\ce{H40} & 669 & 43.56 & 148.50 & 81 & $2.95\times 10^{5}$ & $1.63\times 10^{10}$ & speed & 21 & $1.93\times 10^{-3}$ & $2.43\times 10^{-1}$ \\
\ce{H50} & 894 & 59.68 & 196.40 & 101 & $5.40\times 10^{5}$ & $2.96\times 10^{10}$ & speed & 23 & $1.40\times 10^{-3}$ & $1.93\times 10^{-1}$ \\
\ce{H60} & 1104 & 78.83 & 245.75 & 121 & $8.80\times 10^{5}$ & $4.81\times 10^{10}$ & speed & 23 & $1.07\times 10^{-3}$ & $1.86\times 10^{-1}$ \\
\ce{H70} & 1425 & 95.37 & 297.84 & 141 & $1.35\times 10^{6}$ & $7.41\times 10^{10}$ & speed & 23 & $8.48\times 10^{-4}$ & $1.91\times 10^{-1}$ \\
\ce{H80} & 1648 & 114.36 & 347.80 & 161 & $1.87\times 10^{6}$ & $1.03\times 10^{11}$ & speed & 23 & $7.12\times 10^{-4}$ & $1.97\times 10^{-1}$ \\
    \end{tabular}
\end{ruledtabular}
\end{table*}
\begin{table*}[htbp]
\caption{
    Logical resource requirements for the iron-sulfur cluster models (listed in Table~\ref{tab:fe-s}) across three initial-state infidelity parameter values: (a) $\xi=0.1$, (b) $\xi=0.05$, and (c) $\xi=0.01$.
    Columns are grouped into ``Hamiltonian" ($L_D^*$, $\lambda_R$, $\lambda$), ``Logical circuit" ($N_L$, $G_M$, $G_{\rm total}$), and ``Error correction" (SMM priority, $d$, $\bar{\theta}_L$, $P_{\rm total}$).
    Detailed definitions for these parameters and the underlying assumptions can be found in Appendix~\ref{append:logical_resource}.
}
\label{tab:fe-s_logical}
\begin{ruledtabular}
\begin{tabular}{l c c c c c c c c c c}
    & \multicolumn{3}{c}{Hamiltonian} & \multicolumn{3}{c}{Logical circuit} & \multicolumn{4}{c}{Error correction} \\
    \cmidrule(lr){2-4} \cmidrule(lr){5-7} \cmidrule(lr){8-11}
    Molecule ID & $L_D^*$ & $\lambda_R$ & $\lambda$ & \makecell{$N_{L}$} & \makecell{$G_M$} & $G_{\rm total}$ & SMM priority & \makecell{$d$} & \makecell{$\bar{\theta}_L$} & \makecell{$P_{\rm total}$} \\
    \midrule
    \multicolumn{11}{c}{(a) $\xi=0.1$}\\
    \midrule
$[2\ce{Fe}\mathchar`-2\ce{S}]^{-3}$ & 922 & 12.88 & 45.35 & 41 & $1.71\times 10^{6}$ & $1.71\times 10^{9}$ & accuracy & 23 & $1.02\times 10^{-3}$ & $8.67\times 10^{-2}$ \\
$[2\ce{Fe}\mathchar`-2\ce{S}]^{-2}$ & 1000 & 15.58 & 46.01 & 41 & $2.23\times 10^{6}$ & $2.16\times 10^{9}$ & accuracy & 23 & $7.90\times 10^{-4}$ & $9.07\times 10^{-2}$ \\
$[4\ce{Fe}\mathchar`-4\ce{S}]^{-2}$ & 3695 & 68.61 & 136.89 & 73 & $3.69\times 10^{7}$ & $4.16\times 10^{10}$ & accuracy & 25 & $1.42\times 10^{-4}$ & $3.25\times 10^{-1}$ \\
$[4\ce{Fe}\mathchar`-4\ce{S}]$ & 3817 & 68.44 & 138.73 & 73 & $3.72\times 10^{7}$ & $4.12\times 10^{10}$ & accuracy & 25 & $1.43\times 10^{-4}$ & $3.25\times 10^{-1}$ \\
FeMoco (Sm) & 12176 & 269.15 & 400.52 & 109 & $5.19\times 10^{8}$ & $4.06\times 10^{11}$ & accuracy & 27 & $2.95\times 10^{-5}$ & $1.01\times 10^{0}$ \\
FeMoco (Lg) & 15004 & 273.53 & 521.58 & 153 & $5.92\times 10^{8}$ & $6.59\times 10^{11}$ & accuracy & 29 & $3.37\times 10^{-5}$ & $1.32\times 10^{0}$ \\
    \midrule
    \multicolumn{11}{c}{(b) $\xi=0.05$}\\
    \midrule
$[2\ce{Fe}\mathchar`-2\ce{S}]^{-3}$ & 578 & 14.95 & 43.70 & 41 & $5.50\times 10^{5}$ & $1.42\times 10^{9}$ & accuracy & 23 & $1.52\times 10^{-3}$ & $4.01\times 10^{-2}$ \\
$[2\ce{Fe}\mathchar`-2\ce{S}]^{-2}$ & 759 & 16.89 & 40.77 & 41 & $6.90\times 10^{5}$ & $1.80\times 10^{9}$ & accuracy & 23 & $1.13\times 10^{-3}$ & $3.96\times 10^{-2}$ \\
$[4\ce{Fe}\mathchar`-4\ce{S}]^{-2}$ & 1789 & 73.63 & 141.80 & 73 & $1.03\times 10^{7}$ & $2.70\times 10^{10}$ & accuracy & 25 & $2.64\times 10^{-4}$ & $1.61\times 10^{-1}$ \\
$[4\ce{Fe}\mathchar`-4\ce{S}]$ & 2177 & 74.98 & 135.50 & 73 & $1.10\times 10^{7}$ & $2.92\times 10^{10}$ & accuracy & 25 & $2.37\times 10^{-4}$ & $1.58\times 10^{-1}$ \\
FeMoco (Sm) & 5642 & 286.35 & 396.64 & 109 & $1.42\times 10^{8}$ & $3.37\times 10^{11}$ & accuracy & 27 & $5.35\times 10^{-5}$ & $4.95\times 10^{-1}$ \\
FeMoco (Lg) & 10376 & 295.29 & 522.53 & 153 & $1.81\times 10^{8}$ & $4.79\times 10^{11}$ & accuracy & 27 & $5.53\times 10^{-5}$ & $6.52\times 10^{-1}$ \\
    \midrule
    \multicolumn{11}{c}{(c) $\xi=0.01$}\\
    \midrule
$[2\ce{Fe}\mathchar`-2\ce{S}]^{-3}$ & 196 & 19.75 & 41.26 & 41 & $3.70\times 10^{4}$ & $2.03\times 10^{9}$ & speed & 19 & $4.27\times 10^{-3}$ & $1.52\times 10^{-1}$ \\
$[2\ce{Fe}\mathchar`-2\ce{S}]^{-2}$ & 216 & 22.75 & 43.40 & 41 & $4.68\times 10^{4}$ & $2.57\times 10^{9}$ & speed & 19 & $3.55\times 10^{-3}$ & $1.40\times 10^{-1}$ \\
$[4\ce{Fe}\mathchar`-4\ce{S}]^{-2}$ & 513 & 90.88 & 131.42 & 73 & $6.09\times 10^{5}$ & $3.32\times 10^{10}$ & speed & 21 & $8.27\times 10^{-4}$ & $7.71\times 10^{-2}$ \\
$[4\ce{Fe}\mathchar`-4\ce{S}]$ & 481 & 91.96 & 133.98 & 73 & $6.15\times 10^{5}$ & $3.36\times 10^{10}$ & speed & 21 & $8.34\times 10^{-4}$ & $1.10\times 10^{-1}$ \\
FeMoco (Sm) & 1049 & 326.77 & 385.47 & 109 & $7.05\times 10^{6}$ & $3.85\times 10^{11}$ & speed & 25 & $2.09\times 10^{-4}$ & $1.15\times 10^{-1}$ \\
FeMoco (Lg) & 2321 & 370.23 & 506.14 & 153 & $1.03\times 10^{7}$ & $5.64\times 10^{11}$ & speed & 25 & $1.87\times 10^{-4}$ & $1.40\times 10^{-1}$ \\
    \end{tabular}
\end{ruledtabular}
\end{table*}
\begin{table*}[htbp]
\caption{
    Logical resource requirements for the P450-Cpd I and P450-rest models across three initial-state infidelity parameter values: (a) $\xi=0.1$, (b) $\xi=0.05$, and (c) $\xi=0.01$.
    Columns are grouped into ``Hamiltonian" ($L_D^*$, $\lambda_R$, $\lambda$), ``Logical circuit" ($N_L$, $G_M$, $G_{\rm total}$), and ``Error correction" (SMM priority, $d$, $\bar{\theta}_L$, $P_{\rm total}$).
    Detailed definitions for these parameters and the underlying assumptions can be found in Appendix~\ref{append:logical_resource}.
}
\label{tab:p450_1_logical}
\begin{ruledtabular}
\begin{tabular}{l c c c c c c c c c c}
    & \multicolumn{3}{c}{Hamiltonian} & \multicolumn{3}{c}{Logical circuit} & \multicolumn{4}{c}{Error correction} \\
    \cmidrule(lr){2-4} \cmidrule(lr){5-7} \cmidrule(lr){8-11}
    Molecule ID & $L_D^*$ & $\lambda_R$ & $\lambda$ & \makecell{$N_{L}$} & \makecell{$G_M$} & $G_{\rm total}$ & SMM priority & \makecell{$d$} & \makecell{$\bar{\theta}_L$} & \makecell{$P_{\rm total}$} \\
    \midrule
    \multicolumn{11}{c}{(a) $\xi=0.1$}\\
    \midrule
P450-Cpd I (A) & 16 & 0.13 & 0.87 & 11 & $3.15\times 10^{2}$ & $3.34\times 10^{5}$ & speed & 13 & $1.08\times 10^{-1}$ & $5.66\times 10^{-2}$ \\
P450-Cpd I (B) & 52 & 0.62 & 8.08 & 17 & $9.05\times 10^{3}$ & $1.53\times 10^{7}$ & speed & 17 & $3.42\times 10^{-2}$ & $5.14\times 10^{-1}$ \\
P450-Cpd I (C) & 343 & 4.98 & 24.52 & 31 & $2.89\times 10^{5}$ & $2.86\times 10^{8}$ & accuracy & 23 & $3.25\times 10^{-3}$ & $4.16\times 10^{-2}$ \\
P450-Cpd I (D) & 1447 & 21.93 & 65.08 & 47 & $4.50\times 10^{6}$ & $5.65\times 10^{9}$ & accuracy & 25 & $5.54\times 10^{-4}$ & $1.33\times 10^{-1}$ \\
P450-Cpd I (E) & 3082 & 44.64 & 115.12 & 63 & $1.82\times 10^{7}$ & $2.49\times 10^{10}$ & accuracy & 25 & $2.43\times 10^{-4}$ & $2.57\times 10^{-1}$ \\
P450-Cpd I (F) & 4330 & 61.11 & 153.96 & 83 & $3.42\times 10^{7}$ & $3.57\times 10^{10}$ & accuracy & 25 & $1.72\times 10^{-4}$ & $3.55\times 10^{-1}$ \\
P450-Cpd I (G) & 4582 & 67.55 & 164.61 & 87 & $4.08\times 10^{7}$ & $4.00\times 10^{10}$ & accuracy & 25 & $1.55\times 10^{-4}$ & $3.82\times 10^{-1}$ \\
P450-Cpd I (X) & 7104 & 124.78 & 287.23 & 117 & $1.30\times 10^{8}$ & $1.37\times 10^{11}$ & accuracy & 27 & $8.45\times 10^{-5}$ & $6.90\times 10^{-1}$ \\
P450-rest (A) & 26 & 0.17 & 1.53 & 11 & $7.60\times 10^{2}$ & $9.29\times 10^{5}$ & speed & 15 & $7.73\times 10^{-2}$ & $9.75\times 10^{-2}$ \\
P450-rest (B) & 70 & 0.70 & 5.11 & 17 & $8.64\times 10^{3}$ & $1.11\times 10^{7}$ & speed & 17 & $2.29\times 10^{-2}$ & $3.27\times 10^{-1}$ \\
P450-rest (C) & 253 & 3.33 & 21.58 & 27 & $1.58\times 10^{5}$ & $1.84\times 10^{8}$ & accuracy & 21 & $5.24\times 10^{-3}$ & $3.46\times 10^{-2}$ \\
P450-rest (D) & 893 & 13.71 & 46.14 & 41 & $1.83\times 10^{6}$ & $1.77\times 10^{9}$ & accuracy & 23 & $9.67\times 10^{-4}$ & $8.79\times 10^{-2}$ \\
P450-rest (E) & 1799 & 27.16 & 82.64 & 57 & $7.00\times 10^{6}$ & $7.15\times 10^{9}$ & accuracy & 25 & $4.52\times 10^{-4}$ & $1.71\times 10^{-1}$ \\
P450-rest (F) & 3222 & 46.17 & 129.59 & 81 & $2.02\times 10^{7}$ & $2.49\times 10^{10}$ & accuracy & 25 & $2.46\times 10^{-4}$ & $2.91\times 10^{-1}$ \\
P450-rest (G) & 3215 & 50.93 & 138.03 & 85 & $2.34\times 10^{7}$ & $2.69\times 10^{10}$ & accuracy & 25 & $2.26\times 10^{-4}$ & $3.09\times 10^{-1}$ \\
P450-rest (X) & 5830 & 98.83 & 253.19 & 113 & $8.52\times 10^{7}$ & $1.02\times 10^{11}$ & accuracy & 27 & $1.14\times 10^{-4}$ & $5.94\times 10^{-1}$ \\
    \midrule
    \multicolumn{11}{c}{(b) $\xi=0.05$}\\
    \midrule
P450-Cpd I (A) & 12 & 0.19 & 0.87 & 11 & $1.45\times 10^{2}$ & $3.67\times 10^{5}$ & speed & 13 & $1.26\times 10^{-1}$ & $3.04\times 10^{-2}$ \\
P450-Cpd I (B) & 48 & 0.76 & 8.18 & 17 & $4.09\times 10^{3}$ & $1.25\times 10^{7}$ & speed & 17 & $3.87\times 10^{-2}$ & $2.63\times 10^{-1}$ \\
P450-Cpd I (C) & 233 & 5.79 & 24.08 & 31 & $9.70\times 10^{4}$ & $2.49\times 10^{8}$ & speed & 19 & $4.76\times 10^{-3}$ & $5.13\times 10^{-1}$ \\
P450-Cpd I (D) & 915 & 24.99 & 62.31 & 47 & $1.42\times 10^{6}$ & $3.99\times 10^{9}$ & accuracy & 23 & $8.39\times 10^{-4}$ & $6.26\times 10^{-2}$ \\
P450-Cpd I (E) & 2063 & 49.83 & 113.38 & 63 & $5.78\times 10^{6}$ & $1.66\times 10^{10}$ & accuracy & 25 & $3.76\times 10^{-4}$ & $1.26\times 10^{-1}$ \\
P450-Cpd I (F) & 2690 & 72.92 & 156.44 & 83 & $1.17\times 10^{7}$ & $3.01\times 10^{10}$ & accuracy & 25 & $2.56\times 10^{-4}$ & $1.80\times 10^{-1}$ \\
P450-Cpd I (G) & 2930 & 76.73 & 163.93 & 87 & $1.31\times 10^{7}$ & $3.32\times 10^{10}$ & accuracy & 25 & $2.40\times 10^{-4}$ & $1.90\times 10^{-1}$ \\
P450-Cpd I (X) & 4541 & 137.10 & 286.09 & 117 & $3.99\times 10^{7}$ & $1.04\times 10^{11}$ & accuracy & 27 & $1.37\times 10^{-4}$ & $3.42\times 10^{-1}$ \\
P450-rest (A) & 18 & 0.16 & 1.51 & 11 & $2.59\times 10^{2}$ & $7.01\times 10^{5}$ & speed & 13 & $1.18\times 10^{-1}$ & $5.09\times 10^{-2}$ \\
P450-rest (B) & 59 & 1.06 & 10.24 & 17 & $6.61\times 10^{3}$ & $1.88\times 10^{7}$ & speed & 17 & $2.97\times 10^{-2}$ & $3.26\times 10^{-1}$ \\
P450-rest (C) & 170 & 4.24 & 22.46 & 27 & $5.89\times 10^{4}$ & $1.55\times 10^{8}$ & speed & 19 & $7.32\times 10^{-3}$ & $5.70\times 10^{-1}$ \\
P450-rest (D) & 568 & 15.57 & 46.20 & 41 & $5.87\times 10^{5}$ & $1.48\times 10^{9}$ & accuracy & 23 & $1.51\times 10^{-3}$ & $4.22\times 10^{-2}$ \\
P450-rest (E) & 1180 & 31.04 & 81.66 & 57 & $2.28\times 10^{6}$ & $5.88\times 10^{9}$ & accuracy & 23 & $6.85\times 10^{-4}$ & $8.42\times 10^{-2}$ \\
P450-rest (F) & 1921 & 51.20 & 124.88 & 81 & $6.05\times 10^{6}$ & $1.68\times 10^{10}$ & accuracy & 25 & $3.95\times 10^{-4}$ & $1.38\times 10^{-1}$ \\
P450-rest (G) & 2073 & 57.79 & 137.03 & 85 & $7.52\times 10^{6}$ & $2.02\times 10^{10}$ & accuracy & 25 & $3.49\times 10^{-4}$ & $1.53\times 10^{-1}$ \\
P450-rest (X) & 3498 & 109.62 & 251.50 & 113 & $2.59\times 10^{7}$ & $7.07\times 10^{10}$ & accuracy & 25 & $1.86\times 10^{-4}$ & $2.93\times 10^{-1}$ \\
    \midrule
    \multicolumn{11}{c}{(c) $\xi=0.01$}\\
    \midrule
450-Cpd I (A) & 8 & 0.23 & 0.84 & 11 & $2.44\times 10^{1}$ & $1.33\times 10^{6}$ & speed & 11 & $2.14\times 10^{-1}$ & $8.69\times 10^{-3}$ \\
P450-Cpd I (B) & 25 & 1.49 & 6.92 & 17 & $4.51\times 10^{2}$ & $2.47\times 10^{7}$ & speed & 15 & $6.34\times 10^{-2}$ & $4.74\times 10^{-2}$ \\
P450-Cpd I (C) & 93 & 8.55 & 24.45 & 31 & $8.27\times 10^{3}$ & $4.54\times 10^{8}$ & speed & 17 & $1.14\times 10^{-2}$ & $1.55\times 10^{-1}$ \\
P450-Cpd I (D) & 266 & 33.07 & 60.16 & 47 & $9.28\times 10^{4}$ & $5.07\times 10^{9}$ & speed & 19 & $2.49\times 10^{-3}$ & $1.45\times 10^{-1}$ \\
P450-Cpd I (E) & 554 & 71.03 & 114.20 & 63 & $4.12\times 10^{5}$ & $2.25\times 10^{10}$ & speed & 21 & $1.06\times 10^{-3}$ & $1.36\times 10^{-1}$ \\
P450-Cpd I (F) & 840 & 94.96 & 153.86 & 83 & $7.72\times 10^{5}$ & $4.24\times 10^{10}$ & speed & 23 & $7.65\times 10^{-4}$ & $1.30\times 10^{-1}$ \\
P450-Cpd I (G) & 875 & 102.11 & 163.69 & 87 & $8.81\times 10^{5}$ & $4.83\times 10^{10}$ & speed & 23 & $7.13\times 10^{-4}$ & $1.30\times 10^{-1}$ \\
P450-Cpd I (X) & 1301 & 173.28 & 283.31 & 117 & $2.47\times 10^{6}$ & $1.36\times 10^{11}$ & speed & 23 & $4.40\times 10^{-4}$ & $1.36\times 10^{-1}$ \\
P450-rest (A) & 9 & 0.36 & 1.39 & 11 & $4.27\times 10^{1}$ & $2.34\times 10^{6}$ & speed & 11 & $1.80\times 10^{-1}$ & $1.28\times 10^{-2}$ \\
P450-rest (B) & 37 & 1.36 & 4.87 & 17 & $4.14\times 10^{2}$ & $2.26\times 10^{7}$ & speed & 13 & $4.68\times 10^{-2}$ & $3.21\times 10^{-2}$ \\
P450-rest (C) & 93 & 5.75 & 20.88 & 27 & $5.36\times 10^{3}$ & $2.95\times 10^{8}$ & speed & 17 & $1.52\times 10^{-2}$ & $1.35\times 10^{-1}$ \\
P450-rest (D) & 179 & 21.42 & 43.42 & 41 & $4.05\times 10^{4}$ & $2.23\times 10^{9}$ & speed & 19 & $4.10\times 10^{-3}$ & $1.46\times 10^{-1}$ \\
P450-rest (E) & 411 & 41.16 & 77.35 & 57 & $1.57\times 10^{5}$ & $8.60\times 10^{9}$ & speed & 21 & $1.89\times 10^{-3}$ & $1.33\times 10^{-1}$ \\
P450-rest (F) & 663 & 68.84 & 121.47 & 81 & $4.26\times 10^{5}$ & $2.33\times 10^{10}$ & speed & 21 & $1.09\times 10^{-3}$ & $1.19\times 10^{-1}$ \\
P450-rest (G) & 694 & 74.45 & 130.42 & 85 & $4.92\times 10^{5}$ & $2.68\times 10^{10}$ & speed & 21 & $1.02\times 10^{-3}$ & $1.33\times 10^{-1}$ \\
P450-rest (X) & 1231 & 137.66 & 244.10 & 113 & $1.68\times 10^{6}$ & $9.20\times 10^{10}$ & speed & 23 & $5.56\times 10^{-4}$ & $1.36\times 10^{-1}$ \\
    \end{tabular}
\end{ruledtabular}
\end{table*}
\begin{table*}[htbp]
\caption{
    Logical resource requirements for the P450-inhibited and P450-empty models across three initial-state infidelity parameter values: (a) $\xi=0.1$, (b) $\xi=0.05$, and (c) $\xi=0.01$.
    Columns are grouped into ``Hamiltonian" ($L_D^*$, $\lambda_R$, $\lambda$), ``Logical circuit" ($N_L$, $G_M$, $G_{\rm total}$), and ``Error correction" (SMM priority, $d$, $\bar{\theta}_L$, $P_{\rm total}$).
    Detailed definitions for these parameters and the underlying assumptions can be found in Appendix~\ref{append:logical_resource}.
}
\label{tab:p450_2_logical}
\begin{ruledtabular}
\begin{tabular}{l c c c c c c c c c c}
    & \multicolumn{3}{c}{Hamiltonian} & \multicolumn{3}{c}{Logical circuit} & \multicolumn{4}{c}{Error correction} \\
    \cmidrule(lr){2-4} \cmidrule(lr){5-7} \cmidrule(lr){8-11}
    Molecule ID & $L_D^*$ & $\lambda_R$ & $\lambda$ & \makecell{$N_{L}$} & \makecell{$G_M$} & $G_{\rm total}$ & SMM priority & \makecell{$d$} & \makecell{$\bar{\theta}_L$} & \makecell{$P_{\rm total}$} \\
    \midrule
    \multicolumn{11}{c}{(a) $\xi=0.1$}\\
    \midrule
P450-inhibited (A) & 32 & 0.21 & 1.02 & 11 & $7.51\times 10^{2}$ & $7.21\times 10^{5}$ & speed & 15 & $5.20\times 10^{-2}$ & $6.48\times 10^{-2}$ \\
P450-inhibited (B) & 70 & 1.17 & 10.18 & 19 & $1.98\times 10^{4}$ & $2.42\times 10^{7}$ & speed & 17 & $1.98\times 10^{-2}$ & $6.34\times 10^{-1}$ \\
P450-inhibited (C) & 303 & 3.71 & 21.28 & 29 & $1.90\times 10^{5}$ & $2.16\times 10^{8}$ & accuracy & 21 & $4.29\times 10^{-3}$ & $3.49\times 10^{-2}$ \\
P450-inhibited (D) & 1256 & 15.83 & 48.07 & 43 & $2.53\times 10^{6}$ & $2.38\times 10^{9}$ & accuracy & 23 & $7.27\times 10^{-4}$ & $9.56\times 10^{-2}$ \\
P450-inhibited (E) & 2612 & 42.10 & 109.39 & 65 & $1.56\times 10^{7}$ & $1.28\times 10^{10}$ & accuracy & 25 & $2.68\times 10^{-4}$ & $2.39\times 10^{-1}$ \\
P450-inhibited (F) & 3085 & 45.22 & 129.15 & 89 & $1.93\times 10^{7}$ & $2.39\times 10^{10}$ & accuracy & 25 & $2.56\times 10^{-4}$ & $2.87\times 10^{-1}$ \\
P450-inhibited (G) & 3307 & 49.63 & 136.88 & 93 & $2.28\times 10^{7}$ & $2.65\times 10^{10}$ & accuracy & 25 & $2.30\times 10^{-4}$ & $3.06\times 10^{-1}$ \\
P450-inhibited (X) & 6860 & 98.87 & 261.40 & 121 & $9.13\times 10^{7}$ & $1.07\times 10^{11}$ & accuracy & 27 & $1.10\times 10^{-4}$ & $6.20\times 10^{-1}$ \\
P450-empty (A) & 25 & 0.20 & 1.58 & 11 & $8.37\times 10^{2}$ & $9.50\times 10^{5}$ & speed & 15 & $7.31\times 10^{-2}$ & $1.02\times 10^{-1}$ \\
P450-empty (B) & 101 & 0.89 & 7.08 & 17 & $1.61\times 10^{4}$ & $1.72\times 10^{7}$ & speed & 17 & $1.68\times 10^{-2}$ & $4.48\times 10^{-1}$ \\
P450-empty (C) & 166 & 2.08 & 14.30 & 23 & $6.50\times 10^{4}$ & $6.33\times 10^{7}$ & accuracy & 21 & $8.44\times 10^{-3}$ & $2.21\times 10^{-2}$ \\
P450-empty (D) & 915 & 11.60 & 36.15 & 37 & $1.36\times 10^{6}$ & $1.60\times 10^{9}$ & accuracy & 23 & $1.02\times 10^{-3}$ & $6.95\times 10^{-2}$ \\
P450-empty (E) & 2147 & 25.20 & 74.83 & 53 & $6.61\times 10^{6}$ & $7.45\times 10^{9}$ & accuracy & 25 & $4.34\times 10^{-4}$ & $1.60\times 10^{-1}$ \\
P450-empty (F) & 2312 & 34.52 & 106.47 & 75 & $1.15\times 10^{7}$ & $9.89\times 10^{9}$ & accuracy & 25 & $3.55\times 10^{-4}$ & $2.28\times 10^{-1}$ \\
P450-empty (G) & 4129 & 44.20 & 135.12 & 79 & $2.17\times 10^{7}$ & $2.64\times 10^{10}$ & accuracy & 25 & $2.38\times 10^{-4}$ & $3.08\times 10^{-1}$ \\
P450-empty (X) & 5338 & 90.53 & 247.71 & 111 & $7.31\times 10^{7}$ & $9.03\times 10^{10}$ & accuracy & 27 & $1.30\times 10^{-4}$ & $5.82\times 10^{-1}$ \\
    \midrule
    \multicolumn{11}{c}{(b) $\xi=0.05$}\\
    \midrule
P450-inhibited (A) & 26 & 0.28 & 0.94 & 11 & $2.96\times 10^{2}$ & $7.42\times 10^{5}$ & speed & 13 & $6.17\times 10^{-2}$ & $3.03\times 10^{-2}$ \\
P450-inhibited (B) & 48 & 1.54 & 9.82 & 19 & $7.41\times 10^{3}$ & $2.05\times 10^{7}$ & speed & 17 & $2.56\times 10^{-2}$ & $3.15\times 10^{-1}$ \\
P450-inhibited (C) & 214 & 4.35 & 20.52 & 29 & $6.51\times 10^{4}$ & $1.76\times 10^{8}$ & speed & 19 & $6.05\times 10^{-3}$ & $4.79\times 10^{-1}$ \\
P450-inhibited (D) & 770 & 18.26 & 47.26 & 43 & $8.08\times 10^{5}$ & $2.02\times 10^{9}$ & accuracy & 23 & $1.12\times 10^{-3}$ & $4.53\times 10^{-2}$ \\
P450-inhibited (E) & 1698 & 44.86 & 105.44 & 65 & $4.58\times 10^{6}$ & $1.11\times 10^{10}$ & accuracy & 25 & $4.40\times 10^{-4}$ & $1.15\times 10^{-1}$ \\
P450-inhibited (F) & 1930 & 51.06 & 126.01 & 89 & $6.05\times 10^{6}$ & $1.68\times 10^{10}$ & accuracy & 25 & $3.98\times 10^{-4}$ & $1.39\times 10^{-1}$ \\
P450-inhibited (G) & 2138 & 56.99 & 137.71 & 93 & $7.48\times 10^{6}$ & $2.01\times 10^{10}$ & accuracy & 25 & $3.53\times 10^{-4}$ & $1.54\times 10^{-1}$ \\
P450-inhibited (X) & 3749 & 113.76 & 260.04 & 121 & $2.82\times 10^{7}$ & $7.61\times 10^{10}$ & accuracy & 25 & $1.76\times 10^{-4}$ & $3.07\times 10^{-1}$ \\
P450-empty (A) & 19 & 0.27 & 1.62 & 11 & $3.70\times 10^{2}$ & $9.60\times 10^{5}$ & speed & 13 & $8.95\times 10^{-2}$ & $5.50\times 10^{-2}$ \\
P450-empty (B) & 60 & 1.05 & 10.14 & 17 & $6.60\times 10^{3}$ & $1.90\times 10^{7}$ & speed & 17 & $2.94\times 10^{-2}$ & $3.23\times 10^{-1}$ \\
P450-empty (C) & 130 & 2.46 & 13.05 & 23 & $2.29\times 10^{4}$ & $5.92\times 10^{7}$ & speed & 17 & $1.09\times 10^{-2}$ & $3.90\times 10^{-1}$ \\
P450-empty (D) & 478 & 14.16 & 36.94 & 37 & $4.48\times 10^{5}$ & $1.20\times 10^{9}$ & accuracy & 23 & $1.58\times 10^{-3}$ & $3.39\times 10^{-2}$ \\
P450-empty (E) & 1427 & 29.33 & 74.87 & 53 & $2.22\times 10^{6}$ & $5.90\times 10^{9}$ & accuracy & 23 & $6.46\times 10^{-4}$ & $7.92\times 10^{-2}$ \\
P450-empty (F) & 1534 & 38.12 & 103.98 & 75 & $3.58\times 10^{6}$ & $8.80\times 10^{9}$ & accuracy & 25 & $5.56\times 10^{-4}$ & $1.10\times 10^{-1}$ \\
P450-empty (G) & 2478 & 51.76 & 129.99 & 79 & $6.89\times 10^{6}$ & $1.90\times 10^{10}$ & accuracy & 25 & $3.61\times 10^{-4}$ & $1.47\times 10^{-1}$ \\
    \midrule
    \multicolumn{11}{c}{(c) $\xi=0.01$}\\
    \midrule
P450-inhibited (A) & 13 & 0.42 & 0.90 & 11 & $5.03\times 10^{1}$ & $2.74\times 10^{6}$ & speed & 11 & $1.04\times 10^{-1}$ & $8.68\times 10^{-3}$ \\
P450-inhibited (B) & 22 & 2.37 & 9.54 & 19 & $7.02\times 10^{2}$ & $3.83\times 10^{7}$ & speed & 15 & $5.35\times 10^{-2}$ & $6.23\times 10^{-2}$ \\
P450-inhibited (C) & 86 & 7.54 & 19.21 & 29 & $6.30\times 10^{3}$ & $3.47\times 10^{8}$ & speed & 17 & $1.19\times 10^{-2}$ & $1.24\times 10^{-1}$ \\
P450-inhibited (D) & 220 & 25.13 & 44.75 & 43 & $5.50\times 10^{4}$ & $3.02\times 10^{9}$ & speed & 19 & $3.14\times 10^{-3}$ & $1.20\times 10^{-1}$ \\
P450-inhibited (E) & 547 & 57.76 & 97.77 & 65 & $2.94\times 10^{5}$ & $1.61\times 10^{10}$ & speed & 21 & $1.28\times 10^{-3}$ & $1.12\times 10^{-1}$ \\
P450-inhibited (F) & 700 & 68.85 & 123.61 & 89 & $4.37\times 10^{5}$ & $2.39\times 10^{10}$ & speed & 21 & $1.08\times 10^{-3}$ & $1.25\times 10^{-1}$ \\
P450-inhibited (G) & 757 & 75.23 & 134.12 & 93 & $5.22\times 10^{5}$ & $2.85\times 10^{10}$ & speed & 21 & $9.87\times 10^{-4}$ & $1.22\times 10^{-1}$ \\
P450-inhibited (X) & 1283 & 144.82 & 253.32 & 121 & $1.85\times 10^{6}$ & $1.01\times 10^{11}$ & speed & 23 & $5.25\times 10^{-4}$ & $1.20\times 10^{-1}$ \\
P450-empty (A) & 8 & 0.36 & 1.44 & 11 & $3.83\times 10^{1}$ & $2.10\times 10^{6}$ & speed & 11 & $2.00\times 10^{-1}$ & $1.27\times 10^{-2}$ \\
P450-empty (B) & 36 & 1.60 & 6.09 & 17 & $5.19\times 10^{2}$ & $2.85\times 10^{7}$ & speed & 15 & $4.61\times 10^{-2}$ & $3.98\times 10^{-2}$ \\
P450-empty (C) & 55 & 4.26 & 13.58 & 23 & $2.37\times 10^{3}$ & $1.30\times 10^{8}$ & speed & 15 & $2.24\times 10^{-2}$ & $8.82\times 10^{-2}$ \\
P450-empty (D) & 189 & 18.15 & 34.90 & 37 & $3.12\times 10^{4}$ & $1.70\times 10^{9}$ & speed & 19 & $4.32\times 10^{-3}$ & $1.22\times 10^{-1}$ \\
P450-empty (E) & 451 & 43.16 & 73.53 & 53 & $1.69\times 10^{5}$ & $9.30\times 10^{9}$ & speed & 21 & $1.67\times 10^{-3}$ & $1.06\times 10^{-1}$ \\
P450-empty (F) & 589 & 52.56 & 98.79 & 75 & $2.69\times 10^{5}$ & $1.47\times 10^{10}$ & speed & 21 & $1.41\times 10^{-3}$ & $1.15\times 10^{-1}$ \\
P450-empty (G) & 777 & 75.50 & 127.92 & 79 & $5.20\times 10^{5}$ & $2.84\times 10^{10}$ & speed & 21 & $9.45\times 10^{-4}$ & $1.08\times 10^{-1}$ \\
P450-empty (X) & 1225 & 132.07 & 237.01 & 111 & $1.58\times 10^{6}$ & $8.61\times 10^{10}$ & speed & 23 & $5.77\times 10^{-4}$ & $1.11\times 10^{-1}$ \\
    \end{tabular}
\end{ruledtabular}
\end{table*}
\begin{table*}[htbp]
\caption{
    Logical resource requirements for the ruthenium-based catalyst models, listed for structures I, II, II–III, and V (first half of the catalytic cycle~\cite{Von_Burg2021-du}), across three initial-state infidelity parameter values: (a) $\xi=0.1$, (b) $\xi=0.05$, and (c) $\xi=0.01$.
    Columns are grouped into ``Hamiltonian" ($L_D^*$, $\lambda_R$, $\lambda$), ``Logical circuit" ($N_L$, $G_M$, $G_{\rm total}$), and ``Error correction" (SMM priority, $d$, $\bar{\theta}_L$, $P_{\rm total}$).
    Detailed definitions for these parameters and the underlying assumptions can be found in Appendix~\ref{append:logical_resource}.
}
\label{tab:co2-cat_1_logical}
\begin{ruledtabular}
\begin{tabular}{l c c c c c c c c c c}
    & \multicolumn{3}{c}{Hamiltonian} & \multicolumn{3}{c}{Logical circuit} & \multicolumn{4}{c}{Error correction} \\
    \cmidrule(lr){2-4} \cmidrule(lr){5-7} \cmidrule(lr){8-11}
    Molecule ID & $L_D^*$ & $\lambda_R$ & $\lambda$ & \makecell{$N_{L}$} & \makecell{$G_M$} & $G_{\rm total}$ & SMM priority & \makecell{$d$} & \makecell{$\bar{\theta}_L$} & \makecell{$P_{\rm total}$} \\
    \midrule
    \multicolumn{11}{c}{(a) $\xi=0.1$}\\
    \midrule
Ru-I (Sm) & 61 & 0.67 & 4.59 & 11 & $7.14\times 10^{3}$ & $1.01\times 10^{7}$ & speed & 17 & $2.48\times 10^{-2}$ & $2.93\times 10^{-1}$ \\
Ru-I (Md) & 576 & 8.88 & 26.85 & 33 & $7.29\times 10^{5}$ & $6.41\times 10^{8}$ & accuracy & 23 & $1.41\times 10^{-3}$ & $5.00\times 10^{-2}$ \\
Ru-I (Lg) & 8151 & 192.63 & 270.04 & 105 & $2.59\times 10^{8}$ & $2.73\times 10^{11}$ & accuracy & 27 & $3.99\times 10^{-5}$ & $6.73\times 10^{-1}$ \\
Ru-II (Sm) & 61 & 0.76 & 5.49 & 13 & $8.80\times 10^{3}$ & $1.08\times 10^{7}$ & speed & 17 & $2.41\times 10^{-2}$ & $3.49\times 10^{-1}$ \\
Ru-II (Md) & 1354 & 26.36 & 77.56 & 53 & $5.96\times 10^{6}$ & $6.23\times 10^{9}$ & accuracy & 25 & $4.98\times 10^{-4}$ & $1.57\times 10^{-1}$ \\
Ru-II (Lg) & 11060 & 319.91 & 454.93 & 125 & $6.98\times 10^{8}$ & $8.42\times 10^{11}$ & accuracy & 29 & $2.50\times 10^{-5}$ & $1.14\times 10^{0}$ \\
Ru-II-III (Sm) & 71 & 0.67 & 5.92 & 13 & $9.41\times 10^{3}$ & $1.14\times 10^{7}$ & speed & 17 & $2.42\times 10^{-2}$ & $3.78\times 10^{-1}$ \\
Ru-II-III (Md) & 2273 & 42.90 & 108.56 & 59 & $1.53\times 10^{7}$ & $1.24\times 10^{10}$ & accuracy & 25 & $2.72\times 10^{-4}$ & $2.33\times 10^{-1}$ \\
Ru-II-III (Lg) & 13041 & 381.19 & 538.84 & 131 & $9.89\times 10^{8}$ & $1.01\times 10^{12}$ & accuracy & 29 & $2.08\times 10^{-5}$ & $1.36\times 10^{0}$ \\
Ru-V (Sm) & 460 & 5.03 & 24.35 & 23 & $3.40\times 10^{5}$ & $3.44\times 10^{8}$ & accuracy & 23 & $2.75\times 10^{-3}$ & $4.11\times 10^{-2}$ \\
Ru-V (Md) & 2253 & 26.98 & 85.83 & 49 & $7.76\times 10^{6}$ & $7.97\times 10^{9}$ & accuracy & 25 & $4.24\times 10^{-4}$ & $1.87\times 10^{-1}$ \\
Ru-V (Lg) & 9721 & 234.41 & 383.24 & 121 & $3.94\times 10^{8}$ & $3.17\times 10^{11}$ & accuracy & 27 & $3.72\times 10^{-5}$ & $9.46\times 10^{-1}$ \\
    \midrule
    \multicolumn{11}{c}{(b) $\xi=0.05$}\\
    \midrule
Ru-I (Sm) & 50 & 0.85 & 4.81 & 11 & $3.05\times 10^{3}$ & $8.74\times 10^{6}$ & speed & 15 & $3.09\times 10^{-2}$ & $1.57\times 10^{-1}$ \\
Ru-I (Md) & 376 & 9.98 & 26.88 & 33 & $2.33\times 10^{5}$ & $5.72\times 10^{8}$ & speed & 21 & $2.21\times 10^{-3}$ & $4.14\times 10^{-1}$ \\
Ru-I (Lg) & 3888 & 203.75 & 268.25 & 105 & $7.08\times 10^{7}$ & $1.84\times 10^{11}$ & accuracy & 27 & $7.26\times 10^{-5}$ & $3.32\times 10^{-1}$ \\
Ru-II (Sm) & 55 & 0.99 & 5.50 & 13 & $3.88\times 10^{3}$ & $1.08\times 10^{7}$ & speed & 15 & $2.73\times 10^{-2}$ & $1.76\times 10^{-1}$ \\
Ru-II (Md) & 725 & 29.88 & 78.24 & 53 & $1.82\times 10^{6}$ & $4.70\times 10^{9}$ & accuracy & 23 & $8.24\times 10^{-4}$ & $7.76\times 10^{-2}$ \\
Ru-II (Lg) & 5389 & 332.24 & 453.13 & 125 & $1.86\times 10^{8}$ & $5.07\times 10^{11}$ & accuracy & 27 & $4.67\times 10^{-5}$ & $5.56\times 10^{-1}$ \\
Ru-II-III (Sm) & 46 & 0.97 & 5.83 & 13 & $3.56\times 10^{3}$ & $9.66\times 10^{6}$ & speed & 15 & $3.17\times 10^{-2}$ & $1.87\times 10^{-1}$ \\
Ru-II-III (Md) & 1170 & 47.30 & 106.71 & 59 & $4.42\times 10^{6}$ & $1.06\times 10^{10}$ & accuracy & 25 & $4.62\times 10^{-4}$ & $1.13\times 10^{-1}$ \\
Ru-II-III (Lg) & 6106 & 402.36 & 537.80 & 131 & $2.69\times 10^{8}$ & $6.92\times 10^{11}$ & accuracy & 27 & $3.82\times 10^{-5}$ & $6.64\times 10^{-1}$ \\
Ru-V (Sm) & 229 & 5.43 & 22.78 & 23 & $8.79\times 10^{4}$ & $2.29\times 10^{8}$ & speed & 19 & $4.97\times 10^{-3}$ & $5.06\times 10^{-1}$ \\
Ru-V (Md) & 1291 & 31.12 & 87.14 & 49 & $2.44\times 10^{6}$ & $6.23\times 10^{9}$ & accuracy & 23 & $6.85\times 10^{-4}$ & $9.08\times 10^{-2}$ \\
Ru-V (Lg) & 5167 & 257.74 & 391.86 & 121 & $1.17\times 10^{8}$ & $2.78\times 10^{11}$ & accuracy & 27 & $6.41\times 10^{-5}$ & $4.80\times 10^{-1}$ \\
    \midrule
    \multicolumn{11}{c}{(c) $\xi=0.01$}\\
    \midrule
Ru-I (Sm) & 26 & 1.42 & 4.43 & 11 & $3.01\times 10^{2}$ & $1.65\times 10^{7}$ & speed & 13 & $5.66\times 10^{-2}$ & $2.82\times 10^{-2}$ \\
Ru-I (Md) & 127 & 13.31 & 25.61 & 33 & $1.61\times 10^{4}$ & $8.82\times 10^{8}$ & speed & 17 & $6.13\times 10^{-3}$ & $1.28\times 10^{-1}$ \\
Ru-I (Lg) & 614 & 226.53 & 264.02 & 105 & $3.32\times 10^{6}$ & $1.81\times 10^{11}$ & speed & 23 & $3.05\times 10^{-4}$ & $8.58\times 10^{-2}$ \\
Ru-II (Sm) & 38 & 1.35 & 5.69 & 13 & $5.10\times 10^{2}$ & $2.80\times 10^{7}$ & speed & 13 & $4.71\times 10^{-2}$ & $3.99\times 10^{-2}$ \\
Ru-II (Md) & 243 & 36.67 & 74.27 & 53 & $1.12\times 10^{5}$ & $6.18\times 10^{9}$ & speed & 21 & $2.55\times 10^{-3}$ & $2.00\times 10^{-1}$ \\
Ru-II (Lg) & 827 & 367.03 & 443.02 & 125 & $8.62\times 10^{6}$ & $4.71\times 10^{11}$ & accuracy & 25 & $1.97\times 10^{-4}$ & $1.07\times 10^{-1}$ \\
Ru-II-III (Sm) & 32 & 1.33 & 5.61 & 13 & $3.94\times 10^{2}$ & $2.17\times 10^{7}$ & speed & 13 & $5.50\times 10^{-2}$ & $3.59\times 10^{-2}$ \\
Ru-II-III (Md) & 329 & 57.81 & 101.61 & 59 & $2.57\times 10^{5}$ & $1.41\times 10^{10}$ & speed & 21 & $1.51\times 10^{-3}$ & $1.98\times 10^{-1}$ \\
Ru-II-III (Lg) & 979 & 436.98 & 525.08 & 131 & $1.22\times 10^{7}$ & $6.71\times 10^{11}$ & accuracy & 25 & $1.65\times 10^{-4}$ & $1.28\times 10^{-1}$ \\
Ru-V (Sm) & 103 & 7.95 & 21.83 & 23 & $7.52\times 10^{3}$ & $4.13\times 10^{8}$ & speed & 17 & $1.11\times 10^{-2}$ & $1.37\times 10^{-1}$ \\
Ru-V (Md) & 470 & 43.90 & 88.12 & 49 & $1.89\times 10^{5}$ & $1.04\times 10^{10}$ & speed & 21 & $1.79\times 10^{-3}$ & $1.34\times 10^{-1}$ \\
Ru-V (Lg) & 1125 & 287.24 & 382.91 & 121 & $5.68\times 10^{6}$ & $3.11\times 10^{11}$ & speed & 25 & $2.58\times 10^{-4}$ & $1.30\times 10^{-1}$ \\
    \end{tabular}
\end{ruledtabular}
\end{table*}
\begin{table*}[htbp]
\caption{
    Logical resource requirements for the ruthenium-based catalyst models, listed for structures VIII, VIII-IX, IX, and XVIII (second half of the catalytic cycle~\cite{Von_Burg2021-du}), across three initial-state infidelity parameter values: (a) $\xi=0.1$, (b) $\xi=0.05$, and (c) $\xi=0.01$.
    Columns are grouped into ``Hamiltonian" ($L_D^*$, $\lambda_R$, $\lambda$), ``Logical circuit" ($N_L$, $G_M$, $G_{\rm total}$), and ``Error correction" (SMM priority, $d$, $\bar{\theta}_L$, $P_{\rm total}$).
    Detailed definitions for these parameters and the underlying assumptions can be found in Appendix~\ref{append:logical_resource}.
}
\label{tab:co2-cat_2_logical}
\begin{ruledtabular}
\begin{tabular}{l c c c c c c c c c c}
    & \multicolumn{3}{c}{Hamiltonian} & \multicolumn{3}{c}{Logical circuit} & \multicolumn{4}{c}{Error correction} \\
    \cmidrule(lr){2-4} \cmidrule(lr){5-7} \cmidrule(lr){8-11}
    Molecule ID & $L_D^*$ & $\lambda_R$ & $\lambda$ & \makecell{$N_{L}$} & \makecell{$G_M$} & $G_{\rm total}$ & SMM priority & \makecell{$d$} & \makecell{$\bar{\theta}_L$} & \makecell{$P_{\rm total}$} \\
    \midrule
    \multicolumn{11}{c}{(a) $\xi=0.1$}\\
    \midrule
Ru-VIII (Sm) & 7 & 0.02 & 0.88 & 5 & $9.43\times 10^{1}$ & $1.21\times 10^{5}$ & speed & 11 & $3.61\times 10^{-1}$ & $5.65\times 10^{-2}$ \\
Ru-VIII (Md) & 2374 & 42.89 & 104.81 & 59 & $1.53\times 10^{7}$ & $1.27\times 10^{10}$ & accuracy & 25 & $2.62\times 10^{-4}$ & $2.29\times 10^{-1}$ \\
Ru-VIII (Lg) & 13562 & 410.92 & 573.09 & 131 & $1.14\times 10^{9}$ & $1.10\times 10^{12}$ & accuracy & 29 & $1.92\times 10^{-5}$ & $1.45\times 10^{0}$ \\
Ru-VIII-IX (Sm) & 38 & 0.24 & 2.77 & 9 & $1.98\times 10^{3}$ & $2.57\times 10^{6}$ & speed & 15 & $5.40\times 10^{-2}$ & $1.77\times 10^{-1}$ \\
Ru-VIII-IX (Md) & 1911 & 27.11 & 64.70 & 47 & $6.52\times 10^{6}$ & $8.11\times 10^{9}$ & accuracy & 25 & $3.80\times 10^{-4}$ & $1.42\times 10^{-1}$ \\
Ru-VIII-IX (Lg) & 10865 & 287.02 & 421.44 & 119 & $5.72\times 10^{8}$ & $4.31\times 10^{11}$ & accuracy & 29 & $2.82\times 10^{-5}$ & $1.05\times 10^{0}$ \\
Ru-IX (Sm) & 872 & 12.16 & 45.01 & 33 & $1.56\times 10^{6}$ & $1.57\times 10^{9}$ & accuracy & 23 & $1.11\times 10^{-3}$ & $8.58\times 10^{-2}$ \\
Ru-IX (Md) & 2284 & 35.13 & 103.21 & 53 & $1.15\times 10^{7}$ & $1.00\times 10^{10}$ & accuracy & 25 & $3.43\times 10^{-4}$ & $2.19\times 10^{-1}$ \\
Ru-IX (Lg) & 10766 & 246.59 & 407.31 & 125 & $4.41\times 10^{8}$ & $3.44\times 10^{11}$ & accuracy & 27 & $3.53\times 10^{-5}$ & $1.01\times 10^{0}$ \\
Ru-XVIII (Sm) & 30 & 0.30 & 3.50 & 9 & $2.18\times 10^{3}$ & $2.41\times 10^{6}$ & speed & 15 & $6.21\times 10^{-2}$ & $2.24\times 10^{-1}$ \\
Ru-XVIII (Md) & 961 & 15.83 & 42.57 & 41 & $2.19\times 10^{6}$ & $2.18\times 10^{9}$ & accuracy & 23 & $7.46\times 10^{-4}$ & $8.57\times 10^{-2}$ \\
Ru-XVIII (Lg) & 9280 & 235.84 & 343.63 & 113 & $3.87\times 10^{8}$ & $3.34\times 10^{11}$ & accuracy & 27 & $3.40\times 10^{-5}$ & $8.53\times 10^{-1}$ \\
    \midrule
    \multicolumn{11}{c}{(b) $\xi=0.05$}\\
    \midrule
Ru-VIII (Sm) & 7 & 0.02 & 0.87 & 5 & $5.00\times 10^{1}$ & $1.33\times 10^{5}$ & speed & 11 & $3.66\times 10^{-1}$ & $3.04\times 10^{-2}$ \\
Ru-VIII (Md) & 1245 & 47.94 & 105.81 & 59 & $4.57\times 10^{6}$ & $1.10\times 10^{10}$ & accuracy & 25 & $4.43\times 10^{-4}$ & $1.13\times 10^{-1}$ \\
Ru-VIII (Lg) & 6499 & 428.84 & 574.56 & 131 & $3.06\times 10^{8}$ & $7.73\times 10^{11}$ & accuracy & 27 & $3.59\times 10^{-5}$ & $7.13\times 10^{-1}$ \\
Ru-VIII-IX (Sm) & 23 & 0.35 & 2.77 & 9 & $6.91\times 10^{2}$ & $1.94\times 10^{6}$ & speed & 15 & $7.91\times 10^{-2}$ & $9.07\times 10^{-2}$ \\
Ru-VIII-IX (Md) & 1083 & 30.41 & 66.71 & 47 & $2.00\times 10^{6}$ & $5.45\times 10^{9}$ & accuracy & 23 & $6.38\times 10^{-4}$ & $7.09\times 10^{-2}$ \\
Ru-VIII-IX (Lg) & 5246 & 307.01 & 423.25 & 119 & $1.60\times 10^{8}$ & $4.50\times 10^{11}$ & accuracy & 27 & $5.07\times 10^{-5}$ & $5.18\times 10^{-1}$ \\
Ru-IX (Sm) & 538 & 13.77 & 43.19 & 33 & $4.82\times 10^{5}$ & $1.24\times 10^{9}$ & accuracy & 23 & $1.72\times 10^{-3}$ & $3.92\times 10^{-2}$ \\
Ru-IX (Md) & 1709 & 39.33 & 103.12 & 53 & $3.88\times 10^{6}$ & $9.55\times 10^{9}$ & accuracy & 25 & $5.10\times 10^{-4}$ & $1.12\times 10^{-1}$ \\
Ru-IX (Lg) & 5462 & 265.68 & 406.07 & 125 & $1.25\times 10^{8}$ & $2.96\times 10^{11}$ & accuracy & 27 & $6.22\times 10^{-5}$ & $5.00\times 10^{-1}$ \\
Ru-XVIII (Sm) & 25 & 0.38 & 3.28 & 9 & $8.76\times 10^{2}$ & $2.32\times 10^{6}$ & speed & 15 & $7.32\times 10^{-2}$ & $1.06\times 10^{-1}$ \\
Ru-XVIII (Md) & 539 & 17.87 & 40.96 & 41 & $6.63\times 10^{5}$ & $1.70\times 10^{9}$ & accuracy & 23 & $1.19\times 10^{-3}$ & $3.98\times 10^{-2}$ \\
Ru-XVIII (Lg) & 4147 & 249.26 & 337.38 & 113 & $1.04\times 10^{8}$ & $2.55\times 10^{11}$ & accuracy & 27 & $6.18\times 10^{-5}$ & $4.13\times 10^{-1}$ \\
    \midrule
    \multicolumn{11}{c}{(c) $\xi=0.01$}\\
    \midrule
Ru-VIII (Sm) & 6 & 0.07 & 0.84 & 5 & $1.28\times 10^{1}$ & $6.99\times 10^{5}$ & speed & 9 & $4.06\times 10^{-1}$ & $8.65\times 10^{-3}$ \\
Ru-VIII (Md) & 324 & 59.28 & 102.57 & 59 & $2.67\times 10^{5}$ & $1.46\times 10^{10}$ & speed & 21 & $1.47\times 10^{-3}$ & $2.06\times 10^{-1}$ \\
Ru-VIII (Lg) & 999 & 468.19 & 555.35 & 131 & $1.39\times 10^{7}$ & $7.65\times 10^{11}$ & accuracy & 25 & $1.53\times 10^{-4}$ & $1.35\times 10^{-1}$ \\
Ru-VIII-IX (Sm) & 16 & 0.68 & 2.82 & 9 & $1.16\times 10^{2}$ & $6.35\times 10^{6}$ & speed & 13 & $1.07\times 10^{-1}$ & $2.06\times 10^{-2}$ \\
Ru-VIII-IX (Md) & 223 & 41.03 & 66.78 & 47 & $1.27\times 10^{5}$ & $6.93\times 10^{9}$ & speed & 21 & $2.03\times 10^{-3}$ & $1.41\times 10^{-1}$ \\
Ru-VIII-IX (Lg) & 955 & 336.57 & 414.35 & 119 & $7.41\times 10^{6}$ & $4.05\times 10^{11}$ & accuracy & 25 & $2.14\times 10^{-4}$ & $1.00\times 10^{-1}$ \\
Ru-IX (Sm) & 215 & 18.84 & 41.56 & 33 & $3.68\times 10^{4}$ & $2.02\times 10^{9}$ & speed & 19 & $4.36\times 10^{-3}$ & $1.43\times 10^{-1}$ \\
Ru-IX (Md) & 439 & 53.35 & 99.07 & 53 & $2.46\times 10^{5}$ & $1.35\times 10^{10}$ & speed & 21 & $1.54\times 10^{-3}$ & $1.67\times 10^{-1}$ \\
Ru-IX (Lg) & 1361 & 298.91 & 397.15 & 125 & $6.28\times 10^{6}$ & $3.43\times 10^{11}$ & speed & 25 & $2.42\times 10^{-4}$ & $1.24\times 10^{-1}$ \\
Ru-XVIII (Sm) & 16 & 0.68 & 3.31 & 9 & $1.33\times 10^{2}$ & $7.26\times 10^{6}$ & speed & 13 & $1.11\times 10^{-1}$ & $2.45\times 10^{-2}$ \\
Ru-XVIII (Md) & 159 & 22.79 & 39.39 & 41 & $4.18\times 10^{4}$ & $2.30\times 10^{9}$ & speed & 19 & $3.63\times 10^{-3}$ & $1.14\times 10^{-1}$ \\
Ru-XVIII (Lg) & 694 & 275.20 & 338.47 & 113 & $4.90\times 10^{6}$ & $2.68\times 10^{11}$ & speed & 25 & $2.65\times 10^{-4}$ & $1.23\times 10^{-1}$ \\
    \end{tabular}
\end{ruledtabular}
\end{table*}

\subsubsection{Physical resource estimates \label{append:physical_resource}}
Physical resource estimates, derived from the logical resource estimates, are summarized in Tables~\ref{tab:h_chain_physical} (hydrogen chain models), \ref{tab:fe-s_physical} (iron-sulfur clusters), \ref{tab:p450_1_physical}, \ref{tab:p450_2_physical} (cytochrome P450 active-site models), and \ref{tab:co2-cat_1_physical}, \ref{tab:co2-cat_2_physical} (ruthenium-based catalyst series). For consistency, each table presents columns for:
\begin{itemize}
    \item \textbf{Physical qubits per QPU}: Required physical qubit count per QPU determined as $\mathcal{Q}_{\rm QPU}=N_{\rm patch}\times 2d^2$, where $N_{\rm patch}=2N_L+\sqrt{8N_L}+11$ is the number of surface code patches under the fast block layout~\cite{Litinski2019-qf,Toshio2026}.
    \item \textbf{Maximum per-shot runtime}: Execution time for the deepest circuit $\mathcal{T}_{\rm max}=\mathcal{T}_M$ as defined in Eq.~\eqref{eq:circuit_runtime}.
    \item \textbf{Time-to-solution (single QPU)}: Total runtime $\mathcal{T}_{\rm total}$ as defined in Eq.~\eqref{eq:total_runtime}.
    \item \textbf{Maximum coherent spacetime volume}: Maximum spacetime footprint of a single uninterrupted circuit execution $\mathcal{V}_{\rm coh}^{\rm max}$ defined in Eq.~\eqref{eq:volumes}.
    \item \textbf{Total spacetime volume}: Total spacetime cost of the computation $\mathcal{V}_{\rm total}$ defined in Eq.~\eqref{eq:volumes}.
    \item \textbf{QPU parallelism}: Achievable QPU parallelism $k^\ast = \lfloor \mathcal{Q}_{\rm budget}/\mathcal{Q}_{\rm QPU} \rfloor$ under a fixed physical qubit budget of $\mathcal{Q}_{\rm budget}=5 \times 10^5$.
    \item \textbf{Multi-QPU time-to-solution}: Total runtime under parallel execution using $k^\ast$ QPUs.
\end{itemize}
Selected results are shown in Table~\ref{tab:resource_summary} in the main text.

\begin{table*}[htbp]
\caption{
    Physical resource estimates for the hydrogen chain models across three initial-state infidelity parameter values: (a) $\xi=0.1$, (b) $\xi=0.05$, and (c) $\xi=0.01$. 
    Columns include physical qubits per QPU ($\mathcal{Q}_{\rm QPU}$), maximum per-shot runtime ($\mathcal{T}_{\rm max}$), single-QPU time-to-solution ($\mathcal{T}_{\rm total}$), maximum coherent spacetime volume ($\mathcal{V}_{\rm coh}^{\rm max}$), total  spacetime volume ($\mathcal{V}_{\rm total}$), QPU parallelism ($k^\ast$) under $\mathcal{Q}_{\rm budget}=5\times10^5$, and multi-QPU time-to-solution using $k^\ast$ QPUs.
    Detailed definitions for these parameters and the underlying assumptions can be found in Appendix~\ref{append:physical_resource}.
}
\label{tab:h_chain_physical}
\begin{ruledtabular}
\begin{tabular}{l c c c c c c c}
    Molecule ID & $\mathcal{Q}_{\rm QPU}$ & \makecell{$\mathcal{T}_{\rm max}$ [\si{\second}]} & \makecell{$\mathcal{T}_{\rm total}$ [days]} & \makecell{$\mathcal{V}_{\rm coh}^{\rm max}$\\$[$qubit-hour$]$} & \makecell{$\mathcal{V}_{\rm total}$\\$[$qubit-hour$]$} &  \makecell{$k^*$} & \makecell{Time-to-solution\\($k^\ast$ QPUs) [days]} \\
    \midrule
    \multicolumn{8}{c}{(a) $\xi=0.1$}\\
    \midrule
\ce{H10} & $6.98\times 10^{4}$ & $1.75\times 10^{2}$ & $1.93\times 10^{0}$ & $3.39\times 10^{3}$ & $3.23\times 10^{6}$ & 10 & $2.57\times 10^{-1}$ \\
\ce{H20} & $1.18\times 10^{5}$ & $7.89\times 10^{2}$ & $1.68\times 10^{1}$ & $2.58\times 10^{4}$ & $4.74\times 10^{7}$ & 5 & $4.13\times 10^{0}$ \\
\ce{H30} & $1.94\times 10^{5}$ & $1.74\times 10^{3}$ & $3.67\times 10^{1}$ & $9.36\times 10^{4}$ & $1.71\times 10^{8}$ & 3 & $1.75\times 10^{1}$ \\
\ce{H40} & $2.48\times 10^{5}$ & $2.83\times 10^{3}$ & $1.16\times 10^{2}$ & $1.95\times 10^{5}$ & $6.92\times 10^{8}$ & 2 & $5.81\times 10^{1}$ \\
\ce{H50} & $3.02\times 10^{5}$ & $3.80\times 10^{3}$ & $2.18\times 10^{2}$ & $3.19\times 10^{5}$ & $1.58\times 10^{9}$ & 1 & $2.17\times 10^{2}$ \\
\ce{H60} & $4.14\times 10^{5}$ & $5.15\times 10^{3}$ & $6.46\times 10^{2}$ & $5.93\times 10^{5}$ & $6.43\times 10^{9}$ & 1 & $6.46\times 10^{2}$ \\
\ce{H70} & $4.76\times 10^{5}$ & $6.77\times 10^{3}$ & $1.25\times 10^{3}$ & $8.96\times 10^{5}$ & $1.43\times 10^{10}$ & 1 & $1.25\times 10^{3}$ \\
\ce{H80} & $5.38\times 10^{5}$ & $8.45\times 10^{3}$ & $2.48\times 10^{3}$ & $1.26\times 10^{6}$ & $3.20\times 10^{10}$ & 1 & $2.48\times 10^{3}$ \\
    \midrule
    \multicolumn{8}{c}{(b) $\xi=0.05$}\\
    \midrule
\ce{H10} & $4.76\times 10^{4}$ & $3.27\times 10^{0}$ & $1.88\times 10^{0}$ & $4.33\times 10^{1}$ & $2.15\times 10^{6}$ & 10 & $1.88\times 10^{-1}$ \\
\ce{H20} & $1.18\times 10^{5}$ & $3.52\times 10^{2}$ & $1.30\times 10^{1}$ & $1.15\times 10^{4}$ & $3.67\times 10^{7}$ & 5 & $3.21\times 10^{0}$ \\
\ce{H30} & $1.94\times 10^{5}$ & $8.72\times 10^{2}$ & $3.52\times 10^{1}$ & $4.70\times 10^{4}$ & $1.64\times 10^{8}$ & 3 & $1.72\times 10^{1}$ \\
\ce{H40} & $2.48\times 10^{5}$ & $1.40\times 10^{3}$ & $7.95\times 10^{1}$ & $9.67\times 10^{4}$ & $4.73\times 10^{8}$ & 2 & $3.97\times 10^{1}$ \\
\ce{H50} & $3.02\times 10^{5}$ & $1.78\times 10^{3}$ & $1.26\times 10^{2}$ & $1.49\times 10^{5}$ & $9.10\times 10^{8}$ & 1 & $1.25\times 10^{2}$ \\
\ce{H60} & $3.55\times 10^{5}$ & $2.31\times 10^{3}$ & $2.37\times 10^{2}$ & $2.28\times 10^{5}$ & $2.02\times 10^{9}$ & 1 & $2.37\times 10^{2}$ \\
\ce{H70} & $4.08\times 10^{5}$ & $2.97\times 10^{3}$ & $3.92\times 10^{2}$ & $3.37\times 10^{5}$ & $3.84\times 10^{9}$ & 1 & $3.92\times 10^{2}$ \\
\ce{H80} & $5.38\times 10^{5}$ & $3.95\times 10^{3}$ & $6.71\times 10^{2}$ & $5.91\times 10^{5}$ & $8.67\times 10^{9}$ & 1 & $6.71\times 10^{2}$ \\
    \midrule
    \multicolumn{8}{c}{(c) $\xi=0.01$}\\
    \midrule
\ce{H10} & $3.81\times 10^{4}$ & $3.42\times 10^{-1}$ & $4.29\times 10^{-1}$ & $3.62\times 10^{0}$ & $3.92\times 10^{5}$ & 13 & $3.30\times 10^{-2}$ \\
\ce{H20} & $8.02\times 10^{4}$ & $3.01\times 10^{0}$ & $4.86\times 10^{0}$ & $6.70\times 10^{1}$ & $9.37\times 10^{6}$ & 6 & $8.11\times 10^{-1}$ \\
\ce{H30} & $1.37\times 10^{5}$ & $1.05\times 10^{1}$ & $1.76\times 10^{1}$ & $3.98\times 10^{2}$ & $5.77\times 10^{7}$ & 3 & $5.85\times 10^{0}$ \\
\ce{H40} & $1.75\times 10^{5}$ & $2.34\times 10^{1}$ & $3.97\times 10^{1}$ & $1.14\times 10^{3}$ & $1.67\times 10^{8}$ & 2 & $1.98\times 10^{1}$ \\
\ce{H50} & $2.55\times 10^{5}$ & $4.78\times 10^{1}$ & $8.23\times 10^{1}$ & $3.39\times 10^{3}$ & $5.04\times 10^{8}$ & 1 & $8.21\times 10^{1}$ \\
\ce{H60} & $3.01\times 10^{5}$ & $7.92\times 10^{1}$ & $1.38\times 10^{2}$ & $6.61\times 10^{3}$ & $9.93\times 10^{8}$ & 1 & $1.38\times 10^{2}$ \\
\ce{H70} & $3.46\times 10^{5}$ & $1.23\times 10^{2}$ & $2.19\times 10^{2}$ & $1.18\times 10^{4}$ & $1.81\times 10^{9}$ & 1 & $2.19\times 10^{2}$ \\
\ce{H80} & $3.90\times 10^{5}$ & $1.73\times 10^{2}$ & $3.12\times 10^{2}$ & $1.87\times 10^{4}$ & $2.92\times 10^{9}$ & 1 & $3.12\times 10^{2}$ \\
    \end{tabular}
\end{ruledtabular}
\end{table*}
\begin{table*}[htbp]
\caption{
    Physical resource estimates for the iron-sulfur cluster models across three initial-state infidelity parameter values: (a) $\xi=0.1$, (b) $\xi=0.05$, and (c) $\xi=0.01$. 
    Columns include physical qubits per QPU ($\mathcal{Q}_{\rm QPU}$), maximum per-shot runtime ($\mathcal{T}_{\rm max}$), single-QPU time-to-solution ($\mathcal{T}_{\rm total}$), maximum coherent spacetime volume ($\mathcal{V}_{\rm coh}^{\rm max}$), total  spacetime volume ($\mathcal{V}_{\rm total}$), QPU parallelism ($k^\ast$) under $\mathcal{Q}_{\rm budget}=5\times10^5$, and multi-QPU time-to-solution using $k^\ast$ QPUs.
    Detailed definitions for these parameters and the underlying assumptions can be found in Appendix~\ref{append:physical_resource}.
}
\label{tab:fe-s_physical}
\begin{ruledtabular}
\begin{tabular}{l c c c c c c c}
    Molecule ID & $\mathcal{Q}_{\rm QPU}$ & \makecell{$\mathcal{T}_{\rm max}$ [\si{\second}]} & \makecell{$\mathcal{T}_{\rm total}$ [days]} & \makecell{$\mathcal{V}_{\rm coh}^{\rm max}$\\$[$qubit-hour$]$} & \makecell{$\mathcal{V}_{\rm total}$\\$[$qubit-hour$]$} &  \makecell{$k^*$} & \makecell{Time-to-solution\\($k^\ast$ QPUs) [days]} \\
    \midrule
    \multicolumn{8}{c}{(a) $\xi=0.1$}\\
    \midrule
$\mathrm{[2Fe\mathchar`-2S]}^{-3}$ & $1.18\times 10^{5}$ & $7.01\times 10^{2}$ & $1.04\times 10^{1}$ & $2.29\times 10^{4}$ & $2.94\times 10^{7}$ & 4 & $2.54\times 10^{0}$ \\
$\mathrm{[2Fe\mathchar`-2S]}^{-2}$ & $1.18\times 10^{5}$ & $7.78\times 10^{2}$ & $1.23\times 10^{1}$ & $2.54\times 10^{4}$ & $3.47\times 10^{7}$ & 4 & $2.81\times 10^{0}$ \\
$\mathrm{[4Fe\mathchar`-4S]}^{-2}$ & $2.26\times 10^{5}$ & $3.88\times 10^{3}$ & $1.91\times 10^{2}$ & $2.44\times 10^{5}$ & $1.04\times 10^{9}$ & 2 & $9.52\times 10^{1}$ \\
$\mathrm{[4Fe\mathchar`-4S]}$ & $2.26\times 10^{5}$ & $3.95\times 10^{3}$ & $1.92\times 10^{2}$ & $2.49\times 10^{5}$ & $1.04\times 10^{9}$ & 2 & $9.59\times 10^{1}$ \\
FeMoco (Sm) & $3.77\times 10^{5}$ & $3.36\times 10^{4}$ & $2.51\times 10^{4}$ & $3.52\times 10^{6}$ & $2.27\times 10^{11}$ & 1 & $2.51\times 10^{4}$ \\
FeMoco (Lg) & $5.92\times 10^{5}$ & $4.25\times 10^{4}$ & $1.49\times 10^{5}$ & $6.99\times 10^{6}$ & $2.12\times 10^{12}$ & 1 & $1.49\times 10^{5}$ \\
    \midrule
    \multicolumn{8}{c}{(b) $\xi=0.05$}\\
    \midrule
$\mathrm{[2Fe\mathchar`-2S]}^{-3}$ & $9.80\times 10^{4}$ & $4.48\times 10^{1}$ & $8.98\times 10^{0}$ & $1.22\times 10^{3}$ & $2.11\times 10^{7}$ & 5 & $1.79\times 10^{0}$ \\
$\mathrm{[2Fe\mathchar`-2S]}^{-2}$ & $9.80\times 10^{4}$ & $5.69\times 10^{1}$ & $5.93\times 10^{0}$ & $1.55\times 10^{3}$ & $1.39\times 10^{7}$ & 5 & $1.18\times 10^{0}$ \\
$\mathrm{[4Fe\mathchar`-4S]}^{-2}$ & $2.26\times 10^{5}$ & $1.60\times 10^{3}$ & $9.81\times 10^{1}$ & $1.01\times 10^{5}$ & $5.33\times 10^{8}$ & 2 & $4.88\times 10^{1}$ \\
$\mathrm{[4Fe\mathchar`-4S]}$ & $2.26\times 10^{5}$ & $1.61\times 10^{3}$ & $9.91\times 10^{1}$ & $1.01\times 10^{5}$ & $5.39\times 10^{8}$ & 2 & $4.95\times 10^{1}$ \\
FeMoco (Sm) & $3.77\times 10^{5}$ & $1.06\times 10^{4}$ & $2.77\times 10^{3}$ & $1.11\times 10^{6}$ & $2.51\times 10^{10}$ & 1 & $2.77\times 10^{3}$ \\
FeMoco (Lg) & $5.13\times 10^{5}$ & $1.38\times 10^{4}$ & $7.85\times 10^{3}$ & $1.96\times 10^{6}$ & $9.67\times 10^{10}$ & 1 & $7.85\times 10^{3}$ \\
    \midrule
    \multicolumn{8}{c}{(c) $\xi=0.01$}\\
    \midrule
$\mathrm{[2Fe\mathchar`-2S]}^{-3}$ & $8.02\times 10^{4}$ & $2.51\times 10^{0}$ & $2.81\times 10^{0}$ & $5.59\times 10^{1}$ & $5.41\times 10^{6}$ & 6 & $4.68\times 10^{-1}$ \\
$\mathrm{[2Fe\mathchar`-2S]}^{-2}$ & $8.02\times 10^{4}$ & $3.21\times 10^{0}$ & $3.37\times 10^{0}$ & $7.16\times 10^{1}$ & $6.49\times 10^{6}$ & 6 & $5.62\times 10^{-1}$ \\
$\mathrm{[4Fe\mathchar`-4S]}^{-2}$ & $1.60\times 10^{5}$ & $5.07\times 10^{1}$ & $4.84\times 10^{1}$ & $2.25\times 10^{3}$ & $1.86\times 10^{8}$ & 3 & $1.61\times 10^{1}$ \\
$\mathrm{[4Fe\mathchar`-4S]}$ & $1.60\times 10^{5}$ & $5.12\times 10^{1}$ & $5.15\times 10^{1}$ & $2.27\times 10^{3}$ & $1.98\times 10^{8}$ & 3 & $1.72\times 10^{1}$ \\
FeMoco (Sm) & $3.23\times 10^{5}$ & $7.53\times 10^{2}$ & $8.12\times 10^{2}$ & $6.76\times 10^{4}$ & $6.30\times 10^{9}$ & 1 & $8.12\times 10^{2}$ \\
FeMoco (Lg) & $4.40\times 10^{5}$ & $1.10\times 10^{3}$ & $1.34\times 10^{3}$ & $1.34\times 10^{5}$ & $1.42\times 10^{10}$ & 1 & $1.34\times 10^{3}$ \\
    \end{tabular}
\end{ruledtabular}
\end{table*}
\begin{table*}[htbp]
\caption{
    Physical resource estimates for the P450-Cpd I and P450-rest models across three initial-state infidelity parameter values: (a) $\xi=0.1$, (b) $\xi=0.05$, and (c) $\xi=0.01$. 
    Columns include physical qubits per QPU ($\mathcal{Q}_{\rm QPU}$), maximum per-shot runtime ($\mathcal{T}_{\rm max}$), single-QPU time-to-solution ($\mathcal{T}_{\rm total}$), maximum coherent spacetime volume ($\mathcal{V}_{\rm coh}^{\rm max}$), total  spacetime volume ($\mathcal{V}_{\rm total}$), QPU parallelism ($k^\ast$) under $\mathcal{Q}_{\rm budget}=5\times10^5$, and multi-QPU time-to-solution using $k^\ast$ QPUs.
    Detailed definitions for these parameters and the underlying assumptions can be found in Appendix~\ref{append:physical_resource}.
}
\label{tab:p450_1_physical}
\begin{ruledtabular}
\begin{tabular}{l c c c c c c c}
    Molecule ID & $\mathcal{Q}_{\rm QPU}$ & \makecell{$\mathcal{T}_{\rm max}$ [\si{\second}]} & \makecell{$\mathcal{T}_{\rm total}$ [days]} & \makecell{$\mathcal{V}_{\rm coh}^{\rm max}$\\$[$qubit-hour$]$} & \makecell{$\mathcal{V}_{\rm total}$\\$[$qubit-hour$]$} &  \makecell{$k^*$} & \makecell{Time-to-solution\\($k^\ast$ QPUs) [days]} \\
    \midrule
    \multicolumn{8}{c}{(a) $\xi=0.1$}\\
    \midrule
P450-Cpd I (A) & $1.91\times 10^{4}$ & $3.82\times 10^{-2}$ & $5.74\times 10^{-4}$ & $2.02\times 10^{-1}$ & $2.63\times 10^{2}$ & 26 & $1.93\times 10^{-5}$ \\
P450-Cpd I (B) & $3.28\times 10^{4}$ & $7.43\times 10^{-1}$ & $7.67\times 10^{-2}$ & $6.76\times 10^{0}$ & $6.03\times 10^{4}$ & 15 & $5.03\times 10^{-3}$ \\
P450-Cpd I (C) & $9.39\times 10^{4}$ & $2.04\times 10^{2}$ & $2.15\times 10^{0}$ & $5.33\times 10^{3}$ & $4.84\times 10^{6}$ & 6 & $4.00\times 10^{-1}$ \\
P450-Cpd I (D) & $1.55\times 10^{5}$ & $1.20\times 10^{3}$ & $2.91\times 10^{1}$ & $5.18\times 10^{4}$ & $1.09\times 10^{8}$ & 3 & $9.54\times 10^{0}$ \\
P450-Cpd I (E) & $1.99\times 10^{5}$ & $2.71\times 10^{3}$ & $1.23\times 10^{2}$ & $1.50\times 10^{5}$ & $5.89\times 10^{8}$ & 2 & $6.12\times 10^{1}$ \\
P450-Cpd I (F) & $2.53\times 10^{5}$ & $4.11\times 10^{3}$ & $2.05\times 10^{2}$ & $2.89\times 10^{5}$ & $1.25\times 10^{9}$ & 1 & $2.01\times 10^{2}$ \\
P450-Cpd I (G) & $2.64\times 10^{5}$ & $4.59\times 10^{3}$ & $2.43\times 10^{2}$ & $3.37\times 10^{5}$ & $1.54\times 10^{9}$ & 1 & $2.39\times 10^{2}$ \\
P450-Cpd I (X) & $4.02\times 10^{5}$ & $1.18\times 10^{4}$ & $2.43\times 10^{3}$ & $1.31\times 10^{6}$ & $2.34\times 10^{10}$ & 1 & $2.43\times 10^{3}$ \\
P450-rest (A) & $1.91\times 10^{4}$ & $7.43\times 10^{-2}$ & $1.46\times 10^{-3}$ & $3.94\times 10^{-1}$ & $6.70\times 10^{2}$ & 26 & $5.19\times 10^{-5}$ \\
P450-rest (B) & $3.28\times 10^{4}$ & $6.32\times 10^{-1}$ & $2.64\times 10^{-2}$ & $5.75\times 10^{0}$ & $2.07\times 10^{4}$ & 15 & $1.68\times 10^{-3}$ \\
P450-rest (C) & $8.43\times 10^{4}$ & $1.47\times 10^{2}$ & $1.66\times 10^{0}$ & $3.43\times 10^{3}$ & $3.36\times 10^{6}$ & 8 & $2.90\times 10^{-1}$ \\
P450-rest (D) & $1.18\times 10^{5}$ & $7.04\times 10^{2}$ & $1.05\times 10^{1}$ & $2.30\times 10^{4}$ & $2.97\times 10^{7}$ & 4 & $2.55\times 10^{0}$ \\
P450-rest (E) & $1.83\times 10^{5}$ & $1.63\times 10^{3}$ & $3.84\times 10^{1}$ & $8.28\times 10^{4}$ & $1.69\times 10^{8}$ & 3 & $1.80\times 10^{1}$ \\
P450-rest (F) & $2.48\times 10^{5}$ & $3.09\times 10^{3}$ & $1.42\times 10^{2}$ & $2.13\times 10^{5}$ & $8.46\times 10^{8}$ & 2 & $7.10\times 10^{1}$ \\
P450-rest (G) & $2.59\times 10^{5}$ & $3.36\times 10^{3}$ & $1.56\times 10^{2}$ & $2.42\times 10^{5}$ & $9.67\times 10^{8}$ & 1 & $1.52\times 10^{2}$ \\
P450-rest (X) & $3.89\times 10^{5}$ & $8.88\times 10^{3}$ & $1.41\times 10^{3}$ & $9.60\times 10^{5}$ & $1.32\times 10^{10}$ & 1 & $1.41\times 10^{3}$ \\
    \midrule
    \multicolumn{8}{c}{(b) $\xi=0.05$}\\
    \midrule
P450-Cpd I (A) & $1.43\times 10^{4}$ & $1.76\times 10^{-2}$ & $5.86\times 10^{-4}$ & $7.02\times 10^{-2}$ & $2.01\times 10^{2}$ & 34 & $1.71\times 10^{-5}$ \\
P450-Cpd I (B) & $3.28\times 10^{4}$ & $3.34\times 10^{-1}$ & $3.10\times 10^{-2}$ & $3.04\times 10^{0}$ & $2.43\times 10^{4}$ & 15 & $2.02\times 10^{-3}$ \\
P450-Cpd I (C) & $6.41\times 10^{4}$ & $6.78\times 10^{0}$ & $1.68\times 10^{0}$ & $1.21\times 10^{2}$ & $2.58\times 10^{6}$ & 7 & $2.39\times 10^{-1}$ \\
P450-Cpd I (D) & $1.32\times 10^{5}$ & $5.23\times 10^{2}$ & $2.09\times 10^{1}$ & $1.91\times 10^{4}$ & $6.62\times 10^{7}$ & 3 & $6.88\times 10^{0}$ \\
P450-Cpd I (E) & $1.99\times 10^{5}$ & $1.21\times 10^{3}$ & $6.85\times 10^{1}$ & $6.71\times 10^{4}$ & $3.28\times 10^{8}$ & 2 & $3.40\times 10^{1}$ \\
P450-Cpd I (F) & $2.53\times 10^{5}$ & $1.86\times 10^{3}$ & $1.21\times 10^{2}$ & $1.31\times 10^{5}$ & $7.39\times 10^{8}$ & 1 & $1.19\times 10^{2}$ \\
P450-Cpd I (G) & $2.64\times 10^{5}$ & $1.99\times 10^{3}$ & $1.35\times 10^{2}$ & $1.46\times 10^{5}$ & $8.57\times 10^{8}$ & 1 & $1.33\times 10^{2}$ \\
P450-Cpd I (X) & $4.02\times 10^{5}$ & $4.60\times 10^{3}$ & $6.35\times 10^{2}$ & $5.14\times 10^{5}$ & $6.12\times 10^{9}$ & 1 & $6.35\times 10^{2}$ \\
P450-rest (A) & $1.91\times 10^{4}$ & $3.54\times 10^{-2}$ & $1.30\times 10^{-3}$ & $1.87\times 10^{-1}$ & $5.95\times 10^{2}$ & 26 & $4.76\times 10^{-5}$ \\
P450-rest (B) & $3.28\times 10^{4}$ & $5.23\times 10^{-1}$ & $5.80\times 10^{-2}$ & $4.76\times 10^{0}$ & $4.56\times 10^{4}$ & 15 & $3.81\times 10^{-3}$ \\
P450-rest (C) & $5.75\times 10^{4}$ & $4.08\times 10^{0}$ & $1.32\times 10^{0}$ & $6.53\times 10^{1}$ & $1.82\times 10^{6}$ & 8 & $1.64\times 10^{-1}$ \\
P450-rest (D) & $1.18\times 10^{5}$ & $3.61\times 10^{2}$ & $1.14\times 10^{1}$ & $1.18\times 10^{4}$ & $3.22\times 10^{7}$ & 5 & $2.81\times 10^{0}$ \\
P450-rest (E) & $1.83\times 10^{5}$ & $7.59\times 10^{2}$ & $3.13\times 10^{1}$ & $3.85\times 10^{4}$ & $1.37\times 10^{8}$ & 3 & $1.50\times 10^{1}$ \\
P450-rest (F) & $2.48\times 10^{5}$ & $1.34\times 10^{3}$ & $7.71\times 10^{1}$ & $9.25\times 10^{4}$ & $4.59\times 10^{8}$ & 2 & $3.85\times 10^{1}$ \\
P450-rest (G) & $2.59\times 10^{5}$ & $1.51\times 10^{3}$ & $9.06\times 10^{1}$ & $1.09\times 10^{5}$ & $5.63\times 10^{8}$ & 1 & $8.84\times 10^{1}$ \\
P450-rest (X) & $3.34\times 10^{5}$ & $3.35\times 10^{3}$ & $3.93\times 10^{2}$ & $3.10\times 10^{5}$ & $3.15\times 10^{9}$ & 1 & $3.92\times 10^{2}$ \\
    \midrule
    \multicolumn{8}{c}{(c) $\xi=0.01$}\\
    \midrule
P450-Cpd I (A) & $1.43\times 10^{4}$ & $3.95\times 10^{-3}$ & $2.56\times 10^{-3}$ & $1.57\times 10^{-2}$ & $8.79\times 10^{2}$ & 34 & $7.52\times 10^{-5}$ \\
P450-Cpd I (B) & $2.55\times 10^{4}$ & $4.45\times 10^{-2}$ & $3.43\times 10^{-2}$ & $3.15\times 10^{-1}$ & $2.10\times 10^{4}$ & 19 & $1.80\times 10^{-3}$ \\
P450-Cpd I (C) & $5.13\times 10^{4}$ & $4.94\times 10^{-1}$ & $5.62\times 10^{-1}$ & $7.04\times 10^{0}$ & $6.92\times 10^{5}$ & 9 & $6.24\times 10^{-2}$ \\
P450-Cpd I (D) & $8.98\times 10^{4}$ & $6.55\times 10^{0}$ & $7.74\times 10^{0}$ & $1.64\times 10^{2}$ & $1.67\times 10^{7}$ & 5 & $1.55\times 10^{0}$ \\
P450-Cpd I (E) & $1.41\times 10^{5}$ & $3.39\times 10^{1}$ & $3.74\times 10^{1}$ & $1.32\times 10^{3}$ & $1.26\times 10^{8}$ & 3 & $1.25\times 10^{1}$ \\
P450-Cpd I (F) & $2.15\times 10^{5}$ & $7.08\times 10^{1}$ & $7.64\times 10^{1}$ & $4.22\times 10^{3}$ & $3.94\times 10^{8}$ & 2 & $3.82\times 10^{1}$ \\
P450-Cpd I (G) & $2.24\times 10^{5}$ & $8.11\times 10^{1}$ & $8.74\times 10^{1}$ & $5.04\times 10^{3}$ & $4.69\times 10^{8}$ & 2 & $4.37\times 10^{1}$ \\
P450-Cpd I (X) & $2.92\times 10^{5}$ & $2.33\times 10^{2}$ & $2.79\times 10^{2}$ & $1.89\times 10^{4}$ & $1.95\times 10^{9}$ & 1 & $2.79\times 10^{2}$ \\
P450-rest (A) & $1.43\times 10^{4}$ & $6.56\times 10^{-3}$ & $4.32\times 10^{-3}$ & $2.61\times 10^{-2}$ & $1.48\times 10^{3}$ & 34 & $1.27\times 10^{-4}$ \\
P450-rest (B) & $2.55\times 10^{4}$ & $3.43\times 10^{-2}$ & $2.52\times 10^{-2}$ & $2.43\times 10^{-1}$ & $1.54\times 10^{4}$ & 19 & $1.33\times 10^{-3}$ \\
P450-rest (C) & $4.61\times 10^{4}$ & $3.20\times 10^{-1}$ & $3.63\times 10^{-1}$ & $4.10\times 10^{0}$ & $4.02\times 10^{5}$ & 10 & $3.63\times 10^{-2}$ \\
P450-rest (D) & $8.02\times 10^{4}$ & $2.78\times 10^{0}$ & $3.17\times 10^{0}$ & $6.19\times 10^{1}$ & $6.10\times 10^{6}$ & 6 & $5.28\times 10^{-1}$ \\
P450-rest (E) & $1.29\times 10^{5}$ & $1.24\times 10^{1}$ & $1.43\times 10^{1}$ & $4.45\times 10^{2}$ & $4.42\times 10^{7}$ & 3 & $4.75\times 10^{0}$ \\
P450-rest (F) & $1.75\times 10^{5}$ & $3.50\times 10^{1}$ & $3.77\times 10^{1}$ & $1.70\times 10^{3}$ & $1.59\times 10^{8}$ & 2 & $1.89\times 10^{1}$ \\
P450-rest (G) & $1.83\times 10^{5}$ & $4.05\times 10^{1}$ & $4.44\times 10^{1}$ & $2.06\times 10^{3}$ & $1.95\times 10^{8}$ & 2 & $2.22\times 10^{1}$ \\
P450-rest (X) & $2.83\times 10^{5}$ & $1.57\times 10^{2}$ & $1.88\times 10^{2}$ & $1.23\times 10^{4}$ & $1.27\times 10^{9}$ & 1 & $1.88\times 10^{2}$ \\
    \end{tabular}
\end{ruledtabular}
\end{table*}
\begin{table*}[htbp]
\caption{
    Physical resource estimates for the P450-inhibited and P450-empty models across three initial-state infidelity parameter values: (a) $\xi=0.1$, (b) $\xi=0.05$, and (c) $\xi=0.01$. 
    Columns include physical qubits per QPU ($\mathcal{Q}_{\rm QPU}$), maximum per-shot runtime ($\mathcal{T}_{\rm max}$), single-QPU time-to-solution ($\mathcal{T}_{\rm total}$), maximum coherent spacetime volume ($\mathcal{V}_{\rm coh}^{\rm max}$), total  spacetime volume ($\mathcal{V}_{\rm total}$), QPU parallelism ($k^\ast$) under $\mathcal{Q}_{\rm budget}=5\times10^5$, and multi-QPU time-to-solution using $k^\ast$ QPUs.
    Detailed definitions for these parameters and the underlying assumptions can be found in Appendix~\ref{append:physical_resource}.
}
\label{tab:p450_2_physical}
\begin{ruledtabular}
\begin{tabular}{l c c c c c c c}
    Molecule ID & $\mathcal{Q}_{\rm QPU}$ & \makecell{$\mathcal{T}_{\rm max}$ [\si{\second}]} & \makecell{$\mathcal{T}_{\rm total}$ [days]} & \makecell{$\mathcal{V}_{\rm coh}^{\rm max}$\\$[$qubit-hour$]$} & \makecell{$\mathcal{V}_{\rm total}$\\$[$qubit-hour$]$} &  \makecell{$k^*$} & \makecell{Time-to-solution\\($k^\ast$ QPUs) [days]} \\
    \midrule
    \multicolumn{8}{c}{(a) $\xi=0.1$}\\
    \midrule
P450-inhibited (A) & $1.91\times 10^{4}$ & $6.76\times 10^{-2}$ & $1.03\times 10^{-3}$ & $3.58\times 10^{-1}$ & $4.70\times 10^{2}$ & 26 & $3.70\times 10^{-5}$ \\
P450-inhibited (B) & $3.54\times 10^{4}$ & $1.37\times 10^{0}$ & $1.20\times 10^{-1}$ & $1.35\times 10^{1}$ & $1.02\times 10^{5}$ & 14 & $8.48\times 10^{-3}$ \\
P450-inhibited (C) & $8.91\times 10^{4}$ & $1.55\times 10^{2}$ & $1.75\times 10^{0}$ & $3.84\times 10^{3}$ & $3.75\times 10^{6}$ & 6 & $3.19\times 10^{-1}$ \\
P450-inhibited (D) & $1.22\times 10^{5}$ & $8.03\times 10^{2}$ & $1.20\times 10^{1}$ & $2.73\times 10^{4}$ & $3.52\times 10^{7}$ & 4 & $2.98\times 10^{0}$ \\
P450-inhibited (E) & $2.05\times 10^{5}$ & $2.52\times 10^{3}$ & $6.34\times 10^{1}$ & $1.43\times 10^{5}$ & $3.12\times 10^{8}$ & 2 & $3.14\times 10^{1}$ \\
P450-inhibited (F) & $2.70\times 10^{5}$ & $3.05\times 10^{3}$ & $1.40\times 10^{2}$ & $2.29\times 10^{5}$ & $9.03\times 10^{8}$ & 1 & $1.36\times 10^{2}$ \\
P450-inhibited (G) & $2.80\times 10^{5}$ & $3.32\times 10^{3}$ & $1.55\times 10^{2}$ & $2.59\times 10^{5}$ & $1.04\times 10^{9}$ & 1 & $1.51\times 10^{2}$ \\
P450-inhibited (X) & $4.14\times 10^{5}$ & $9.30\times 10^{3}$ & $1.59\times 10^{3}$ & $1.07\times 10^{6}$ & $1.58\times 10^{10}$ & 1 & $1.59\times 10^{3}$ \\
P450-empty (A) & $1.91\times 10^{4}$ & $7.32\times 10^{-2}$ & $1.40\times 10^{-3}$ & $3.88\times 10^{-1}$ & $6.40\times 10^{2}$ & 26 & $4.98\times 10^{-5}$ \\
P450-empty (B) & $3.28\times 10^{4}$ & $1.03\times 10^{0}$ & $5.52\times 10^{-2}$ & $9.41\times 10^{0}$ & $4.34\times 10^{4}$ & 15 & $3.64\times 10^{-3}$ \\
P450-empty (C) & $6.22\times 10^{4}$ & $7.71\times 10^{1}$ & $5.97\times 10^{-1}$ & $1.33\times 10^{3}$ & $8.91\times 10^{5}$ & 9 & $7.25\times 10^{-2}$ \\
P450-empty (D) & $1.08\times 10^{5}$ & $5.52\times 10^{2}$ & $8.87\times 10^{0}$ & $1.66\times 10^{4}$ & $2.30\times 10^{7}$ & 4 & $2.14\times 10^{0}$ \\
P450-empty (E) & $1.72\times 10^{5}$ & $1.61\times 10^{3}$ & $3.82\times 10^{1}$ & $7.69\times 10^{4}$ & $1.58\times 10^{8}$ & 3 & $1.79\times 10^{1}$ \\
P450-empty (F) & $2.32\times 10^{5}$ & $2.29\times 10^{3}$ & $5.64\times 10^{1}$ & $1.48\times 10^{5}$ & $3.14\times 10^{8}$ & 2 & $2.80\times 10^{1}$ \\
P450-empty (G) & $2.43\times 10^{5}$ & $3.32\times 10^{3}$ & $1.58\times 10^{2}$ & $2.24\times 10^{5}$ & $9.18\times 10^{8}$ & 2 & $7.88\times 10^{1}$ \\
P450-empty (X) & $3.83\times 10^{5}$ & $8.09\times 10^{3}$ & $1.27\times 10^{3}$ & $8.61\times 10^{5}$ & $1.17\times 10^{10}$ & 1 & $1.27\times 10^{3}$ \\
    \midrule
    \multicolumn{8}{c}{(b) $\xi=0.05$}\\
    \midrule
P450-inhibited (A) & $1.91\times 10^{4}$ & $3.00\times 10^{-2}$ & $1.00\times 10^{-3}$ & $1.59\times 10^{-1}$ & $4.60\times 10^{2}$ & 26 & $3.72\times 10^{-5}$ \\
P450-inhibited (B) & $3.54\times 10^{4}$ & $5.25\times 10^{-1}$ & $4.80\times 10^{-2}$ & $5.17\times 10^{0}$ & $4.08\times 10^{4}$ & 14 & $3.37\times 10^{-3}$ \\
P450-inhibited (C) & $6.08\times 10^{4}$ & $4.47\times 10^{0}$ & $1.15\times 10^{0}$ & $7.54\times 10^{1}$ & $1.67\times 10^{6}$ & 8 & $1.43\times 10^{-1}$ \\
P450-inhibited (D) & $1.02\times 10^{5}$ & $6.71\times 10^{1}$ & $1.20\times 10^{1}$ & $1.90\times 10^{3}$ & $2.94\times 10^{7}$ & 4 & $3.00\times 10^{0}$ \\
P450-inhibited (E) & $2.05\times 10^{5}$ & $1.11\times 10^{3}$ & $5.03\times 10^{1}$ & $6.34\times 10^{4}$ & $2.47\times 10^{8}$ & 2 & $2.50\times 10^{1}$ \\
P450-inhibited (F) & $2.70\times 10^{5}$ & $1.35\times 10^{3}$ & $7.80\times 10^{1}$ & $1.01\times 10^{5}$ & $5.05\times 10^{8}$ & 1 & $7.60\times 10^{1}$ \\
P450-inhibited (G) & $2.80\times 10^{5}$ & $1.52\times 10^{3}$ & $9.16\times 10^{1}$ & $1.18\times 10^{5}$ & $6.16\times 10^{8}$ & 1 & $8.96\times 10^{1}$ \\
P450-inhibited (X) & $4.14\times 10^{5}$ & $3.81\times 10^{3}$ & $4.63\times 10^{2}$ & $4.39\times 10^{5}$ & $4.61\times 10^{9}$ & 1 & $4.63\times 10^{2}$ \\
P450-empty (A) & $1.91\times 10^{4}$ & $3.84\times 10^{-2}$ & $1.42\times 10^{-3}$ & $2.03\times 10^{-1}$ & $6.51\times 10^{2}$ & 26 & $5.23\times 10^{-5}$ \\
P450-empty (B) & $3.28\times 10^{4}$ & $4.96\times 10^{-1}$ & $6.27\times 10^{-2}$ & $4.52\times 10^{0}$ & $4.93\times 10^{4}$ & 15 & $4.12\times 10^{-3}$ \\
P450-empty (C) & $4.08\times 10^{4}$ & $1.40\times 10^{0}$ & $2.03\times 10^{-1}$ & $1.59\times 10^{1}$ & $1.98\times 10^{5}$ & 12 & $1.69\times 10^{-2}$ \\
P450-empty (D) & $9.01\times 10^{4}$ & $3.65\times 10^{1}$ & $6.64\times 10^{0}$ & $9.13\times 10^{2}$ & $1.44\times 10^{7}$ & 5 & $1.32\times 10^{0}$ \\
P450-empty (E) & $1.46\times 10^{5}$ & $7.16\times 10^{2}$ & $2.93\times 10^{1}$ & $2.89\times 10^{4}$ & $1.02\times 10^{8}$ & 3 & $9.67\times 10^{0}$ \\
P450-empty (F) & $2.32\times 10^{5}$ & $1.05\times 10^{3}$ & $4.67\times 10^{1}$ & $6.77\times 10^{4}$ & $2.60\times 10^{8}$ & 2 & $2.32\times 10^{1}$ \\
P450-empty (G) & $2.43\times 10^{5}$ & $1.46\times 10^{3}$ & $8.63\times 10^{1}$ & $9.85\times 10^{4}$ & $5.03\times 10^{8}$ & 2 & $4.31\times 10^{1}$ \\
P450-empty (X) & $3.28\times 10^{5}$ & $3.15\times 10^{3}$ & $3.62\times 10^{2}$ & $2.88\times 10^{5}$ & $2.85\times 10^{9}$ & 1 & $3.62\times 10^{2}$ \\
    \midrule
    \multicolumn{8}{c}{(c) $\xi=0.01$}\\
    \midrule
P450-inhibited (A) & $1.43\times 10^{4}$ & $7.50\times 10^{-3}$ & $4.86\times 10^{-3}$ & $2.98\times 10^{-2}$ & $1.67\times 10^{3}$ & 34 & $1.43\times 10^{-4}$ \\
P450-inhibited (B) & $2.76\times 10^{4}$ & $5.59\times 10^{-2}$ & $4.57\times 10^{-2}$ & $4.28\times 10^{-1}$ & $3.02\times 10^{4}$ & 18 & $2.53\times 10^{-3}$ \\
P450-inhibited (C) & $4.87\times 10^{4}$ & $3.66\times 10^{-1}$ & $3.83\times 10^{-1}$ & $4.94\times 10^{0}$ & $4.47\times 10^{5}$ & 10 & $3.82\times 10^{-2}$ \\
P450-inhibited (D) & $8.34\times 10^{4}$ & $3.81\times 10^{0}$ & $3.88\times 10^{0}$ & $8.83\times 10^{1}$ & $7.76\times 10^{6}$ & 5 & $7.75\times 10^{-1}$ \\
P450-inhibited (E) & $1.44\times 10^{5}$ & $2.39\times 10^{1}$ & $2.44\times 10^{1}$ & $9.59\times 10^{2}$ & $8.48\times 10^{7}$ & 3 & $8.15\times 10^{0}$ \\
P450-inhibited (F) & $1.90\times 10^{5}$ & $3.59\times 10^{1}$ & $3.80\times 10^{1}$ & $1.90\times 10^{3}$ & $1.74\times 10^{8}$ & 2 & $1.90\times 10^{1}$ \\
P450-inhibited (G) & $1.98\times 10^{5}$ & $4.31\times 10^{1}$ & $4.58\times 10^{1}$ & $2.37\times 10^{3}$ & $2.17\times 10^{8}$ & 2 & $2.29\times 10^{1}$ \\
P450-inhibited (X) & $3.01\times 10^{5}$ & $1.73\times 10^{2}$ & $2.01\times 10^{2}$ & $1.44\times 10^{4}$ & $1.45\times 10^{9}$ & 1 & $2.01\times 10^{2}$ \\
P450-empty (A) & $1.43\times 10^{4}$ & $6.06\times 10^{-3}$ & $3.99\times 10^{-3}$ & $2.41\times 10^{-2}$ & $1.37\times 10^{3}$ & 34 & $1.17\times 10^{-4}$ \\
P450-empty (B) & $2.55\times 10^{4}$ & $4.26\times 10^{-2}$ & $3.26\times 10^{-2}$ & $3.02\times 10^{-1}$ & $1.99\times 10^{4}$ & 19 & $1.71\times 10^{-3}$ \\
P450-empty (C) & $3.18\times 10^{4}$ & $1.34\times 10^{-1}$ & $1.29\times 10^{-1}$ & $1.18\times 10^{0}$ & $9.84\times 10^{4}$ & 15 & $8.61\times 10^{-3}$ \\
P450-empty (D) & $7.38\times 10^{4}$ & $2.12\times 10^{0}$ & $2.13\times 10^{0}$ & $4.35\times 10^{1}$ & $3.77\times 10^{6}$ & 6 & $3.54\times 10^{-1}$ \\
P450-empty (E) & $1.21\times 10^{5}$ & $1.35\times 10^{1}$ & $1.35\times 10^{1}$ & $4.55\times 10^{2}$ & $3.95\times 10^{7}$ & 4 & $3.38\times 10^{0}$ \\
P450-empty (F) & $1.64\times 10^{5}$ & $2.17\times 10^{1}$ & $2.29\times 10^{1}$ & $9.86\times 10^{2}$ & $8.97\times 10^{7}$ & 3 & $7.62\times 10^{0}$ \\
P450-empty (G) & $1.71\times 10^{5}$ & $4.30\times 10^{1}$ & $4.38\times 10^{1}$ & $2.05\times 10^{3}$ & $1.80\times 10^{8}$ & 2 & $2.19\times 10^{1}$ \\
P450-empty (X) & $2.78\times 10^{5}$ & $1.47\times 10^{2}$ & $1.66\times 10^{2}$ & $1.13\times 10^{4}$ & $1.11\times 10^{9}$ & 1 & $1.66\times 10^{2}$ \\
    \end{tabular}
\end{ruledtabular}
\end{table*}
\begin{table*}[htbp]
\caption{
    Physical resource estimates for the ruthenium-based catalyst models, listed for structures I, II, II–III, and V (first half of the catalytic cycle~\cite{Von_Burg2021-du}), across three initial-state infidelity parameter values: (a) $\xi=0.1$, (b) $\xi=0.05$, and (c) $\xi=0.01$. 
    Columns include physical qubits per QPU ($\mathcal{Q}_{\rm QPU}$), maximum per-shot runtime ($\mathcal{T}_{\rm max}$), single-QPU time-to-solution ($\mathcal{T}_{\rm total}$), maximum coherent spacetime volume ($\mathcal{V}_{\rm coh}^{\rm max}$), total  spacetime volume ($\mathcal{V}_{\rm total}$), QPU parallelism ($k^\ast$) under $\mathcal{Q}_{\rm budget}=5\times10^5$, and multi-QPU time-to-solution using $k^\ast$ QPUs.
    Detailed definitions for these parameters and the underlying assumptions can be found in Appendix~\ref{append:physical_resource}.
}
\label{tab:co2-cat_1_physical}
\begin{ruledtabular}
\begin{tabular}{l c c c c c c c}
    Molecule ID & $\mathcal{Q}_{\rm QPU}$ & \makecell{$\mathcal{T}_{\rm max}$ [\si{\second}]} & \makecell{$\mathcal{T}_{\rm total}$ [days]} & \makecell{$\mathcal{V}_{\rm coh}^{\rm max}$\\$[$qubit-hour$]$} & \makecell{$\mathcal{V}_{\rm total}$\\$[$qubit-hour$]$} &  \makecell{$k^*$} & \makecell{Time-to-solution\\($k^\ast$ QPUs) [days]} \\
    \midrule
    \multicolumn{8}{c}{(a) $\xi=0.1$}\\
    \midrule
Ru-I (Sm) & $2.45\times 10^{4}$ & $5.32\times 10^{-1}$ & $2.16\times 10^{-2}$ & $3.62\times 10^{0}$ & $1.27\times 10^{4}$ & 20 & $1.02\times 10^{-3}$ \\
Ru-I (Md) & $9.87\times 10^{4}$ & $4.03\times 10^{2}$ & $4.10\times 10^{0}$ & $1.10\times 10^{4}$ & $9.70\times 10^{6}$ & 6 & $7.78\times 10^{-1}$ \\
Ru-I (Lg) & $3.64\times 10^{5}$ & $1.78\times 10^{4}$ & $3.58\times 10^{3}$ & $1.80\times 10^{6}$ & $3.13\times 10^{10}$ & 1 & $3.58\times 10^{3}$ \\
Ru-II (Sm) & $2.73\times 10^{4}$ & $6.43\times 10^{-1}$ & $2.93\times 10^{-2}$ & $4.87\times 10^{0}$ & $1.92\times 10^{4}$ & 18 & $1.56\times 10^{-3}$ \\
Ru-II (Md) & $1.72\times 10^{5}$ & $1.34\times 10^{3}$ & $3.02\times 10^{1}$ & $6.40\times 10^{4}$ & $1.25\times 10^{8}$ & 3 & $1.38\times 10^{1}$ \\
Ru-II (Lg) & $4.92\times 10^{5}$ & $4.69\times 10^{4}$ & $9.20\times 10^{4}$ & $6.41\times 10^{6}$ & $1.09\times 10^{12}$ & 1 & $9.20\times 10^{4}$ \\
Ru-II-III (Sm) & $2.73\times 10^{4}$ & $6.83\times 10^{-1}$ & $3.22\times 10^{-2}$ & $5.18\times 10^{0}$ & $2.11\times 10^{4}$ & 18 & $1.72\times 10^{-3}$ \\
Ru-II-III (Md) & $1.88\times 10^{5}$ & $2.33\times 10^{3}$ & $5.58\times 10^{1}$ & $1.22\times 10^{5}$ & $2.52\times 10^{8}$ & 2 & $2.67\times 10^{1}$ \\
Ru-II-III (Lg) & $5.14\times 10^{5}$ & $6.48\times 10^{4}$ & $2.70\times 10^{5}$ & $9.25\times 10^{6}$ & $3.33\times 10^{12}$ & 1 & $2.70\times 10^{5}$ \\
Ru-V (Sm) & $7.47\times 10^{4}$ & $2.65\times 10^{2}$ & $2.84\times 10^{0}$ & $5.49\times 10^{3}$ & $5.08\times 10^{6}$ & 8 & $4.19\times 10^{-1}$ \\
Ru-V (Md) & $1.61\times 10^{5}$ & $1.81\times 10^{3}$ & $4.25\times 10^{1}$ & $8.11\times 10^{4}$ & $1.64\times 10^{8}$ & 3 & $1.40\times 10^{1}$ \\
Ru-V (Lg) & $4.14\times 10^{5}$ & $2.68\times 10^{4}$ & $1.51\times 10^{4}$ & $3.08\times 10^{6}$ & $1.50\times 10^{11}$ & 1 & $1.51\times 10^{4}$ \\
    \midrule
    \multicolumn{8}{c}{(b) $\xi=0.05$}\\
    \midrule
Ru-I (Sm) & $1.91\times 10^{4}$ & $2.04\times 10^{-1}$ & $1.30\times 10^{-2}$ & $1.08\times 10^{0}$ & $5.94\times 10^{3}$ & 26 & $4.96\times 10^{-4}$ \\
Ru-I (Md) & $8.22\times 10^{4}$ & $1.85\times 10^{1}$ & $2.87\times 10^{0}$ & $4.23\times 10^{2}$ & $5.66\times 10^{6}$ & 6 & $4.74\times 10^{-1}$ \\
Ru-I (Lg) & $3.64\times 10^{5}$ & $5.88\times 10^{3}$ & $7.79\times 10^{2}$ & $5.95\times 10^{5}$ & $6.82\times 10^{9}$ & 1 & $7.79\times 10^{2}$ \\
Ru-II (Sm) & $2.73\times 10^{4}$ & $2.70\times 10^{-1}$ & $1.87\times 10^{-2}$ & $2.04\times 10^{0}$ & $1.23\times 10^{4}$ & 18 & $1.00\times 10^{-3}$ \\
Ru-II (Md) & $1.46\times 10^{5}$ & $5.95\times 10^{2}$ & $2.39\times 10^{1}$ & $2.40\times 10^{4}$ & $8.35\times 10^{7}$ & 3 & $7.86\times 10^{0}$ \\
Ru-II (Lg) & $4.27\times 10^{5}$ & $1.33\times 10^{4}$ & $5.23\times 10^{3}$ & $1.57\times 10^{6}$ & $5.36\times 10^{10}$ & 1 & $5.23\times 10^{3}$ \\
Ru-II-III (Sm) & $2.73\times 10^{4}$ & $2.79\times 10^{-1}$ & $1.87\times 10^{-2}$ & $2.12\times 10^{0}$ & $1.22\times 10^{4}$ & 18 & $1.00\times 10^{-3}$ \\
Ru-II-III (Md) & $1.88\times 10^{5}$ & $9.97\times 10^{2}$ & $4.38\times 10^{1}$ & $5.22\times 10^{4}$ & $1.98\times 10^{8}$ & 3 & $2.13\times 10^{1}$ \\
Ru-II-III (Lg) & $4.45\times 10^{5}$ & $1.84\times 10^{4}$ & $1.12\times 10^{4}$ & $2.27\times 10^{6}$ & $1.19\times 10^{11}$ & 1 & $1.12\times 10^{4}$ \\
Ru-V (Sm) & $5.09\times 10^{4}$ & $6.07\times 10^{0}$ & $1.72\times 10^{0}$ & $8.59\times 10^{1}$ & $2.10\times 10^{6}$ & 9 & $1.90\times 10^{-1}$ \\
Ru-V (Md) & $1.36\times 10^{5}$ & $7.59\times 10^{2}$ & $3.17\times 10^{1}$ & $2.87\times 10^{4}$ & $1.04\times 10^{8}$ & 3 & $1.05\times 10^{1}$ \\
Ru-V (Lg) & $4.14\times 10^{5}$ & $9.31\times 10^{3}$ & $2.28\times 10^{3}$ & $1.07\times 10^{6}$ & $2.26\times 10^{10}$ & 1 & $2.28\times 10^{3}$ \\
    \midrule
    \multicolumn{8}{c}{(c) $\xi=0.01$}\\
    \midrule
Ru-I (Sm) & $1.91\times 10^{4}$ & $2.91\times 10^{-2}$ & $2.07\times 10^{-2}$ & $1.54\times 10^{-1}$ & $9.49\times 10^{3}$ & 26 & $7.96\times 10^{-4}$ \\
Ru-I (Md) & $5.39\times 10^{4}$ & $9.58\times 10^{-1}$ & $9.53\times 10^{-1}$ & $1.43\times 10^{1}$ & $1.23\times 10^{6}$ & 9 & $1.06\times 10^{-1}$ \\
Ru-I (Lg) & $2.64\times 10^{5}$ & $3.20\times 10^{2}$ & $3.11\times 10^{2}$ & $2.35\times 10^{4}$ & $1.97\times 10^{9}$ & 1 & $3.11\times 10^{2}$ \\
Ru-II (Sm) & $2.12\times 10^{4}$ & $4.25\times 10^{-2}$ & $3.26\times 10^{-2}$ & $2.51\times 10^{-1}$ & $1.66\times 10^{4}$ & 23 & $1.41\times 10^{-3}$ \\
Ru-II (Md) & $1.21\times 10^{5}$ & $8.79\times 10^{0}$ & $1.42\times 10^{1}$ & $2.96\times 10^{2}$ & $4.13\times 10^{7}$ & 4 & $3.54\times 10^{0}$ \\
Ru-II (Lg) & $3.66\times 10^{5}$ & $1.16\times 10^{3}$ & $1.20\times 10^{3}$ & $1.18\times 10^{5}$ & $1.05\times 10^{10}$ & 1 & $1.20\times 10^{3}$ \\
Ru-II-III (Sm) & $2.12\times 10^{4}$ & $3.76\times 10^{-2}$ & $2.81\times 10^{-2}$ & $2.22\times 10^{-1}$ & $1.43\times 10^{4}$ & 23 & $1.22\times 10^{-3}$ \\
Ru-II-III (Md) & $1.33\times 10^{5}$ & $2.08\times 10^{1}$ & $3.05\times 10^{1}$ & $7.68\times 10^{2}$ & $9.75\times 10^{7}$ & 3 & $1.02\times 10^{1}$ \\
Ru-II-III (Lg) & $3.82\times 10^{5}$ & $1.44\times 10^{3}$ & $1.66\times 10^{3}$ & $1.53\times 10^{5}$ & $1.52\times 10^{10}$ & 1 & $1.66\times 10^{3}$ \\
Ru-V (Sm) & $4.08\times 10^{4}$ & $4.37\times 10^{-1}$ & $4.81\times 10^{-1}$ & $4.95\times 10^{0}$ & $4.70\times 10^{5}$ & 12 & $4.00\times 10^{-2}$ \\
Ru-V (Md) & $1.14\times 10^{5}$ & $1.50\times 10^{1}$ & $1.68\times 10^{1}$ & $4.74\times 10^{2}$ & $4.58\times 10^{7}$ & 4 & $4.20\times 10^{0}$ \\
Ru-V (Lg) & $3.55\times 10^{5}$ & $6.02\times 10^{2}$ & $7.26\times 10^{2}$ & $5.94\times 10^{4}$ & $6.19\times 10^{9}$ & 1 & $7.26\times 10^{2}$ \\
    \end{tabular}
\end{ruledtabular}
\end{table*}
\begin{table*}[htbp]
\caption{
    Physical resource estimates for the ruthenium-based catalyst models, listed for structures VIII, VIII-IX, IX, and XVIII (second half of the catalytic cycle~\cite{Von_Burg2021-du}), across three initial-state infidelity parameter values: (a) $\xi=0.1$, (b) $\xi=0.05$, and (c) $\xi=0.01$. 
    Columns include physical qubits per QPU ($\mathcal{Q}_{\rm QPU}$), maximum per-shot runtime ($\mathcal{T}_{\rm max}$), single-QPU time-to-solution ($\mathcal{T}_{\rm total}$), maximum coherent spacetime volume ($\mathcal{V}_{\rm coh}^{\rm max}$), total  spacetime volume ($\mathcal{V}_{\rm total}$), QPU parallelism ($k^\ast$) under $\mathcal{Q}_{\rm budget}=5\times10^5$, and multi-QPU time-to-solution using $k^\ast$ QPUs.
    Detailed definitions for these parameters and the underlying assumptions can be found in Appendix~\ref{append:physical_resource}.
}
\label{tab:co2-cat_2_physical}
\begin{ruledtabular}
\begin{tabular}{l c c c c c c c}
    Molecule ID & $\mathcal{Q}_{\rm QPU}$ & \makecell{$\mathcal{T}_{\rm max}$ [\si{\second}]} & \makecell{$\mathcal{T}_{\rm total}$ [days]} & \makecell{$\mathcal{V}_{\rm coh}^{\rm max}$\\$[$qubit-hour$]$} & \makecell{$\mathcal{V}_{\rm total}$\\$[$qubit-hour$]$} &  \makecell{$k^*$} & \makecell{Time-to-solution\\($k^\ast$ QPUs) [days]} \\
    \midrule
    \multicolumn{8}{c}{(a) $\xi=0.1$}\\
    \midrule
Ru-VIII (Sm) & $9.24\times 10^{3}$ & $1.53\times 10^{-2}$ & $2.47\times 10^{-4}$ & $3.93\times 10^{-2}$ & $5.47\times 10^{1}$ & 54 & $4.37\times 10^{-6}$ \\
Ru-VIII (Md) & $1.88\times 10^{5}$ & $2.30\times 10^{3}$ & $5.57\times 10^{1}$ & $1.21\times 10^{5}$ & $2.52\times 10^{8}$ & 2 & $2.66\times 10^{1}$ \\
Ru-VIII (Lg) & $5.14\times 10^{5}$ & $7.39\times 10^{4}$ & $4.73\times 10^{5}$ & $1.05\times 10^{7}$ & $5.83\times 10^{12}$ & 1 & $4.73\times 10^{5}$ \\
Ru-VIII-IX (Sm) & $1.69\times 10^{4}$ & $1.74\times 10^{-1}$ & $4.58\times 10^{-3}$ & $8.16\times 10^{-1}$ & $1.86\times 10^{3}$ & 29 & $1.51\times 10^{-4}$ \\
Ru-VIII-IX (Md) & $1.55\times 10^{5}$ & $1.33\times 10^{3}$ & $3.37\times 10^{1}$ & $5.75\times 10^{4}$ & $1.26\times 10^{8}$ & 3 & $1.11\times 10^{1}$ \\
Ru-VIII-IX (Lg) & $4.71\times 10^{5}$ & $3.93\times 10^{4}$ & $3.42\times 10^{4}$ & $5.14\times 10^{6}$ & $3.86\times 10^{11}$ & 1 & $3.42\times 10^{4}$ \\
Ru-IX (Sm) & $9.87\times 10^{4}$ & $6.73\times 10^{2}$ & $9.97\times 10^{0}$ & $1.84\times 10^{4}$ & $2.36\times 10^{7}$ & 5 & $1.95\times 10^{0}$ \\
Ru-IX (Md) & $1.72\times 10^{5}$ & $2.25\times 10^{3}$ & $5.44\times 10^{1}$ & $1.08\times 10^{5}$ & $2.25\times 10^{8}$ & 2 & $2.62\times 10^{1}$ \\
Ru-IX (Lg) & $4.27\times 10^{5}$ & $2.97\times 10^{4}$ & $2.25\times 10^{4}$ & $3.53\times 10^{6}$ & $2.31\times 10^{11}$ & 1 & $2.25\times 10^{4}$ \\
Ru-XVIII (Sm) & $1.69\times 10^{4}$ & $2.08\times 10^{-1}$ & $5.19\times 10^{-3}$ & $9.74\times 10^{-1}$ & $2.10\times 10^{3}$ & 29 & $1.73\times 10^{-4}$ \\
Ru-XVIII (Md) & $1.18\times 10^{5}$ & $6.65\times 10^{2}$ & $1.04\times 10^{1}$ & $2.17\times 10^{4}$ & $2.93\times 10^{7}$ & 4 & $2.51\times 10^{0}$ \\
Ru-XVIII (Lg) & $3.89\times 10^{5}$ & $2.57\times 10^{4}$ & $9.76\times 10^{3}$ & $2.78\times 10^{6}$ & $9.12\times 10^{10}$ & 1 & $9.76\times 10^{3}$ \\
    \midrule
    \multicolumn{8}{c}{(b) $\xi=0.05$}\\
    \midrule
Ru-VIII (Sm) & $9.24\times 10^{3}$ & $8.12\times 10^{-3}$ & $2.68\times 10^{-4}$ & $2.08\times 10^{-2}$ & $5.94\times 10^{1}$ & 54 & $4.84\times 10^{-6}$ \\
Ru-VIII (Md) & $1.88\times 10^{5}$ & $1.00\times 10^{3}$ & $4.41\times 10^{1}$ & $5.25\times 10^{4}$ & $1.99\times 10^{8}$ & 3 & $2.14\times 10^{1}$ \\
Ru-VIII (Lg) & $4.45\times 10^{5}$ & $2.06\times 10^{4}$ & $1.54\times 10^{4}$ & $2.54\times 10^{6}$ & $1.64\times 10^{11}$ & 1 & $1.54\times 10^{4}$ \\
Ru-VIII-IX (Sm) & $1.69\times 10^{4}$ & $7.82\times 10^{-2}$ & $3.50\times 10^{-3}$ & $3.66\times 10^{-1}$ & $1.42\times 10^{3}$ & 29 & $1.17\times 10^{-4}$ \\
Ru-VIII-IX (Md) & $1.32\times 10^{5}$ & $5.87\times 10^{2}$ & $2.40\times 10^{1}$ & $2.15\times 10^{4}$ & $7.58\times 10^{7}$ & 3 & $7.90\times 10^{0}$ \\
Ru-VIII-IX (Lg) & $4.08\times 10^{5}$ & $1.17\times 10^{4}$ & $3.98\times 10^{3}$ & $1.33\times 10^{6}$ & $3.90\times 10^{10}$ & 1 & $3.98\times 10^{3}$ \\
Ru-IX (Sm) & $9.87\times 10^{4}$ & $3.03\times 10^{2}$ & $9.53\times 10^{0}$ & $8.30\times 10^{3}$ & $2.26\times 10^{7}$ & 6 & $1.88\times 10^{0}$ \\
Ru-IX (Md) & $1.46\times 10^{5}$ & $3.66\times 10^{2}$ & $5.96\times 10^{1}$ & $1.48\times 10^{4}$ & $2.08\times 10^{8}$ & 3 & $1.98\times 10^{1}$ \\
Ru-IX (Lg) & $4.27\times 10^{5}$ & $9.86\times 10^{3}$ & $2.64\times 10^{3}$ & $1.17\times 10^{6}$ & $2.70\times 10^{10}$ & 1 & $2.64\times 10^{3}$ \\
Ru-XVIII (Sm) & $1.69\times 10^{4}$ & $8.98\times 10^{-2}$ & $4.05\times 10^{-3}$ & $4.21\times 10^{-1}$ & $1.64\times 10^{3}$ & 29 & $1.36\times 10^{-4}$ \\
Ru-XVIII (Md) & $9.80\times 10^{4}$ & $5.49\times 10^{1}$ & $1.27\times 10^{1}$ & $1.49\times 10^{3}$ & $2.98\times 10^{7}$ & 5 & $2.52\times 10^{0}$ \\
Ru-XVIII (Lg) & $3.89\times 10^{5}$ & $8.14\times 10^{3}$ & $1.50\times 10^{3}$ & $8.81\times 10^{5}$ & $1.40\times 10^{10}$ & 1 & $1.50\times 10^{3}$ \\
    \midrule
    \multicolumn{8}{c}{(c) $\xi=0.01$}\\
    \midrule
Ru-VIII (Sm) & $6.61\times 10^{3}$ & $1.76\times 10^{-3}$ & $1.14\times 10^{-3}$ & $3.24\times 10^{-3}$ & $1.81\times 10^{2}$ & 75 & $1.52\times 10^{-5}$ \\
Ru-VIII (Md) & $1.33\times 10^{5}$ & $2.17\times 10^{1}$ & $3.61\times 10^{1}$ & $8.01\times 10^{2}$ & $1.15\times 10^{8}$ & 3 & $1.20\times 10^{1}$ \\
Ru-VIII (Lg) & $3.82\times 10^{5}$ & $1.57\times 10^{3}$ & $1.87\times 10^{3}$ & $1.67\times 10^{5}$ & $1.71\times 10^{10}$ & 1 & $1.87\times 10^{3}$ \\
Ru-VIII-IX (Sm) & $1.27\times 10^{4}$ & $1.58\times 10^{-2}$ & $1.07\times 10^{-2}$ & $5.55\times 10^{-2}$ & $3.26\times 10^{3}$ & 39 & $2.75\times 10^{-4}$ \\
Ru-VIII-IX (Md) & $1.10\times 10^{5}$ & $1.01\times 10^{1}$ & $1.22\times 10^{1}$ & $3.06\times 10^{2}$ & $3.20\times 10^{7}$ & 4 & $3.04\times 10^{0}$ \\
Ru-VIII-IX (Lg) & $3.50\times 10^{5}$ & $1.05\times 10^{3}$ & $1.05\times 10^{3}$ & $1.02\times 10^{5}$ & $8.84\times 10^{9}$ & 1 & $1.05\times 10^{3}$ \\
Ru-IX (Sm) & $6.73\times 10^{4}$ & $2.49\times 10^{0}$ & $2.87\times 10^{0}$ & $4.65\times 10^{1}$ & $4.64\times 10^{6}$ & 7 & $4.10\times 10^{-1}$ \\
Ru-IX (Md) & $1.21\times 10^{5}$ & $1.98\times 10^{1}$ & $2.31\times 10^{1}$ & $6.67\times 10^{2}$ & $6.72\times 10^{7}$ & 4 & $5.77\times 10^{0}$ \\
Ru-IX (Lg) & $3.66\times 10^{5}$ & $6.67\times 10^{2}$ & $7.79\times 10^{2}$ & $6.78\times 10^{4}$ & $6.84\times 10^{9}$ & 1 & $7.79\times 10^{2}$ \\
Ru-XVIII (Sm) & $1.27\times 10^{4}$ & $1.72\times 10^{-2}$ & $1.19\times 10^{-2}$ & $6.07\times 10^{-2}$ & $3.61\times 10^{3}$ & 39 & $3.05\times 10^{-4}$ \\
Ru-XVIII (Md) & $8.02\times 10^{4}$ & $2.87\times 10^{0}$ & $3.05\times 10^{0}$ & $6.39\times 10^{1}$ & $5.88\times 10^{6}$ & 6 & $5.09\times 10^{-1}$ \\
Ru-XVIII (Lg) & $3.34\times 10^{5}$ & $5.19\times 10^{2}$ & $6.06\times 10^{2}$ & $4.81\times 10^{4}$ & $4.85\times 10^{9}$ & 1 & $6.05\times 10^{2}$ \\
    \end{tabular}
\end{ruledtabular}
\end{table*}

\section{Detailed performance of STAR-magic mutation \label{append:resource_assumptions}}

Here, we describe some technical details regarding the STAR-magic mutation (SMM) protocol~\cite{Toshio2026}, necessary to derive our resource estimates in Sec.~\ref{sec:resource_estimation}.

\begin{figure}[tbp]
  \centering
  \includegraphics[width=0.48\textwidth]{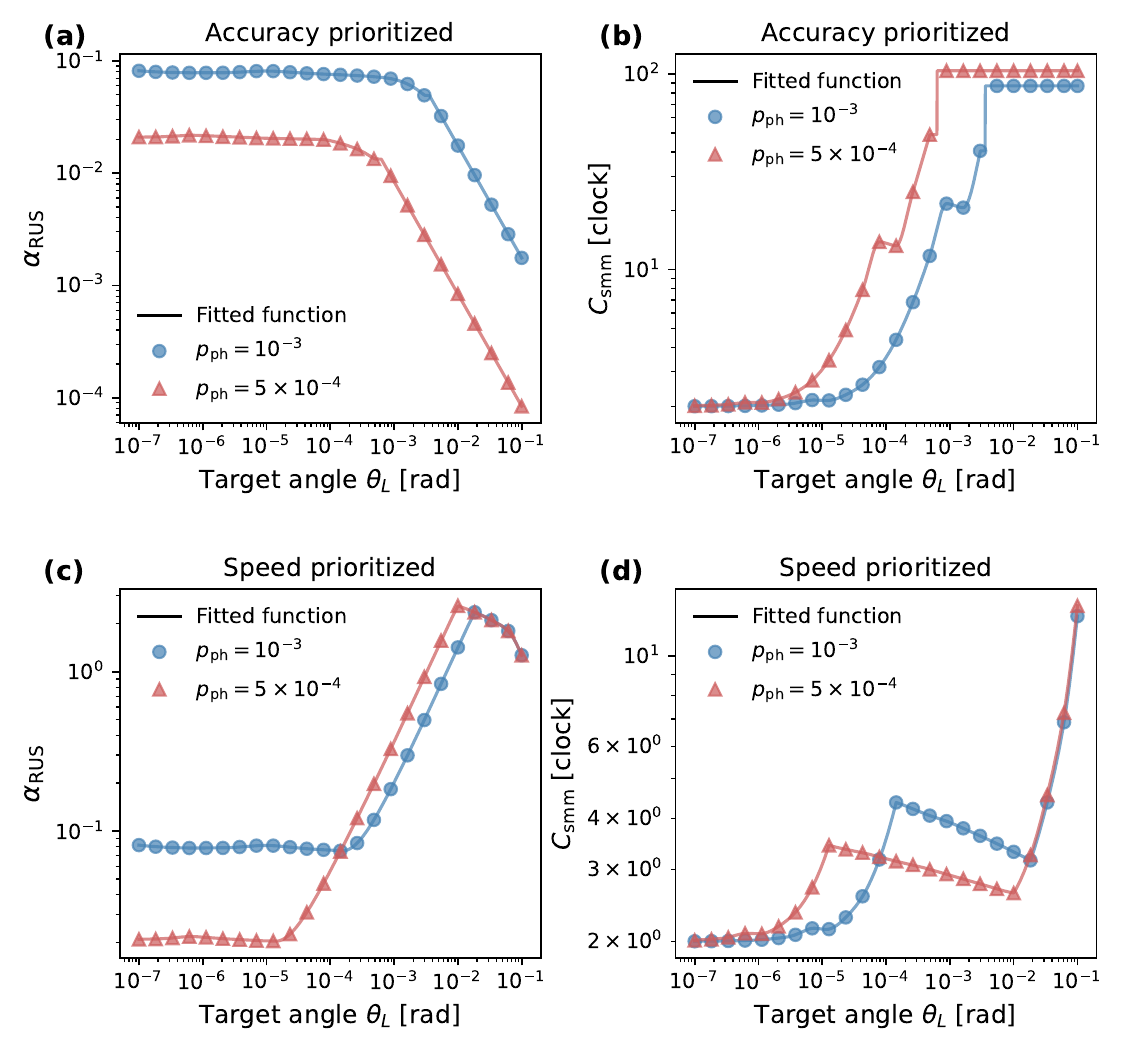}
    \caption{Angular dependence of performance parameters for the SMM protocol. All data are obtained for a code distance of $d=21$, with similar trends expected for other code distances. The numerical calculations utilized performance data for magic state cultivation in Ref.~\cite{Gidney2024-lq}. Blue circles (red triangles) represent the calculated data points for $p_{\rm ph}=10^{-3}$ ($p_{\rm ph}=5\times10^{-4}$), and solid lines indicate the cubic spline fitting curves.
    Panels (a) and (b) show the logical error factor $\alpha_{\rm RUS}(\theta_L)$ and execution time $C_{\rm smm}(\theta_L)$ in clock cycles, respectively, for the ``accuracy prioritized" setting. Panels (c) and (d) display $\alpha_{\rm RUS}(\theta_L)$ and $C_{\rm smm}(\theta_L)$ in clock cycles for the ``speed prioritized" setting.}
    \label{fig:star_magic_mutation}
\end{figure}

As discussed in Sec.~\ref{subsec:STAR}, both the logical error factor $\alpha_{\rm RUS}(\theta_L)$ and the execution time $C_{\rm smm}(\theta_L)$ associated with a single analog rotation gate $\hat{R}_P(\theta_L)=e^{i\theta_L\hat{P}}$ are sensitive to the adjustment of the threshold angle $\theta_{\rm th}$ on the SMM architecture.
Consequently, while optimizing performance specifically for each rotation angle $\theta_L$ is desirable, such fine-grained optimization is computationally demanding. For simplicity in this study, we therefore conducted resource estimation using two distinct patterns of threshold angle adjustment. The functional dependencies of the performance parameters ($\alpha_{\rm RUS}(\theta_L)$ and $C_{\rm smm}(\theta_L)$) for each pattern are shown in Fig.~\ref{fig:star_magic_mutation}.

The first pattern, termed ``accuracy prioritized," involves an implementation that emphasizes minimizing the logical error rate $P_L(\theta_L)=\alpha_{\rm RUS}(\theta_L)\cdot \theta_L \cdot p_{\rm ph}$. In this setting, the threshold angle $\theta_{\rm th}$ is determined for each $\theta_L$ by selecting the optimal integer $n$ (from $\theta_{\rm th}=2^n\theta_L$, $n\in \mathbb{N}$) that yields the lowest $P_L(\theta_L)$. 
The second pattern, ``speed prioritized," focuses on accelerating the execution of large-angle rotation gates. For this setting, the threshold angle is chosen to satisfy $\theta_{\rm th}/\theta_L \geq 2^5$ when $p_{\rm ph}=10^{-3}$, and $\theta_{\rm th}/\theta_L \geq 2^6$ when $p_{\rm ph}=5\times10^{-4}$. By actively setting the threshold angle through these criteria, the RUS protocol is configured to delay the transition to the slow $T$-gate execution, effectively accelerating the overall execution time at the cost of accuracy~\cite{Toshio2026}.
Figure~\ref{fig:star_magic_mutation} clearly demonstrates the trade-off existing between the error rate and execution speed, particularly in the regime with relatively large rotation angles, i.e., $\theta_L \gtrsim 10^{-4}$ for $p_{\rm ph}=10^{-3}$ and $\theta_L \gtrsim 10^{-5}$ for $p_{\rm ph}=5\times10^{-4}$.

During resource estimation in Sec.~\ref{sec:resource_estimation}, we estimate the physical space-time cost for both patterns and ultimately select the configuration that yields the shorter time-to-solution. A smaller logical error rate can occasionally result in a shorter runtime, as the sampling overhead from the error mitigation protocol depends on $P_L(\theta_L)$ (see Eq.~\eqref{eq:pec_gamma_total}).
For both patterns, the angular dependencies of $\alpha_{\rm RUS}(\theta_L)$ and $C_{\rm smm}(\theta_L)$ were determined by fitting the data presented in Fig.~\ref{fig:star_magic_mutation} using the cubic spline method implemented in the \texttt{SciPy} package~\cite{2020SciPy-NMeth}.

\section{Resource scaling at reduced physical error rates \label{append:error_rate_sensitivity}}

\begin{figure}[htbp]
  \centering
  \includegraphics[width=0.49\textwidth]{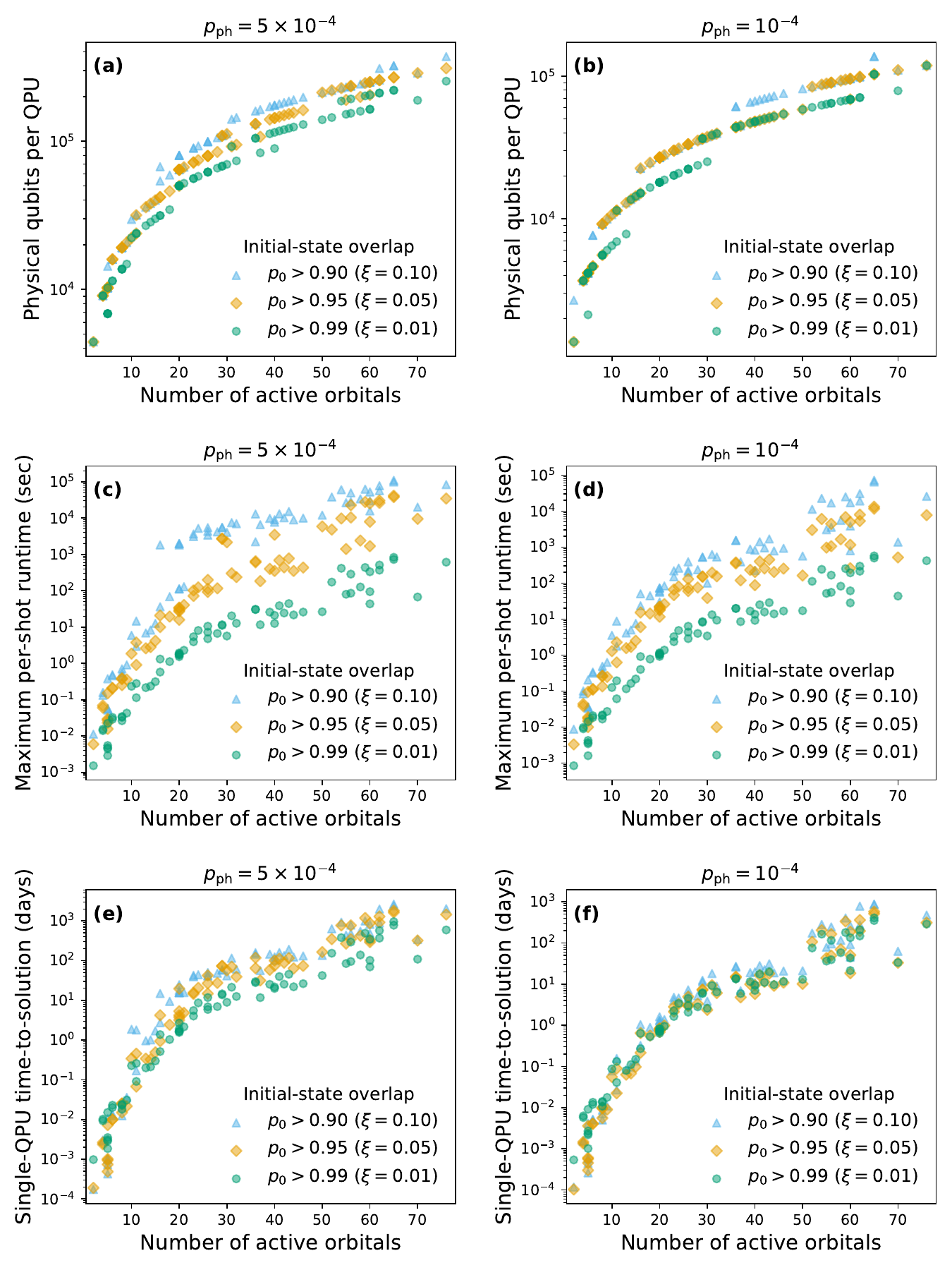}
    \caption{Scaling of core resource components for partially randomized RPE on SMM-based partially fault-tolerant quantum computing architecture. The plots show (a,b) the number of physical qubits required per QPU, (c,d) the maximum per-shot runtime $\mathcal{T}_{\rm max}$, and (e,f) the total time-to-solution assuming execution on a single QPU $\mathcal{T}_{\rm total}$, all as a function of the number of active orbitals. For each pair, the first panel (a,c,e) corresponds to a physical error rate of $p_{\rm ph}=5\times10^{-4}$, and the second (b,d,f) to $p_{\rm ph}=10^{-4}$. In all panels, the shape of the markers indicates the initial-state infidelity parameter $\xi$, shown together with the corresponding lower bound on the initial-state overlap $p_0=|\braket{\psi_0|\psi}|^2$. Panels (c,d) and (e,f) use different time units to emphasize the distinct physical meanings of per-circuit execution cost and overall computational effort. All data points are obtained for the UWC-optimized Hamiltonian representations of benchmark molecular active-space models detailed in Appendix~\ref{append:models_detail}, including hydrogen chains, iron-sulfur clusters, P450 active sites, and ruthenium-based catalysts. The target accuracy $\epsilon$ is set to be $\epsilon=1.6$ mHa. 
    }
  \label{fig:resource_scaling_small_p}
\end{figure}

In this Appendix, we present detailed resource estimates assuming reduced physical error rates of $p_{\rm ph} = 5\times10^{-4}$ and $p_{\rm ph} = 10^{-4}$. These results complement the representative values shown in Table~\ref{tab:error_rate_sensitivity} and provide complete scaling trends across all benchmark systems.

Figure~\ref{fig:resource_scaling_small_p} displays the corresponding physical resource estimates for the full benchmark set as a function of the number of active orbitals $N$. These estimates are provided for a range of initial-state infidelity parameters $\xi \in \{0.1, 0.05, 0.01 \}$.
Similar to Fig.~\ref{fig:resource_scaling} (which shows results for $p_{\rm ph}=10^{-3}$), we plot the physical qubit requirement per QPU ($\mathcal{Q}_{\rm QPU}$), the maximum per-shot circuit execution time ($\mathcal{T}_{\rm max}$), and the single-QPU time-to-solution ($\mathcal{T}_{\rm total}$) for all molecular instances considered in this work, employing $p_{\rm ph}=5\times10^{-4}$ or $p_{\rm ph}=10^{-4}$.
The overall scaling behaviors exhibit patterns similar to those observed in the baseline results for $p_{\rm ph}=10^{-3}$ (Fig.~\ref{fig:resource_scaling}).
However, the absolute values of $\mathcal{Q}_{\rm QPU}$ and $\mathcal{T}_{\rm total}$ are substantially lower compared to those obtained with $p_{\rm ph}=10^{-3}$.
In contrast, the value of $\mathcal{T}_{\rm max}$ shows only a marginal change under reduced physical error rates. This is because the per-shot runtime is primarily determined by the gate count, a factor largely independent of $p_{\rm ph}$.

Furthermore, we highlight a crucial difference in $\mathcal{T}_{\rm total}$ for larger orbital regimes ($N \gtrsim 50$) under higher infidelity ($\xi=0.1,0.05$).
Unlike the exponential increase observed in Fig.~\ref{fig:resource_scaling}(c) for $p_{\rm ph}=10^{-3}$, $\mathcal{T}_{\rm total}$ in Figs.~\ref{fig:resource_scaling_small_p}(c) and~\ref{fig:resource_scaling_small_p}(f) do not exhibit such a dramatic rise in this regime. This improvement is attributed to the total logical error rate ($P_{\rm total}$, Eq.~\eqref{eq:P_total_star}) being at least one order of magnitude smaller than in the $p_{\rm ph}=10^{-3}$ case.
Consequently, the sampling overhead of SMM (Eq.~\eqref{eq:pec_gamma_total}) does not increase exponentially, even for the deepest circuit case of $\xi=0.1$.
These results strongly suggest that reducing the physical error rate can extend the problem size solvable within realistic time-scales and under restricted qubit budgets, potentially broadening the feasibility window shown in Fig.~\ref{fig:feasibility_map}.

\section{Resource estimation for full-fledged fault-tolerant quantum computing \label{append:ftqc}}

\begin{table*}[t]
\caption{Reference resource estimates for qubitization-based QPE, utilizing spectral amplification with the DFTHC block encoding~\cite{Low2025-tb} under a full-fledged FTQC architecture. 
Physical qubits and time-to-solution are estimated using \texttt{Qualtran}~\cite{harrigan2024qualtran} with four CCZ factories and runtime-optimized code distances. Spacetime volume is obtained by multiplying the physical qubit count and the time-to-solution.  
Literature values are shown in parentheses. 
For physical qubits, the main value denotes the total count, with a breakdown into data and factory qubits given in brackets.  
Minor discrepancies with literature values may arise from simplifying assumptions in \texttt{Qualtran}.
}
\label{tab:ftqc_resource}
\begin{ruledtabular}
\begin{tabular}{l c c c c c c c}
    Molecule ID & \makecell{Orbitals} & \makecell{Logical\\qubits} & Toffoli gates & \makecell{Code distance \\ $(d_1,\,d_2 ,\, d)$} & \makecell{Physical qubits \\ \small{[ Data / Factory} ]} & \makecell{Time-to-solution\\ $[$hours$]$ } & \makecell{Spacetime volume\\$[$qubit-hour$]$}  \\
    \colrule
    $[\mathrm{2Fe\mathchar`-2S}]^{-2}$ & 20 & 463 & $3.99\times10^{7}$ & (15, 23, 27) & \makecell{$1.49 \times 10^6 $  \\ \small{[ $1.01\times10^6$ / $4.81\times10^5$ ]} } & 0.47 & $7.06\times10^5$ \\
    $[\mathrm{4Fe\mathchar`-4S}]^{-2}$ & 36 & 868 & $1.73\times10^{8}$ & (15, 27, 31) & \makecell{$3.03\times10^6$ \\ \small{[ $2.50\times10^6$ / $5.32\times10^5$]}} & 2.05 & $6.22\times10^6$\\
    FeMoco (Sm) & 54 & 1132 & $3.42\times10^{8}$ & (17, 25, 29) & \makecell{$3.46\times10^6$ ($3.2\times10^6$~\cite{Low2025-tb}) \\ \small{[ $2.86\times10^6$ / $6.04\times10^5$ ]}} & \makecell{4.57\\(2.7~\cite{Low2025-tb})} & $1.58\times10^7$\\
    FeMoco (Lg) & 76 & 1454 & $1.00\times10^{9}$ & (17, 27, 31) & \makecell{$4.82\times 10^6$ ($4.5\times10^6$~\cite{Low2025-tb}) \\ \small{[ $4.19\times10^6$ / $6.31\times10^5$} ]} & \makecell{13.37\\(8.6~\cite{Low2025-tb})} & $6.45\times10^7$\\
    \end{tabular}
\end{ruledtabular}
\end{table*} 

In this Appendix, we present resource estimation results for a full-fledged FTQC setting, serving as a reference for our findings in a partially fault-tolerant setting. Specifically, we conduct resource estimation using \texttt{Qualtran}~\cite{harrigan2024qualtran} for the state-of-the-art qubitization approach for molecular systems, proposed in Ref.~\cite{Low2025-tb}.

Table~\ref{tab:ftqc_resource} summarizes the estimated resources required to obtain ground-state energy estimates for representative iron-sulfur cluster models. The logical qubit count and Toffoli gate count are cited from Ref.~\cite{Low2025-tb}. These results were obtained by adopting the spectral amplification method for the sum-of-squares Hamiltonian representations, derived by combining double-factorized tensor hypercontraction and BLISS. The physical space-time costs (i.e., physical qubit count and time-to-solution) are estimated using \texttt{Qualtran}~\cite{harrigan2024qualtran} under the AutoCCZ cost model~\cite{Gidney2019-zv}. Specifically, we assume the use of four CCZ factories with a physical error rate of $p_{\rm ph}=10^{-3}$, a surface code cycle time of 1 \si{\micro\second}, and a control system reaction time of 10 \si{\micro\second}. This setup is consistent with that used in Refs.~\cite{Lee2021-tz,Low2025-tb}. The code distance for both distillation and data blocks is determined to minimize the total runtime cost through a grid search protocol, as implemented in \texttt{Qualtran}~\cite{harrigan2024qualtran}.
Here, $d_1$ and $d_2$ denote the surface-code distances used in the level-1 and level-2 magic-state distillation blocks, respectively, while $d$ denotes the code distance for the logical data qubits implementing the algorithm. 

We observe that over 1 million physical qubits are required even for the smallest $[\mathrm{2Fe\mathchar`-2S}]^{-2}$ model with 20 orbitals. 
Notably, the majority of these are allocated to the data block, including a large amount of ancilla qubits for the qubitization protocol, which alone requires approximately one million physical qubits. This highlights that even with future improvements in magic state distillation that could reduce factory footprints, such as exemplified by magic state cultivation~\cite{Gidney2024-lq,Hirano2025-mz,Hetenyi2026-ga}, the qubit requirement for data remains a formidable challenge, far exceeding tens of thousands of qubits.
This indicates that conventional full-fledged FTQC necessitates millions of qubits, a significantly larger hardware scale than that of the early-FTQC setting considered in this work, as discussed in the main text. On the other hand, the time-to-solution is on the order of hours, even for the largest 76-orbital FeMoco model. This is significantly lower than that for the early fault-tolerant setting, primarily reflecting the difference in the QPE algorithms employed. The QPE protocol for the full-fledged fault-tolerant setting is executed using a single-shot measurement of a quantum Fourier transform circuit with a good input state~\cite{kitaev1995quantum}. In contrast, the single-ancilla QPE  algorithms are executed via repeated measurements of the Hadamard test circuit, as described in Sec.~\ref{sec:single_ancilla_prpe}, thus generally requiring a longer computational time than full-fledged fault-tolerant QPE algorithms.

\FloatBarrier
\nocite{*}
\bibliography{main}

\end{document}